\documentclass[11pt]{article}
\usepackage{epsfig} 
\newcommand{\sect}[1]{ \section{#1} \setcounter{equation}{0} }

\newcommand{\Dslash}{D \! \! \! \! /}

\newcommand{\half}{\mbox{\small{$\frac{1}{2}$}}} 
\newcommand{\quarter}{\mbox{\small{$\frac{1}{4}$}}} 
\newcommand{\threequarters}{\mbox{\small{$\frac{3}{4}$}}} 
\newcommand{\MSbar}{\overline{\mbox{MS}}} 
 
\newcommand{\Nf}{N_{\!f}}
\newcommand{\NF}{N_{\!F}}
\newcommand{\NA}{N_{\!A}}

\begin{document}
\title{The one loop $\MSbar$ static potential in the Gribov-Zwanziger 
Lagrangian}
\author{J.A. Gracey, \\ Theoretical Physics Division, \\ 
Department of Mathematical Sciences, \\ University of Liverpool, \\ P.O. Box 
147, \\ Liverpool, \\ L69 3BX, \\ United Kingdom.} 
\date{} 
\maketitle 

\vspace{5cm} 
\noindent 
{\bf Abstract.} We compute the static potential in the Gribov-Zwanziger
Lagrangian as a function of the Gribov mass, $\gamma$, in the $\MSbar$ scheme 
in the Landau gauge at one loop. The usual gauge independent one loop 
perturbative static potential is recovered in the limit as 
$\gamma$~$\rightarrow$~$0$. By contrast the Gribov-Zwanziger static potential 
contains the term $\gamma^2/(p^2)^2$. However, the linearly rising potential in
coordinate space as a function of the radial variable $r$ does not emerge due 
to a compensating behaviour as $r$~$\rightarrow$~$\infty$. Though in the short
distance limit a dipole behaviour is present. We also demonstrate enhancement 
in the propagator of the bosonic localizing Zwanziger ghost field when the one
loop Gribov gap equation is satisfied. The explicit form of the one loop gap 
equation for the Gribov mass parameter is also computed in the MOM scheme and 
the zero momentum value of the renormalization group invariant effective 
coupling constant is shown to be the same value as that in the $\MSbar$ scheme. 

\vspace{-19cm}
\hspace{13.5cm}
{\bf LTH 842}

\newpage

\sect{Introduction.}

One of the main outstanding problems in quantum field theory is to understand
the confinement of the quarks and gluons associated with the strong nuclear
force. The underlying field theory is Yang-Mills or Quantum Chromodynamics
(QCD) which is a non-abelian gauge theory based on the $SU(3)$ colour group.
Whilst its ultraviolet properties such as asymptotic freedom, \cite{1,2}, are 
relatively straightforward to extract from perturbative calculations, it is the
infrared properties which are difficult to truly access. These lie at the heart
of the confinement problem. In early analyses of trying to understand how 
quarks and gluons are confined, it was recognised that the force separating 
these fundamental quanta of QCD behaved as a constant at large distances. In 
other words the potential energy at a colour charge separation distance of $r$ 
was proportional to $r$. Known as the linearly rising potential it has been 
studied using lattice regularization which goes beyond perturbation theory. 
Indeed there is widespread acceptance that such numerical analyses do produce 
a linearly rising potential. See, for example, \cite{3,4,5}. Another aspect of 
early attempts to understand such a potential has been clearly given in 
Mandelstam's review, \cite{6}, which emphasises the infrared behaviour of the 
gluon field, as well as in the work of others, \cite{7,8,9}. Briefly, in the 
ultraviolet the gluon behaves as a standard fundamental massless particle with 
a simple propagator proportional to $1/p^2$ where $p$ is the gluon momentum. 
However, such a behaviour can really only be valid asymptotically at high 
energy. Unlike the photon and overlooking for the moment the fact that the 
gluon is not a gauge invariant concept, if the gluon propagator's behaviour was
of this form {\em for all} momenta then it could naively but erroneously be 
identified as a real physical object contrary to its accepted confined 
property. Instead due to the non-linear nature of the non-abelian gauge theory 
the gluon's propagator must be modified non-perturbatively or dynamically to a 
non-fundamental form without a simple pole. Given this, one natural form which 
it might take in order to fit with the linear potential is a dipole, 
$1/(p^2)^2$. The justification for this is that considering the force between 
coloured sources, the exchange of such a gluon at low energies will 
{\em naturally} produce a linear potential in coordinate space after taking a 
simple Fourier transform. However, whilst such a behaviour is an ultimate goal,
the evidence for a dipole behaviour in the gluon propagator never actually
manifested itself in subsequent studies. Indeed with the use of more powerful 
computers, lattice gauge theories have been able to probe the infrared 
behaviour of the gluon propagator more deeply at low energy. Without getting 
drawn too deeply for the moment into the current debate over which of the 
scaling, \cite{10,11,12,13,14,15,16,17,18,19,20,21}, or decoupling solutions, 
\cite{22,23,24,25,26,27,28,29}, is correct, it is fair to say that the gluon 
does not appear to have a dipole form at low momenta. Instead the propagator 
freezes either to zero or a finite non-zero value for each of these respective 
current scenarios. Therefore, the extraction of the dipole behaviour is 
currently in abeyance.

The next key breakthrough in the pursuit of gluon confinement effectively
dawned with Gribov's seminal work noting the ambiguity in gauge fixing in the 
Landau gauge, \cite{10}. In essence gauge fixing locally in a non-abelian gauge
theory can be achieved uniquely but globally one will always end up with gauge 
or Gribov copies hindering the process. This was resolved by Gribov, \cite{10},
by restricting the path integral measure to the first Gribov region, $\Omega$, 
defined as that region encompassing the trivial gauge solution,
$A^a_\mu$~$=$~$0$, where $A^a_\mu$ is the gluon field or gauge potential, such
that the Faddeev-Popov operator has strictly positive eigenvalues. This 
restriction of the path integral produces a radically different gluon 
propagator which is clearly non-fundamental and vanishes in the infrared limit.
Moreover, a mass parameter is introduced, known as the Gribov mass and denoted 
by $\gamma$, which is not an independent parameter of the theory. Indeed 
$\gamma$ satisfies a gap equation which can be evaluated at one loop, 
\cite{10}. One major consequence is that the propagator of the Faddeev-Popov 
ghost, which is required for the Landau gauge fixing, also has a modification. 
Instead of a simple $1/p^2$ form, it behaves as $1/(p^2)^2$ as 
$p^2$~$\rightarrow$~$0$. Whilst this is a dipole, a Faddeev-Popov ghost clearly
cannot be exchanged between colour sources as it is Grassmann in nature as it 
is required to restore unitarity. Subsequent progress in this area was via a 
series of articles in the main by Zwanziger and collaborators, 
\cite{11,12,13,14,15,16,17,18,19,20,21}. Briefly, the drawback of the Gribov 
semi-classical analysis is that the resultant effective Lagrangian is non-local
and therefore essentially inadequate for practical quantum computations. 
However, in a series of articles Zwanziger managed to localize the non-locality
at the expense of introducing extra localizing fields, \cite{12,13,17,18}. The 
resulting Lagrangian was renormalizable, \cite{18,30,31}, meaning that it was 
possible to perform calculations with it. Indeed its renormalization is such 
that the known ultraviolet structure of Yang-Mills or QCD is totally unaltered 
by the extra localizing fields. Instead they only effectively play a role in 
the infrared. Subsequently, the two loop $\MSbar$ extension to Gribov's gap 
equation was determined in \cite{32} verifying the Faddeev-Popov ghost 
enhancement at two loops and gluon suppression at one loop, \cite{33}. The 
former property is not inconsistent with the Kugo-Ojima confinement criterion, 
\cite{34,35}, derived for a BRST invariant Yang-Mills theory. Indeed it was 
shown that the Grassmann localizing ghost also was enhanced at two loops, 
\cite{33}, giving a propagator of dipole form but it is equally not directly 
relevant for confinement of coloured sources for the same reason as the 
Faddeev-Popov ghost.

Given the potential for the Gribov-Zwanziger Lagrangian to be a candidate for
understanding confined gluons we now indicate the primary aim of this article.
The explicit calculations of \cite{32,33} at one and two loops have opened the 
possibility of calculating the potential between two coloured sources using the
Gribov-Zwanziger localized Lagrangian, \cite{12,13,17,18}. One aim is to see
whether the behaviour of the potential is significantly different from that
computed in the usual version of QCD. Indeed this was first discussed by
Susskind in \cite{36} and other authors at around the same time, \cite{37,38}. 
They considered the Wilson loop definition and examined the energy between two
coloured sources fixed in time but at a spatial separation. This is known as
the static limit. An advantage of the Wilson loop is that it is gauge invariant
and in principle one can extract the resulting potential in any gauge. 
Originally this was achieved in the Feynman gauge at one loop in the $\MSbar$
scheme, \cite{36,37,38}. Subsequently, the two loop static potential was 
determined in the same gauge in \cite{39,40}. This calculation was subsummed 
into a more general computation by Schr\"{o}der in \cite{41,42}. There the full
two loop potential was constructed in an arbitrary linear covariant gauge in 
ordinary (massless) perturbation theory. It clearly demonstrated the explicit 
cancellation of the linear gauge fixing parameter (and en route corrected a
minor error in the original expression given in \cite{39,40}). The two loop 
static potential for a general colour state emerged later in \cite{43}. More 
recently the three loop static potential has been the subject of interest with 
both the quark, \cite{44,45,46}, and purely gluonic contributions now
available, \cite{47,48}. Since such an extensive formalism already exists (and 
is comprehensively reviewed in \cite{49}) it is a rather straightforward 
exercise to apply it to the Gribov-Zwanziger Lagrangian at one loop. At the 
very least any perturbative potential, which will depend on $\gamma$, is needed
since it would have to match onto the behaviour of the potential beyond the 
perturbative approximation. At this point it is worth noting that Zwanziger 
considered the Wilson loop in the Gribov-Zwanziger context using a lattice 
approach in \cite{19}. There it was argued that a linear rising potential could
emerge if there was ghost enhancement and gluon suppression, including gluon 
propagator freezing to a non-zero value. Moreover, the string tension was 
proportional to a combination of the zero momentum values of the gluon and 
Faddeev-Popov ghost form factors. 

Although the static potential forms a main part of our article, we will also 
consider the structure of all the contributing fields to see whether a dipole 
exchange between coloured sources could somehow emerge in the Gribov-Zwanziger 
Lagrangian. In making this previous statement we have been careful in our 
wording and not mentioned the gluon. The reason for this is simple. The premise
that it is solely the gluon field itself which is responsible for confinement 
in the Gribov-Zwanziger context may need to be refined. This is primarily due 
to the appearance of the Zwanziger localizing ghosts in the work of 
\cite{12,13,17,18}. These play no role in the ultraviolet dynamics of QCD but 
do become important in the infrared. There is a {\em bosonic} Zwanziger ghost 
which is spin-$1$ and carries non-abelian colour charge. In some ways it could 
be regarded as a gluon component since its equation of motion implies it is a 
non-local projection of $A^a_\mu$. However, in the full Gribov-Zwanziger 
Lagrangian we regard it as a separate entity. Therefore, in principle the 
appearance of a dipole behaviour in the infrared between coloured sources could
also derive from other spin-$1$ fields aside from the gluon itself. Indeed this
is indicated very strongly in Zwanziger's recent article, \cite{50}, where 
Dyson Schwinger techniques are applied to the bosonic localizing ghost to 
produce an enhancement akin to that of both Grassmann ghost fields. Though the 
enhancement is actually beyond dipole being $1/(p^2)^3$ as 
$p^2$~$\rightarrow$~$0$. However, since it is non-Grassmann, \cite{50}, it is 
evidently a much better candidate for an exchange field between coloured 
sources. Therefore, in the current context of the full one loop $\MSbar$ static
potential in the Gribov-Zwanziger Lagrangian we will speculate how it might be 
possible to go beyond the one loop potential and qualitatively produce a 
linearly rising potential. En route we will record some new properties of the 
underlying Gribov gap equation. As emphasised by Zwanziger, in the 
Gribov-Zwanziger context the theory can only be regarded as a gauge theory when
the Gribov mass, $\gamma$, explicitly satisfies this equation. In the original 
calculations of Gribov, \cite{10}, to produce the Faddeev-Popov ghost 
enhancement the main lesson was that the gap equation was central to seeing 
towards the infrared. To a lesser extent it also played a role in qualitatively
producing the freezing to a non-zero value, \cite{33}, of a renormalization 
group invariant definition of the coupling constant based on the Landau gauge 
properties of the gluon ghost vertex, \cite{51}. Therefore, we believe that the
gap equation, which is derived from the condition defining the first Gribov 
region, should be a central requirement to producing the infrared property of a
linearly rising potential. In our calculations we will follow the more recent
formulation of the Gribov-Zwanziger Lagrangian of \cite{50} where the 
localizing bosonic ghosts were resolved into their real and imaginary parts.
Whilst this will be different from the earlier loop computations of 
\cite{32,33}, it transpires that those results are not correct since the 
propagator of the real part of the bosonic localizing ghost was treated 
erroneously. Therefore, throughout the discussion we will re-address some of 
the results of \cite{32,33} and provide the correct details. It turns out that 
the main features such as one loop gluon suppression, two loop Faddeev-Popov 
ghost enhancement and the freezing of a renormalization group invariant 
effective coupling constant to a finite non-zero value remain unaltered. It is 
only the actual numerical details which are revised since the omission affects 
the finite parts of Feynman graphs upon which there are no non-trivial 
independent checks. Finally, given the current interest in the alternative 
infrared structure known as the decoupling solution,
\cite{22,23,24,25,26,27,28,29}, we will briefly discuss its status within the 
static potential approach. Indeed in \cite{52} a novel way of producing the 
decoupling solution within the Gribov-Zwanziger Lagrangian was considered and 
later tested in explicit calculations, \cite{53}.

The paper is organised as follows. In section $2$ we review the static 
potential formalism of \cite{36,37,38,39,40,41,42} as well as the necessary 
ingredients from the Gribov-Zwanziger Lagrangian, \cite{12,13,17,18}. Section
$3$ is devoted to the formal construction of the one loop corrections to the
propagators in the reformulation of the Gribov-Zwanziger Lagrangian in terms of
the real and imaginary parts of the localizing bosonic ghost discussed in 
\cite{50}. The explicit one loop $\MSbar$ static potential is constructed in 
section $4$ prior to considering several new calculational aspects of the 
Gribov gap equation for $\gamma$ in section $5$ including the correct two loop
$\MSbar$ expression. The role the original gap equation plays in the one loop 
enhancement of the bosonic localizing ghosts is discussed in section $6$ and 
its implications for trying to extract the linearly rising potential 
qualitatively in the Gribov-Zwanziger static potential are also considered. 
Section $7$ focuses on considering power corrections to the potential and other
quantities which can be computed in the Gribov-Zwanziger Lagrangian as well as 
elementary ideas on the underlying properties of higher order Feynman graphs. 
We give our conclusions in section $8$. Several appendices are provided. The 
first details the decomposition of products of colour group structure functions
into a standard basis which is required in the discussion of the bosonic ghost 
enhancement. The next two appendices provide the explicit {\em exact} one loop 
corrections respectively to the transverse and longitudinal parts of all the 
$2$-point functions of the spin-$1$ fields in the Gribov-Zwanziger Lagrangian. 
The longitudinal parts are presented as they are relevant to the current debate
on the BRST structure of the Gribov-Zwanziger Lagrangian. The final appendix 
records the explicit values of scalar amplitudes in the decomposition of the 
correlation function of a Lorentz tensor operator involving the field strength.
It extends the power correction computation of the analogous scalar operator 
considered in section $7$.

\sect{Formalism.}

We begin by briefly summarizing the static potential formalism in both the
original QCD and Gribov-Zwanziger contexts. The key is the definition of the
Wilson loop where the loop is taken to be a rectangle. The spatial extent is
length $r$ and the time interval is denoted by $T$. When the temporal side of
the rectangle is very much in excess of the spatial extent, $T$~$\gg$~$r$, then
the potential, $V(r)$, between two heavy static colour sources is given by
\begin{equation}
V(r) ~=~ -~ \lim_{T\rightarrow \infty} \frac{1}{iT} \ln \left\langle 0 \left|
\, \mbox{Tr} \, {\cal P} \, \exp \left( ig \oint dx^\mu \, A^a_\mu T^a \right) 
\right| 0 \right\rangle ~.
\label{potdef}
\end{equation}
Here ${\cal P}$ denotes the path ordering prescription, $A^a_\mu$ is the gluon
field and $T^a$ are the colour group generators satisfying 
\begin{equation}
[ T^a, T^b ] ~=~ i f^{abc} T^c
\end{equation}
where $f^{abc}$ are the structure constants. Throughout our discussion on the
definition and properties of the static potential we make the same general
assumptions as have been discussed extensively in \cite{36,37,38,39,40,41,42}
such as gauge invariance of $V(r)$. Also, for instance, it is accepted that
the potential of (\ref{potdef}) can be reformulated in terms of a functional
integral with an external colour source, $J^a_\mu(x)$. In other words
\begin{equation}
V(r) ~=~ -~ \lim_{T\rightarrow\infty} \frac{1}{iT} 
\frac{\mbox{tr}Z[J]}{\mbox{tr}Z[0]}
\label{potdefz}
\end{equation}
where formally 
\begin{equation}
Z[J] ~=~ \int {\cal D} A_\mu \, {\cal D} \psi \, {\cal D} \bar{\psi} \, 
{\cal D} c \, {\cal D} \bar{c} ~ \exp \left[ -\, \int d^4 x \, \left( L ~+~ 
J^{a\,\mu} A^a_\mu \right) \right]
\label{actj}
\end{equation}
and the Lagrangian $L$ will be discussed later in detail but will either be the
original QCD Lagrangian or that of the Gribov-Zwanziger theory. The 
denominator, $Z[0]$, is included as a normalization. The other main assumption,
which also occurs in the usual QCD case and has been discussed in 
\cite{36,37,38,39,40,41}, is that of the exponentiation of the values of the
actual Feynman diagrams in order to properly extract the static potential
implicit in the definition, (\ref{potdef}). This has been proved to all orders
in the abelian case. For the non-abelian case the exponentiation for the Wilson
loop for ordinary QCD has been demonstrated in \cite{54,55}. Although we assume 
that that analysis extends to the Gribov-Zwanziger case we note that for our 
one loop computation, the exponentiation does indeed occur which is sufficient
for our present purpose.

For the moment we follow the early approach of \cite{36,37,38} for the 
canonical formalism of QCD and the Lagrangian is given by 
\begin{equation} 
L^{\mbox{\footnotesize{QCD}}} ~=~ -~ \frac{1}{4} G_{\mu\nu}^a 
G^{a \, \mu\nu} ~-~ \frac{1}{2\alpha} (\partial^\mu A^a_\mu)^2 ~-~ 
\bar{c}^a \partial^\mu D_\mu c^a ~+~ i \bar{\psi}^{iI} \Dslash \psi^{iI} 
\label{lqcd}
\end{equation} 
where $G^a_{\mu\nu}$ is the field strength, the covariant derivative, $D_\mu$,
is defined by
\begin{eqnarray}
D_\mu c^a &=& \partial_\mu c^a ~-~ g f^{abc} A^b_\mu c^c \nonumber \\
D_\mu \psi^{iI} &=& \partial_\mu \psi^{iI} ~+~ i g T^a_{IJ} A^a_\mu \psi^{iJ} 
\end{eqnarray} 
$g$ is the coupling constant and we have included the usual linear covariant 
gauge fixing term with parameter $\alpha$. It is formally present to allow for 
the non-singular inversion of the quadratic part of the Lagrangian in momentum 
space to determine the propagators but we will set $\alpha$~$=$~$0$ throughout
thereafter. Associated with this are the Faddeev-Popov ghosts, $c^a$ and 
$\bar{c}^a$, and massless quarks, $\psi^{iI}$, are included where the indices 
have the ranges $1$~$\leq$~$a$~$\leq$~$\NA$, $1$~$\leq$~$i$~$\leq$~$\Nf$ and 
$1$~$\leq$~$I$~$\leq$~$\NF$ where $\NF$ and $\NA$ are the respective dimensions
of the fundamental and adjoint representations and $\Nf$ is the number of quark
flavours. To recover the static set-up of the colour sources we take, 
\cite{36,37,38}, 
\begin{equation}
J^a_\mu(x) ~=~ g v_\mu T^a \left[ \delta^{(3)} \left( \mathbf{x} 
- \half \mathbf{r} \right) ~-~ \delta^{(3)} \left( \mathbf{x} - \half
\mathbf{r}^\prime \right) \right]
\label{jdef}
\end{equation}
where $v_\mu$~$=$~$\eta_{\mu 0}$ is a unit vector and we will set
$r$~$=$~$|\mathbf{r}-\mathbf{r}^\prime|$. This vector $v_\mu$ will be present
throughout our calculations and in effect projects out the time component of
the gluon it couples to. Whilst we will be concentrating on the 
Gribov-Zwanziger case throughout, which relates to a theory of a confined
gluon, we note that the group generator present in the source will be taken as
being the fundamental representation initially. Though one can regard the heavy
sources as being gluons in which case one would use the adjoint representation
of the source interaction. The calculation of the one loop static potential is 
the same irrespective of whether the generator is in the fundamental or adjoint
representations. Computationally differences will only arise in the two loop
static potential, \cite{43}. At this point it is worth briefly mentioning that 
in the Coulomb gauge, which Gribov also considered in \cite{10}, it is the time
component of the gluon propagator which receives attention in regard to
confinement. So the vector $v_\mu$ is in effect the bridge between the Landau 
gauge we use here and the Coulomb gauge results in relation to a confining 
potential through the gauge invariant Wilson loop. 

Given this formulation of the static potential, it is possible to evaluate
$V(r)$ either directly in coordinate space, \cite{36,37,38}, or in momentum
space, \cite{39,40,41,42}. For the latter $V(r)$ emerges after a Fourier
transform and in our conventions we take 
\begin{equation}
V(r) ~=~ \int \frac{d^3 \mathbf{k}}{(2\pi)^3} \, e^{i\mathbf{k}.\mathbf{r}} \,
\tilde{V}(\mathbf{k})
\label{foudef}
\end{equation}
where $\tilde{V}(\mathbf{k})$ is the momentum space static potential. 
Performing the angular integrations since $\tilde{V}(\mathbf{k})$ will only
depend on the momentum length $k$ then 
\begin{equation}
V(r) ~=~ \frac{1}{2\pi^2} \int_0^\infty d k ~ k^2 \tilde{V}(k) 
\frac{\sin(kr)}{kr} ~.
\end{equation}
At this point it is perhaps apt to recall differing forms of $\tilde{V}(k)$ 
and examine their implications for coordinate space. Although we will consider
forms relevant for this article, a more general library of mappings between
spaces can be found, for instance, in \cite{56}. Taking
\begin{equation}
\tilde{V}_C(k) ~=~ \frac{A_C}{k^2} ~~~,~~~
\tilde{V}_l(k) ~=~ \frac{A_l}{(k^2)^2} ~~~\mbox{and}~~~ 
\tilde{V}_b(k) ~=~ \frac{A_b}{(k^2)^3}
\label{toypot}
\end{equation} 
then
\begin{equation}
V_C(r) ~=~ \frac{A_C}{4\pi r} ~~~,~~~
V_l(r) ~=~ -~ \frac{A_l}{8\pi} r ~~~\mbox{and}~~~ 
V_b(r) ~=~ \frac{A_b}{96\pi} r^3
\end{equation} 
respectively which result from the elementary integrals 
\begin{equation}
\int_0^\infty dx \, \frac{\sin x}{x} ~=~ \frac{\pi}{2} ~~~,~~~
\int_0^\infty dx \, \frac{\sin x}{x^3} ~=~ -~ \frac{\pi}{4} ~~~\mbox{and}~~~
\int_0^\infty dx \, \frac{\sin x}{x^5} ~=~ \frac{\pi}{96} 
\end{equation}
respectively. So a linearly rising potential will emerge from a simple dipole
term. However, it is amusing to note that an asymptotic linear behaviour as
$r$~$\rightarrow$~$\infty$ can also emerge from other forms of 
$\tilde{V}(k)$ such as that given by including in $\tilde{V}_l(k)$ any
reasonable function of $k$, which is non-singular and does not upset the 
integral convergence, purely on dimensional grounds.

In the early work of \cite{36,37,38} the static potential was calculated at one
loop directly in coordinate space. The presence of the static sources modifies 
the Feynman rules and introduces Heaviside step functions of the source 
interaction positions. Whilst it is relatively straightforward to compute the 
one loop potential, since we will ultimately be dealing with a formulation of 
QCD which is effectively massive, it is more natural to work in momentum space.
The momentum space formalism was developed in \cite{39,40,41,42} and we defer 
to those articles for the justification of the more technical aspects rather 
than unnecessarily repeat them here. Suffice to say that the momentum space 
Feynman rules for the source gluon couplings do not alter from those recorded 
in an appendix of \cite{41} and we note that \cite{41} gives comprehensive 
detail on many of the technical aspects of computing the static potentials. 
Moreover, the advantage of a momentum space approach is that it is in principle
more straightforward to extend to two loops. For completeness in reviewing the 
earlier static potential result for pure QCD, we note that the one loop 
$\MSbar$ result of \cite{36,37,38} is
\begin{eqnarray}
\left. \frac{}{} \tilde{V}( \mathbf{p} ) \right|_{\mbox{\footnotesize{QCD}}}
&=& -~ \frac{16 \pi^2 C_F a}{\mathbf{p}^2} \left[ 1 ~+~ \left[ \left[ 
\frac{31}{9} - \frac{11}{3} \ln \left[ \frac{\mathbf{p}^2}{\mu^2} \right] 
\right] C_A ~+~ \left[ \frac{4}{3} \ln \left[ \frac{\mathbf{p}^2}{\mu^2} 
\right] - \frac{20}{9} \right] T_F \Nf \right] a \right. \nonumber \\
&& \left. ~~~~~~~~~~~~~~~~~+~ O(a^2) \right]
\end{eqnarray} 
where $\mu$ is the mass scale introduced to ensure the coupling constant is
dimensionless when dimensional regularization is used, $a$~$=$~$g^2/(16\pi^2)$ 
and the colour group Casimirs are defined by
\begin{equation}
\mbox{Tr} \left( T^a T^b \right) ~=~ T_F \delta^{ab} ~~~,~~~
T^a T^a ~=~ C_F I ~~~,~~~ f^{acd} f^{bcd} ~=~ C_A \delta^{ab} 
\end{equation} 
and $I$ is the $\NF$ dimensional unit matrix.

To extend the static potential to the Gribov-Zwanziger case we return to the
definition (\ref{potdef}) and (\ref{potdefz}). In Gribov's original work,
\cite{10}, the key observation in the Landau gauge was that of the partition of
configuration space into regions defined by zeroes of the Faddeev-Popov
operator ${\cal M}(A)$~$=$~$-$~$\partial^\mu D_\mu$ which is hermitian. This is
because one has copies of the gauge configuration satisfying the {\em same}
gauge fixing condition. The region containing the origin, $A^a_\mu$~$=$~$0$, is
referred to as the Gribov region and denoted by $\Omega$. Gribov showed that to
exclude copies from the gauge fixing procedure, the path integral measure must 
be restricted to the region $\Omega$ which effectively introduced a cutoff into
the theory. Consequently, the structure of the theory was altered in that the 
gluon propagator ceased to be of a fundamental form in the infrared region. 
This was primarily due to the appearance in the Lagrangian of an additional 
non-local term. Its origin is in defining the boundary of $\Omega$ by the 
no-pole or horizon condition and leads to the presence of a mass parameter, 
$\gamma$, known as the Gribov mass. This is not an independent object but must 
satisfy a gap equation derived from the horizon condition, \cite{10}. One 
outcome of this is that the propagator of the Faddeev-Popov ghost takes a 
dipole form in the infrared. Whilst this is of the form for a linear confining 
potential in coordinate space, the Faddeev-Popov ghost is never directly 
exchanged between gluons or quarks. Given this position it is straightforward 
to see what one must do to determine the static potential in the Gribov 
picture. One simply returns to (\ref{potdefz}) and restricts the integration 
measure to $\Omega$ 
\begin{equation}
Z[J] ~=~ \int_\Omega {\cal D} A_\mu \, {\cal D} \psi \, {\cal D} \bar{\psi} \, 
{\cal D} c \, {\cal D} \bar{c} ~ \exp \left[ \,-\, \int d^4 x \, \left( L ~+~ 
J^{a\,\mu} A^a_\mu \right) \right] ~.
\end{equation}
Implementing the horizon condition this equates to, \cite{12},
\begin{equation}
Z[J] ~=~ \int {\cal D} A_\mu \, {\cal D} \psi \, {\cal D} \bar{\psi} \, 
{\cal D} c \, {\cal D} \bar{c} ~ \exp \left[ \,-\, \int d^4 x \, \left( 
L^{\mbox{\footnotesize{Grib}}} ~+~ J^{a\,\mu} A^a_\mu \right) \right]
\end{equation}
where 
\begin{equation} 
L^{\mbox{\footnotesize{Grib}}} ~=~ L^{\mbox{\footnotesize{QCD}}} ~+~ 
\frac{C_A \gamma^4}{2} A^{a\,\mu} \frac{1}{\partial^\nu D_\nu} A^a_\mu ~-~ 
\frac{d \NA \gamma^4}{2g^2} 
\label{laggrib}
\end{equation} 
where $d$ is the dimension of spacetime.

To proceed we now use the localization of the horizon non-locality introduced
in \cite{12,13,17,18} by Zwanziger. This involves introducing an additional set
of fields called localizing or Zwanziger ghosts, $\{\phi^{ab}_\mu,
\bar{\phi}^{ab}_\mu\}$ and $\{\omega^{ab}_\mu, \bar{\omega}^{ab}_\mu\}$, where
the former are commuting and the latter are anti-commuting. They are regarded
as {\em internal} fields, similar to the Faddeev-Popov ghosts, in that they do 
not couple directly to quarks. The resulting path integral becomes
\begin{equation}
Z[J] ~=~ \int {\cal D} A_\mu \, {\cal D} \psi \, {\cal D} \bar{\psi} \, 
{\cal D} c \, {\cal D} \bar{c} \, {\cal D} \xi \, {\cal D} \rho \, 
{\cal D} \omega \, {\cal D} \bar{\omega} ~ \exp \left[ -\, \int d^4 x \, 
\left( L^{\mbox{\footnotesize{GZ}}} ~+~ J^{a\,\mu} A^a_\mu \right) \right]
\end{equation}
where 
\begin{eqnarray}
L^{\mbox{\footnotesize{GZ}}} &=& L^{\mbox{\footnotesize{QCD}}} ~+~ 
\frac{1}{2} \rho^{ab \, \mu} \partial^\nu \left( D_\nu \rho_\mu 
\right)^{ab} ~+~ \frac{i}{2} \rho^{ab \, \mu} \partial^\nu 
\left( D_\nu \xi_\mu \right)^{ab} ~-~ \frac{i}{2} \xi^{ab \, \mu} 
\partial^\nu \left( D_\nu \rho_\mu \right)^{ab} \nonumber \\
&& +~ \frac{1}{2} \xi^{ab \, \mu} \partial^\nu \left( D_\nu \xi_\mu 
\right)^{ab} ~-~ \bar{\omega}^{ab \, \mu} \partial^\nu \left( D_\nu \omega_\mu 
\right)^{ab} ~-~ \frac{1}{\sqrt{2}} g f^{abc} \partial^\nu 
\bar{\omega}^{ae}_\mu \left( D_\nu c \right)^b \rho^{ec \, \mu} \nonumber \\
&& -~ \frac{i}{\sqrt{2}} g f^{abc} \partial^\nu \bar{\omega}^{ae}_\mu 
\left( D_\nu c \right)^b \xi^{ec \, \mu} ~-~ i \gamma^2 f^{abc} A^{a \, \mu} 
\xi^{bc}_\mu ~-~ \frac{d \NA \gamma^4}{2g^2} 
\label{laggz}
\end{eqnarray} 
and we have introduced the real and imaginary fields $\rho^{ab}_\mu$ and
$\xi^{ab}_\mu$ similar to \cite{50} given by
\begin{equation}
\phi^{ab}_\mu ~=~ \frac{1}{\sqrt{2}} \left( \rho^{ab}_\mu ~+~ i \xi^{ab}_\mu 
\right) ~~,~~ 
\bar{\phi}^{ab}_\mu ~=~ \frac{1}{\sqrt{2}} \left( \rho^{ab}_\mu ~-~ 
i \xi^{ab}_\mu \right) ~.
\end{equation} 
In choosing to work with the real fields $\rho^{ab}_\mu$ and $\xi^{ab}_\mu$ we
will en route be correcting the error in earlier loop calculations,
\cite{32,33}, where the $\rho^{ab}_\mu$ propagator was erroneously treated in 
the initial computer algebra derivation. One main benefit is that (\ref{laggz})
is renormalizable, \cite{18,30,31}, which allows us to do explicit 
calculations. We briefly note that in using (\ref{laggz}) to perform 
computations the gauge symmetry is broken in much the same fashion as in the 
$\gamma^2$~$=$~$0$ situation via the (local) gauge fixing criterion in 
(\ref{lqcd}). However, the usefulness of (\ref{laggz}) in determining 
quantities of physical interest, in addition to renormalizability, resides 
essentially on two criteria. These are the gauge invariance of the object which
is apparent, for instance, in the local or ultraviolet limit and the fact that 
$\gamma$ is constrained in the Gribov gap equation to be a (non-perturbative) 
function of the coupling constant and is not an independent parameter of the 
theory. In other words (\ref{laggz}) has no meaning as a gauge theory unless 
$\gamma$ satisfies the gap equation defined by the horizon condition, 
\cite{11,12,13,14,15,16,17,18,19}. For completeness we note the Landau gauge 
propagators of the gauge sector required for that computation and the current 
one are, 
\begin{eqnarray}
\langle A^a_\mu(p) A^b_\nu(-p) \rangle &=& -~ 
\frac{\delta^{ab}p^2}{[(p^2)^2+C_A\gamma^4]} P_{\mu\nu}(p) \nonumber \\
\langle A^a_\mu(p) \xi^{bc}_\nu(-p) \rangle &=& 
\frac{i f^{abc}\gamma^2}{[(p^2)^2+C_A\gamma^4]} P_{\mu\nu}(p) 
\nonumber \\
\langle A^a_\mu(p) \rho^{bc}_\nu(-p) \rangle &=& 0 \nonumber \\ 
\langle \xi^{ab}_\mu(p) \xi^{cd}_\nu(-p) \rangle &=& -~ 
\frac{\delta^{ac}\delta^{bd}}{p^2}\eta_{\mu\nu} ~+~
\frac{f^{abe}f^{cde}\gamma^4}{p^2[(p^2)^2+C_A\gamma^4]} P_{\mu\nu}(p) 
\nonumber \\ 
\langle \xi^{ab}_\mu(p) \rho^{cd}_\nu(-p) \rangle &=& 0 \nonumber \\ 
\langle \rho^{ab}_\mu(p) \rho^{cd}_\nu(-p) \rangle &=& 
\langle \omega^{ab}_\mu(p) \bar{\omega}^{cd}_\nu(-p) \rangle ~=~ -~ 
\frac{\delta^{ac}\delta^{bd}}{p^2} \eta_{\mu\nu} 
\label{propdef}
\end{eqnarray} 
where 
\begin{equation}
P_{\mu\nu}(p) ~=~ \eta_{\mu\nu} ~-~ \frac{p_\mu p_\nu}{p^2} 
\end{equation}
is the transverse projector and we have a procedure similar to that discussed 
in \cite{33} for handling the mixed propagator. The explicit renormalization 
properties of the Gribov mass and fields $\rho^{ab}_\mu$, $\xi^{ab}_\mu$ and 
$\omega^{ab}_\mu$ are predetermined by Slavnov-Taylor identities, 
\cite{18,30,31}, which have been verified by explicit computations in the 
$\MSbar$ scheme, \cite{32,33}, in the ultraviolet situation where one can work 
with the $\gamma$~$=$~$0$ limit. Those calculations were carried out with the 
symbolic manipulation machinery of {\sc Form}, \cite{57}, which we will also 
use here. The absence of a mixed $\rho^{ab}_\mu$-$\xi^{ab}_\mu$ propagator is 
due to the fact that the cross-term of the quadratic part of the Lagrangian is 
a total derivative which can be dropped, \cite{50}. However, we have retained 
the gluon $\rho^{ab}_\mu$-$\xi^{ab}_\mu$ vertex even though it can be written 
with a factor proportional to the Landau gauge condition, 
$\partial^\mu A^a_\mu$, \cite{50}. Whilst this will vanish for transverse 
gluons in the Landau gauge we retain it as part of our Feynman rules since we 
will consider longitudinal corrections to the $2$-point functions at one loop. 

Indeed given this, it is worth recalling how the propagators of (\ref{propdef})
are constructed in practice since it is partly related to the situation with 
regard to other linear covariant gauges and justifies why we focus solely on 
the Landau gauge in the Gribov case. Also these comments are based around the
more detailed analysis of \cite{58}, to which we refer the interested reader, 
where Gribov copies were considered for linear covariant gauges when the gauge 
parameter is small. First, we recall, \cite{33}, that in deriving 
(\ref{propdef}) in order to avoid a singular determinant in Lorentz space in 
inverting the quadratic part of the momentum space Lagrangian, the gauge 
parameter is kept non-zero. The Landau gauge expressions, (\ref{propdef}), 
emerge when $\alpha$ is set to zero, \cite{33}. Specifically, if we work in the
basis of fields  $\{ A^a_\mu, \xi^{ab}_\mu, \rho^{ab}_\mu \}$ then with a 
non-zero $\alpha$ the matrix of quadratic terms in the Lagrangian in momentum 
space is  
\begin{eqnarray}
\Lambda^{\{ab|cd\}}_{\mu\nu}(p) &=&
\left(
\begin{array}{ccc}
- \, p^2 \delta^{ac} & - \, i \gamma^2 f^{acd} & 0 \\
- \, i \gamma^2 f^{cab} & - \, p^2 \delta^{ac} \delta^{bd} & 0 \\
0 & 0 & - \, p^2 \delta^{ac} \delta^{bd} \\
\end{array}
\right) P_{\mu\nu}(p) \nonumber \\
&& +~ \left(
\begin{array}{ccc}
- \, \frac{p^2}{\alpha} \delta^{ac} & - \, i \gamma^2 f^{acd} & 0 \\
- \, i \gamma^2 f^{cab} & - \, p^2 \delta^{ac} \delta^{bd} & 0 \\
0 & 0 & - \, p^2 \delta^{ac} \delta^{bd} \\
\end{array}
\right) L_{\mu\nu}(p) 
\end{eqnarray}
where 
\begin{equation}
L_{\mu\nu}(p) ~=~ \frac{p_\mu p_\nu}{p^2} 
\end{equation}
is the longitudinal projector which satisfies the trivial relation
$P_{\mu\nu}(p)$~$+$~$L_{\mu\nu}(p)$~$=$~$\eta_{\mu\nu}$. If we denote the 
matrix of propagators by $\Pi^{\{ab|cd\}}_{\mu\nu}(p)$ then the propagators 
must satisfy 
\begin{equation}
\Lambda^{\{ab|cd\}}_{\mu\sigma}(p) \Pi^{\{cd|pq\}\sigma}_{~~~~~~~~\,\nu}(p) ~=~ 
\left(
\begin{array}{ccc}
\delta^{cp} & 0 & 0 \\
0 & \delta^{cp} \delta^{dq} & 0 \\
0 & 0 & \delta^{cp} \delta^{dq} \\
\end{array}
\right) \eta_{\mu\nu} 
\label{matid}
\end{equation} 
and the matrix on the right hand side is the unit matrix on the colour space 
for this sector of fields. Equipped with this the non-zero $\alpha$ propagators
are 
\begin{eqnarray}
\langle A^a_\mu(p) A^b_\nu(-p) \rangle &=& -~ 
\frac{\delta^{ab}p^2}{[(p^2)^2+C_A\gamma^4]} P_{\mu\nu}(p) ~-~ 
\frac{\alpha\delta^{ab}p^2}{[(p^2)^2+\alpha C_A\gamma^4]} L_{\mu\nu}(p) 
\nonumber \\
\langle A^a_\mu(p) \xi^{bc}_\nu(-p) \rangle &=& 
\frac{i f^{abc}\gamma^2}{[(p^2)^2+C_A\gamma^4]} P_{\mu\nu}(p) ~+~ 
\frac{i \alpha f^{abc}\gamma^2}{[(p^2)^2+ \alpha C_A\gamma^4]} L_{\mu\nu}(p) 
\nonumber \\
\langle A^a_\mu(p) \rho^{bc}_\nu(-p) \rangle &=& 0 \nonumber \\ 
\langle \xi^{ab}_\mu(p) \xi^{cd}_\nu(-p) \rangle &=& -~ 
\frac{\delta^{ac}\delta^{bd}}{p^2}\eta_{\mu\nu} ~+~
\frac{f^{abe}f^{cde}\gamma^4}{p^2[(p^2)^2+C_A\gamma^4]} P_{\mu\nu}(p) ~+~
\frac{\alpha f^{abe}f^{cde}\gamma^4}{p^2[(p^2)^2+\alpha C_A\gamma^4]} 
L_{\mu\nu}(p) \nonumber \\ 
\langle \xi^{ab}_\mu(p) \rho^{cd}_\nu(-p) \rangle &=& 0 \nonumber \\ 
\langle \rho^{ab}_\mu(p) \rho^{cd}_\nu(-p) \rangle &=& 
\langle \omega^{ab}_\mu(p) \bar{\omega}^{cd}_\nu(-p) \rangle ~=~ -~ 
\frac{\delta^{ac}\delta^{bd}}{p^2} \eta_{\mu\nu} ~. 
\label{propdefal}
\end{eqnarray} 
These reduce to those of (\ref{propdef}) in the $\alpha$~$\rightarrow$~$0$
limit. Though we have included the non-zero $\alpha$ forms since the 
longitudinal pieces play an important role in determining the one loop 
corrections to the gluon propagator. However, the propagators of
(\ref{propdefal}) have no meaning as such and must not be regarded as the 
propagators for the extension of the Gribov-Zwanziger Lagrangian to other 
linear covariant gauges. The Gribov-Zwanziger Lagrangian is purely a Landau 
gauge construction. To clarify this we recall certain properties in the 
construction of (\ref{laggz}) based on \cite{58}. First, in the set of linear 
covariant gauges the Faddeev-Popov operator is only hermitian in the Landau 
gauge which allows for the classification of its eigenvalues into positive and 
negative values in that case only, \cite{10,11,12,13,14,15,15,16,17,18,19}. 
This gives a natural way of dividing configuration space into definite regions.
For other linear covariant gauges the loss of hermiticity means the eigenvalues
can be complex and therefore there is not a natural or straightforward way of 
partitioning configuration space to even examine how Gribov copies are mapped 
between regions as one can do in the Landau gauge. Therefore, for non-Landau 
linear covariant gauges it does not seem clear how the path integral could be 
cut-off and the analogous first Gribov region defined by a no-pole condition as
in the Landau case. Moreover, as the latter condition manifests itself as the 
non-locality to be localized in the Landau gauge one might simply assume that 
the natural extension to other linear covariant gauges is to merely use 
(\ref{laggrib}) and proceed with a non-zero gauge parameter. However, this 
overlooks the fact that in the Landau gauge the gluon is transverse but in 
other linear covariant gauges it is not. Therefore, in the non-local term of 
(\ref{laggrib}) the gauge field in the denominator covariant derivative itself 
actually involves an additional non-local projection, \cite{58}. Thus, for 
non-zero $\alpha$ the analogous Gribov-Zwanziger Lagrangian, (\ref{laggz}), 
would have additional non-localities which would also need to be localized 
before any computations could proceed assuming any final local Lagrangian would
actually be renormalizable, \cite{58}. So, for instance, the intermediate 
propagators of (\ref{propdefal}) that exist for non-zero $\alpha$ in deriving 
(\ref{propdef}) have no true meaning.

\sect{$2$-point function corrections.}

Given that we are now using a version of the Gribov-Zwanziger Lagrangian which
correctly involves the real and imaginary parts of the bosonic localizing 
ghosts, it is appropriate to discuss the one loop corrections to all the 
$2$-point functions and propagators of the fields. Though we will concentrate 
primarily on the mixed sector for reasons which will become apparent later. 
Whilst the main details of this analysis parallels that given in \cite{33}, 
which was incorrect, we include it here for completeness. We define the 
$3$~$\times$~$3$ matrix of one loop $2$-point functions formally by  
\begin{eqnarray}
\Lambda^{\{ab|cd\}}_{\mu\nu}(p) &=&
\left(
\begin{array}{ccc}
- \, p^2 \delta^{ac} & - \, i \gamma^2 f^{acd} & 0 \\
- \, i \gamma^2 f^{cab} & - \, p^2 \delta^{ac} \delta^{bd} & 0 \\
0 & 0 & - \, p^2 \delta^{ac} \delta^{bd} \\
\end{array}
\right) P_{\mu\nu}(p) \nonumber \\
&& +~ \left(
\begin{array}{ccc}
- \, \frac{p^2}{\alpha} \delta^{ac} & - \, i \gamma^2 f^{acd} & 0 \\
- \, i \gamma^2 f^{cab} & - \, p^2 \delta^{ac} \delta^{bd} & 0 \\
0 & 0 & - \, p^2 \delta^{ac} \delta^{bd} \\
\end{array}
\right) L_{\mu\nu}(p) \nonumber \\ 
&& +~ 
\left(
\begin{array}{ccc}
X \delta^{ac} & U f^{acd} & V f^{acd} \\
U f^{cab} & Q^{abcd}_\xi & 0 \\
V f^{cab} & 0 & Q^{abcd}_\rho \\
\end{array}
\right) P_{\mu\nu}(p) a \nonumber \\ 
&& +~ 
\left(
\begin{array}{ccc}
X^L \delta^{ac} & U^L f^{acd} & V^L f^{acd} \\
U^L f^{cab} & Q^{L\,abcd}_\xi & 0 \\
V^L f^{cab} & 0 & Q^{L\,abcd}_\rho \\
\end{array}
\right) L_{\mu\nu}(p) a ~+~ O(a^2) 
\label{twoptdef} 
\end{eqnarray} 
with respect to the same basis $\{ A^a_\mu, \xi^{ab}_\mu, \rho^{ab}_\mu \}$ 
where
\begin{eqnarray}
Q^{abcd}_\xi &=& Q_\xi \delta^{ac} \delta^{bd} ~+~ W_\xi f^{ace} f^{bde} ~+~ 
R_\xi f^{abe} f^{cde} ~+~ S_\xi d_A^{abcd} \nonumber \\ 
Q^{abcd}_\rho &=& Q_\rho \delta^{ac} \delta^{bd} ~+~ 
W_\rho f^{ace} f^{bde} ~+~ R_\rho f^{abe} f^{cde} ~+~ S_\rho d_A^{abcd} 
\nonumber \\ 
Q^{L\,abcd}_\xi &=& Q^L_\xi \delta^{ac} \delta^{bd} ~+~ 
W^L_\xi f^{ace} f^{bde} ~+~ R^L_\xi f^{abe} f^{cde} ~+~ S^L_\xi d_A^{abcd} 
\nonumber \\ 
Q^{L\,abcd}_\rho &=& Q^L_\rho \delta^{ac} \delta^{bd} ~+~ 
W^L_\rho f^{ace} f^{bde} ~+~ R^L_\rho f^{abe} f^{cde} ~+~ 
S^L_\rho d_A^{abcd} 
\end{eqnarray}
represents the colour decomposition and
\begin{equation}
d_A^{abcd} ~=~ \frac{1}{6} \mbox{Tr} \left( T_A^a T_A^{(b} T_A^c T_A^{d)}
\right)
\end{equation}
is totally symmetric and $T_A^a$ is the adjoint representation of the colour
group generators. The quantities in the final two matrices of (\ref{twoptdef})
represent the one loop corrections and we have incorporated the fact that there
is no one loop correction to the $\rho^{ab}_\mu$-$\xi^{cd}_\nu$ $2$-point 
functions. The longitudinal $2$-point functions are indicated by the 
superscript $L$. Given (\ref{twoptdef}) we define a similar formal form for the
inverse to one loop by
\begin{eqnarray}
\Pi^{\{cd|pq\}}_{\mu\nu}(p) &=& \left(
\begin{array}{ccc}
- \, \frac{p^2}{[(p^2)^2+C_A\gamma^4]} \delta^{cp} & 
\frac{i\gamma^2}{[(p^2)^2+C_A\gamma^4]} f^{cpq} & 0 \\
\frac{i\gamma^2}{[(p^2)^2+C_A\gamma^4]} f^{pcd} & 
- \, \frac{1}{p^2} \delta^{cp} \delta^{dq} 
\, + \, \frac{\gamma^4}{p^2[(p^2)^2+C_A\gamma^4]} f^{cdr} f^{pqr} & 0 \\
0 & 0 & - \, \frac{\delta^{cp} \delta^{dq}}{p^2} \\
\end{array}
\right) P_{\mu\nu}(p) \nonumber \\
&& + \left(
\begin{array}{ccc}
- \, \frac{\alpha p^2}{[(p^2)^2+\alpha C_A\gamma^4]} \delta^{cp} & 
\frac{i\alpha\gamma^2}{[(p^2)^2+\alpha C_A\gamma^4]} f^{cpq} & 0 \\
\frac{i\alpha\gamma^2}{[(p^2)^2+\alpha C_A\gamma^4]} f^{pcd} & 
- \, \frac{1}{p^2} \delta^{cp} \delta^{dq} 
+ \frac{\alpha\gamma^4}{p^2[(p^2)^2+\alpha C_A\gamma^4]} f^{cdr} f^{pqr} 
& 0 \\
0 & 0 & \! - \, \frac{\delta^{cp} \delta^{dq}}{p^2} \\
\end{array}
\! \right) \! L_{\mu\nu}(p) \nonumber \\
&& + \, \left(
\begin{array}{ccc}
A \delta^{cp} & B f^{cpq} & C f^{cpq} \\
B f^{pcd} & D^{cdpq}_\xi & E^{cdpq} \\
C f^{pcd} & E^{cdpq} & D^{cdpq}_\rho \\
\end{array}
\right) P_{\mu\nu}(p) a \nonumber \\ 
&& + \, \left(
\begin{array}{ccc}
A^L \delta^{cp} & B^L f^{cpq} & C^L f^{cpq} \\
B^L f^{pcd} & D^{\,cdpq}_\xi & E^{L\,cdpq} \\
C^L f^{pcd} & E^{\,cdpq} & D^{\,cdpq}_\rho \\
\end{array}
\right) L_{\mu\nu}(p) a ~+~ O(a^2) 
\label{ffdef} 
\end{eqnarray} 
for non-zero $\alpha$, where
\begin{eqnarray}
E^{cdpq} &=& E \delta^{cp} \delta^{dq} ~+~ F f^{cpe} f^{dqe} ~+~ 
G f^{cde} f^{pqe} ~+~ H d_A^{cdpq} \nonumber \\
D^{cdpq}_\xi &=& D_\xi \delta^{cp} \delta^{dq} ~+~ J_\xi f^{cpe} f^{dqe} ~+~ 
K_\xi f^{cde} f^{pqe} ~+~ L_\xi d_A^{cdpq} \nonumber \\
D^{cdpq}_\rho &=& D_\rho \delta^{cp} \delta^{dq} ~+~ J_\rho f^{cpe} f^{dqe} ~+~ 
K_\rho f^{cde} f^{pqe} ~+~ L_\rho d_A^{cdpq} \nonumber \\ 
E^{L\,cdpq} &=& E^L \delta^{cp} \delta^{dq} ~+~ F^L f^{cpe} f^{dqe} ~+~ 
G^L f^{cde} f^{pqe} ~+~ H^L d_A^{cdpq} \nonumber \\
D^{L\,cdpq}_\xi &=& D^L_\xi \delta^{cp} \delta^{dq} ~+~ 
J^L_\xi f^{cpe} f^{dqe} ~+~ K^L_\xi f^{cde} f^{pqe} ~+~ 
L^L_\xi d_A^{cdpq} \nonumber \\
D^{L\,cdpq}_\rho &=& D^L_\rho \delta^{cp} \delta^{dq} ~+~ 
J^L_\rho f^{cpe} f^{dqe} ~+~ K^L_\rho f^{cde} f^{pqe} ~+~ L^L_\rho d_A^{cdpq} 
\end{eqnarray}
and the inverse satisfies (\ref{matid}) again. The quantities in the final two
matrices of (\ref{ffdef}) represent the one loop corrections to the 
propagators. By multiplying out the elements of the left hand side of 
(\ref{matid}) to one loop we can determine the relation of the one loop 
propagator corrections to those of the $2$-point functions. For the transverse 
sector we find 
\begin{eqnarray} 
A &=& -~ \frac{1}{[(p^2)^2+C_A\gamma^4]^2} \left[ (p^2)^2 X - 2 i C_A \gamma^2
p^2 U - C_A \gamma^4 \left[ Q_\xi + C_A R_\xi + \half C_A W_\xi \right] \right]
\nonumber \\  
B &=& \frac{1}{[(p^2)^2+C_A\gamma^4]^2} \left[ i \gamma^2 p^2 X 
- ( (p^2)^2 - C_A \gamma^4 ) U + i \gamma^2 p^2 \left[ Q_\xi + C_A R_\xi 
+ \half C_A W_\xi \right] \right] \nonumber \\  
C &=& -~ \frac{V}{[(p^2)^2+C_A\gamma^4]} ~~,~~  
D_\xi ~=~ -~ \frac{Q_\xi}{(p^2)^2} ~~,~~ 
J_\xi ~=~ -~ \frac{W_\xi}{(p^2)^2} ~~,~~ 
L_\xi ~=~ -~ \frac{S_\xi}{(p^2)^2} \nonumber \\  
K_\xi &=& \frac{1}{[(p^2)^2+C_A\gamma^4]^2} \left[ \gamma^4 X 
+ 2 i \gamma^2 p^2 U - (p^2)^2 R_\xi + \gamma^4 \left[ Q_\xi + \half C_A W_\xi 
\right] \right] \nonumber \\ 
&& +~ \frac{\gamma^4 \left[ Q_\xi + \half C_A W_\xi \right]}
{(p^2)^2[(p^2)^2+C_A\gamma^4]} \nonumber \\ 
E &=& 0 ~~,~~ F ~=~ 0 ~~,~~ 
G ~=~ \frac{i \gamma^2 V}{p^2[(p^2)^2+C_A\gamma^4]} ~~,~~ H ~=~ 0
\nonumber \\ 
D_\rho &=& -~ \frac{Q_\rho}{(p^2)^2} ~~,~~ 
J_\rho ~=~ -~ \frac{W_\rho}{(p^2)^2} ~~,~~ 
K_\rho ~=~ -~ \frac{R_\rho}{(p^2)^2} ~~,~~ 
L_\rho ~=~ -~ \frac{S_\rho}{(p^2)^2} 
\label{traprop}
\end{eqnarray} 
to one loop. For the longitudinal sector, we retain for the moment the non-zero
$\alpha$ and find formally, 
\begin{eqnarray} 
A^L &=& -~ \frac{\alpha^2}{[(p^2)^2+\alpha C_A\gamma^4]^2} \left[ (p^2)^2 X^L 
- 2 i C_A \gamma^2 p^2 U^L - C_A \gamma^4 \left[ Q^L_\xi + C_A R^L_\xi 
+ \half C_A W^L_\xi \right] \right] \nonumber \\  
B^L &=& \frac{\alpha}{[(p^2)^2+\alpha C_A\gamma^4]^2} \left[ i \alpha \gamma^2 
p^2 X^L - [ (p^2)^2 - \alpha C_A \gamma^4 ] U^L + i \gamma^2 p^2 
\left[ Q^L_\xi + C_A R^L_\xi + \half C_A W^L_\xi \right] \right] \nonumber \\  
C^L &=& -~ \frac{\alpha V^L}{[(p^2)^2+\alpha C_A\gamma^4]} ~~,~~  
D^L_\xi ~=~ -~ \frac{Q^L_\xi}{(p^2)^2} ~~,~~ 
J^L_\xi ~=~ -~ \frac{W^L_\xi}{(p^2)^2} ~~,~~ 
L^L_\xi ~=~ -~ \frac{S^L_\xi}{(p^2)^2} \nonumber \\  
K^L_\xi &=& \frac{\alpha}{[(p^2)^2+\alpha C_A\gamma^4]^2} \left[ \alpha 
\gamma^4 X^L + 2 i \gamma^2 p^2 U^L - \frac{(p^2)^2}{\alpha} R^L_\xi 
+ \gamma^4 \left[ Q^L_\xi + \half C_A W^L_\xi \right] \right] \nonumber \\ 
&& +~ \frac{\alpha \gamma^4 \left[ Q^L_\xi + \half C_A W^L_\xi \right]}
{(p^2)^2[(p^2)^2+\alpha C_A\gamma^4]} \nonumber \\ 
E^L &=& 0 ~~,~~ F^L ~=~ 0 ~~,~~ 
G^L ~=~ \frac{i \alpha \gamma^2 V^L}{p^2[(p^2)^2+\alpha C_A\gamma^4]} ~~,~~ 
H^L ~=~ 0 \nonumber \\ 
D^L_\rho &=& -~ \frac{Q^L_\rho}{(p^2)^2} ~~,~~ 
J^L_\rho ~=~ -~ \frac{W^L_\rho}{(p^2)^2} ~~,~~ 
K^L_\rho ~=~ -~ \frac{R^L_\rho}{(p^2)^2} ~~,~~ 
L^L_\rho ~=~ -~ \frac{S^L_\rho}{(p^2)^2} 
\end{eqnarray} 
to one loop too. We note that the exact one loop expressions for each of these 
$2$-point functions are recorded in appendices B and C. Whilst these represent 
our arbitrary $\alpha$ manipulations, we must restrict $\alpha$ to the Landau 
gauge for the propagators corrections to have any meaning in the 
Gribov-Zwanziger context. Since the one loop corrections are non-singular in
$\alpha$ then taking $\alpha$~$\rightarrow$~$0$ for the longitudinal sector we 
find
\begin{eqnarray}  
A^L &=& B^L ~=~ C^L ~=~ E^L ~=~ F^L ~=~ G^L ~=~ H^L ~=~ 0 \nonumber \\
D^L_\xi &=& -~ \frac{Q^L_\xi}{(p^2)^2} ~~,~~ 
J^L_\xi ~=~ -~ \frac{W^L_\xi}{(p^2)^2} ~~,~~ 
K^L_\xi ~=~ -~ \frac{R^L_\xi}{(p^2)^2} ~~,~~ 
L^L_\xi ~=~ -~ \frac{S^L_\xi}{(p^2)^2} \nonumber \\  
D^L_\rho &=& -~ \frac{Q^L_\rho}{(p^2)^2} ~~,~~ 
J^L_\rho ~=~ -~ \frac{W^L_\rho}{(p^2)^2} ~~,~~ 
K^L_\rho ~=~ -~ \frac{R^L_\rho}{(p^2)^2} ~~,~~ 
L^L_\rho ~=~ -~ \frac{S^L_\rho}{(p^2)^2} ~. 
\end{eqnarray}
Using the explicit values from appendices B and C one can in principle deduce
the formal expressions for the one loop corrections to the propagators. For
instance, given that $A^L$~$=$~$0$ when $\alpha$~$=$~$0$ then the one loop
correction to the gluon propagator is actually transverse as is the mixed
gluon $\xi^{ab}_\mu$ propagator. Given the structures of $D^{L\,abcd}_\xi$
and $D^{L\,abcd}_\rho$ then both the $\xi^{ab}_\mu$ and $\rho^{ab}_\mu$ 
propagator corrections are not transverse like (\ref{propdef}).

For the remaining fields $c^a$ and $\omega^a_\mu$, if we write the ghost form 
factors as 
\begin{equation} 
\langle c^a(p) \bar{c}^b(-p) \rangle ~=~ \frac{D_c(p^2)}{p^2} 
\delta^{ab} ~~~,~~~ \langle \omega^{ab}_\mu(p) \bar{\omega}^{cd}_\nu(-p) 
\rangle ~=~ \delta^{ac}\delta^{bd}\frac{D_\omega(p^2)}{p^2} \eta_{\mu\nu} 
\end{equation}
then
\begin{eqnarray}  
D_c(p^2) ~=~ D_\omega(p^2) &=& \left[ -~ 1 ~+~ \left[ \frac{5}{4} ~-~ 
\frac{3}{8} \ln \left( \frac{C_A \gamma^4}{\mu^4} \right) ~+~ 
\frac{3 \sqrt{C_A} \gamma^2}{4p^2} \tan^{-1} \left[ 
\frac{\sqrt{C_A}\gamma^2}{p^2} \right] \right. \right. \nonumber \\
&& \left. \left. ~~~~~~~~~~~~~-~ \frac{3 \pi \sqrt{C_A} \gamma^2}{8p^2} \,+\, 
\frac{C_A\gamma^4}{8(p^2)^2} \ln \left[ 1 + \frac{(p^2)^2}{C_A\gamma^4} 
\right] \,-\, \frac{3}{8} \ln \left[ 1 + \frac{(p^2)^2}{C_A\gamma^4} \right]
\right. \right. \nonumber \\
&& \left. \left. ~~~~~~~~~~~~~-~ \frac{p^2}{4\sqrt{C_A}\gamma^2} \tan^{-1} 
\left[ \frac{\sqrt{C_A}\gamma^2}{p^2} \right] \right] C_A a \right]^{-1} ~+~ 
O(a^2) ~. 
\end{eqnarray}  
Using the one loop Gribov gap equation, \cite{10}, then $D^{-1}_c(p^2)$ is 
$O((p^2)^2)$ as $p^2$ $\rightarrow$ $0$ and hence both propagators enhance. 
Likewise the $\rho^{ab}_\mu$ propagator enhances whilst one colour component of
the $\xi^{ab}_\mu$ propagator also suggests there will be some sort of
enhancement for that field too, \cite{33}. We will return to this latter point 
in more detail in section $6$. 

Equipped with these $2$-point functions and the formal expressions for the one 
loop corrections to the $\{ A^a_\mu, \xi^{ab}_\mu, \rho^{ab}_\mu \}$ matrix of 
propagators we can study the effective renormalization group invariant coupling
constant. The behaviour of this effective coupling constant in the zero 
momentum limit has been the subject of debate over the years and whether it 
freezes to a finite non-zero or zero value is still unresolved. Defining the 
gluon propagator form factor as
\begin{equation}
\langle A^a_\mu(p) A^b_\nu(-p) \rangle ~=~ -~ \delta^{ab} \frac{D_A(p^2)}{p^2}
P_{\mu\nu}(p)
\end{equation}
then the effective coupling constant is defined by
\begin{equation}
\alpha^{\mbox{\footnotesize{eff}}} (p^2) ~=~ \alpha_s(\mu) D_A(p^2) \left(
D_c(p^2) \right)^2
\label{effccdef}
\end{equation}
where we have introduced the more common strong coupling constant 
$\alpha_s$~$=$~$g^2/(4\pi)$. Using the zero momentum values for the $2$-point
functions recorded in appendix B we have
\begin{equation}
\alpha^{\mbox{\footnotesize{eff}}}_s (0) ~=~ \lim_{p^2 \rightarrow 0} \left[
\frac{ \alpha(\mu) \left[ 1 + C_A \left( \frac{3}{8} \ln \left(
\frac{C_A \gamma^4(\mu)}{\mu^4} \right) - \frac{9}{16} \right) a(\mu)
\right] (p^2)^2 }
{ C_A \gamma^4(\mu) \left[ 1 + C_A \left( \frac{3}{8} \ln \left(
\frac{C_A \gamma^4(\mu)}{\mu^4} \right) - \frac{5}{8}
+ \frac{\pi p^2}{8 \sqrt{C_A} \gamma^2(\mu)} \right) a(\mu) \right]^2 }
\right] ~.
\end{equation}
Hence, evaluating this we find
\begin{equation}
\alpha^{\mbox{\footnotesize{eff}}}_s (0) ~=~ \frac{16}{\pi C_A} 
\label{alfre}
\end{equation}
which is different to that of \cite{33} since the effect of the omitted 
propagator has been correctly included. Numerically for $SU(3)$ we have 
$\alpha^{\mbox{\footnotesize{eff}}}_s (0)$~$=$~$1.698$ which is $4\%$ lower
than the value quoted in \cite{33}.

\sect{Static potential.}

We now return to the problem of computing the static potential. Given the 
reformulation of the basic QCD Lagrangian, (\ref{lqcd}), to incorporate the 
Gribov problem in terms of additional fields $\rho^{ab}_\mu$, $\xi^{ab}_\mu$, 
$\omega^{ab}_\mu$ and $\bar{\omega}^{ab}_\mu$, it is important to stress that 
these localizing fields are regarded as {\em internal} and do not couple to the
static potential source (\ref{jdef}) of (\ref{actj}). Ultimately one begins 
with a Lagrangian, (\ref{laggz}), involving a gluon and it is this quantum 
which is confined. Thus the Feynman rules for the source coupling are the same 
as those for the usual perturbative case. The main difference is that one uses 
the Feynman rules for the Gribov-Zwanziger Lagrangian, (\ref{laggz}). The 
latter is not a trivial point if one considers the Feynman diagrams 
contributing to the static potential. Concentrating, for the moment, on the 
single particle exchange graphs, the leading graph is illustrated in Figure $1$
where the sources are represented by the thick vertical lines and the gluon by 
the spring. However, at the next order there are corrections to the exchanged 
particle and due to the mixing in the $\{A^a_\mu, \xi^{ab}_\mu\}$ sector, there
are extra graphs aside from the gluon self-energy corrections. These are 
illustrated in Figure $2$ where the central blob denotes all possible one 
particle irreducible one loop corrections. The split propagators involving 
gluons and $\xi^{ab}_\mu$, denoted by the double line, are the $A^a_\mu$ 
$\xi^{bc}_\nu$ propagators. One key point is that as there is no direct 
coupling of either bosonic localizing ghost to a source then there is no single
exchange of this field akin to the diagram of Figure $1$. Whilst the exact one 
loop corrections to all the $2$-point functions, which lead to the propagator 
corrections, are known explicitly, we still have to assemble all the pieces for
the full static potential. For instance, in addition to the graphs of Figures 
$1$ and $2$, there are source gluon vertex corrections as well as double gluon 
exchange boxes at one loop.

\begin{figure}[ht]
\hspace{6cm}
\epsfig{file=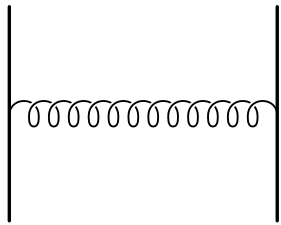,height=3cm}
\vspace{0.5cm}
\caption{Tree contribution to static potential.}
\end{figure}

\begin{figure}[ht]
\hspace{2.2cm}
\epsfig{file=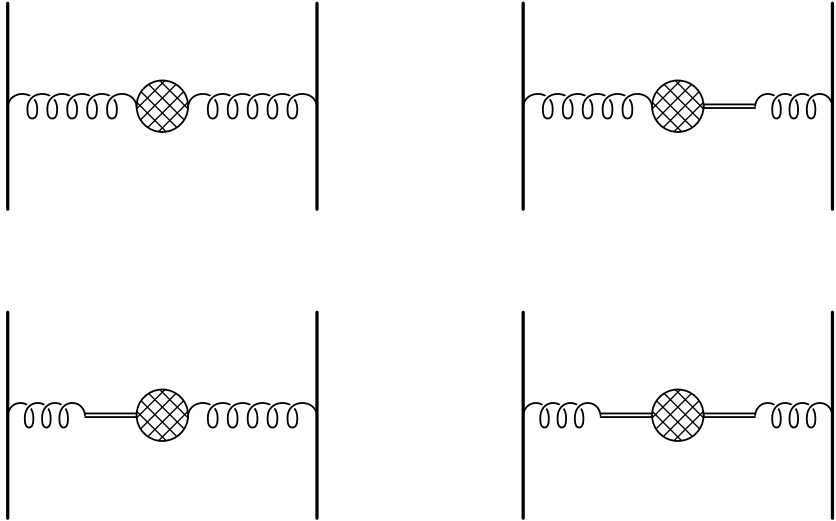,height=7cm}
\vspace{0.5cm}
\caption{Self-energy corrections to single gluon exchange graph.}
\end{figure}

Our calculation proceeds in general terms along the same lines as the earlier
work of \cite{39,40,41,42}. The main difference is that Feynman integrals with
source propagators as well as Gribov type propagators, as opposed to massless
ones, have to be handled for non-single exchange Feynman graphs. For those 
where either the source propagator or vertex is corrected one has the
potential problem of infrared divergences occurring. In the perturbative case
the momentum space resolution of this was given in, for example, 
\cite{39,40,41}. Briefly with massless propagators and a source propagator,
$1/(kv)$, an infrared divergence arises at the vertex correction but this
cancels the double gluon exchange infrared divergence. However, this statement 
needs to be qualified since in dimensional regularization, which we use here, 
the regularizing parameter $\epsilon$, where $d$~$=$~$4$~$-$~$2\epsilon$, 
cannot distinguish between infrared or ultraviolet infinities. Indeed in the 
original one loop Feynman gauge calculation of \cite{37} they actually cancel 
in the one loop correction to each of the vertices of Figure $1$. For other 
linear covariant gauges the infrared infinities cancel in the combination of 
double gluon exchange and source vertex corrections. In \cite{41} this infrared
problem was discussed in detail where a fictitious infrared regularizing mass
was introduced to explicitly isolate the infrared divergences. They were then
shown to cancel for an arbitrary linear covariant gauge. Indeed it was argued
that at one loop these infrared divergences are expected to cancel since 
the full Wilson loop graph that the infrared divergent graphs each actually
originate from, is the same diagram before it is cut open for the static limit.
As that original Wilson loop topology is infrared safe the infrared
cancellation could be regarded as a consistency check and the intermediate
regularizing fictitious mass safely removed. In the Gribov case the presence of
a natural mass arising from the Gribov parameter in the gluon propagators 
serves to act as a non-fictitious infrared regularization. Therefore, one can 
trace the infrared infinity naturally and verify its explicit cancellation.

Having discussed this technical issue then all that remains is to describe the
calculational set-up. The Feynman diagrams contributing to the one loop static
potential are generated using the {\sc Qgraf} package, \cite{59}. In total 
there are $1$ tree graph and $31$ one loop graphs which is more than the 
non-Gribov case due to the extra fields and mixed propagators. This latter 
total is broken down into one snail graph correcting the Gribov parameter of 
Figure $1$, $18$ single exchange graphs, $10$ source vertex corrections and $2$
double (gluon) exchange graphs. The topologies which are additional to those of
Figures $1$ and $2$ are provided in Figure $3$. All the graphs in {\sc Qgraf} 
notation are then converted into {\sc Form} notation in order to harness the 
power necessary to handle all the resulting intermediate algebra. This 
conversion adds all the colour and Lorentz indices to the fields. The Feynman 
rules are automatically inserted to produce the set of Feynman integrals which 
need to be evaluated explicitly. For the single particle exchange diagrams this
amounts to applying the basic integrals which were already treated in \cite{33}
where the exact one loop corrections to all the $2$-point functions in the 
theory were given. 

\begin{figure}[ht]
\hspace{2.8cm}
\epsfig{file=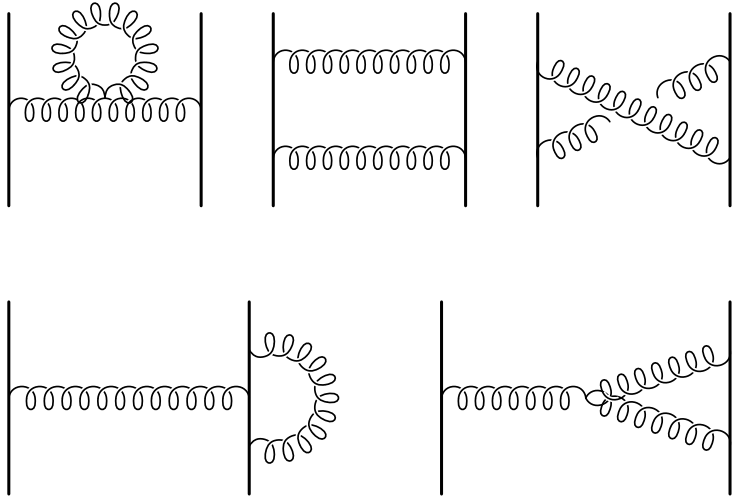,width=10cm}
\vspace{0.5cm}
\caption{Additional one loop topologies for static potential.}
\end{figure}

The final part is to handle the remaining box and source vertex corrections 
which will involve source propagators. Therefore, we define the general scalar
master integral
\begin{equation}
I_1(\alpha,\beta,\lambda,m_1^2,m_2^2,p,v) ~=~ \int \frac{d^dk}{(2\pi)^d}
\frac{1}{[k^2+m_1^2]^\alpha [(k-p)^2+m_2^2]^\beta (kv)^\lambda}
\label{masdef}
\end{equation}
for arbitrary masses $m_1$ and $m_2$ where $p_\mu$ is the external or exchange 
momentum and will satisfy $pv$~$=$~$0$, with $v^2$~$=$~$1$, in the static limit
throughout. Such integrals emerge after repeated substitution of the elementary
relations
\begin{equation}
kp ~=~ \frac{1}{2} \left[ k^2 ~+~ p^2 ~-~ (k-p)^2 \right]
\end{equation}
and
\begin{equation}
\frac{1}{[(k^2)^2+C_A\gamma^4]} ~=~ \frac{1}{2i\sqrt{C_A}\gamma^2} \left[ 
\frac{1}{[k^2-i\sqrt{C_A}\gamma^2]} ~-~ \frac{1}{[k^2+i\sqrt{C_A}\gamma^2]} 
\right] ~.
\end{equation}
Although (\ref{masdef}) has been given in \cite{41}, for completeness we give
several intermediate results we used, in our notation. Using integration by
parts it is possible to write a recurrence relation for (\ref{masdef}) as
\begin{eqnarray}
\lambda I_1(\alpha,\beta,\lambda+1,m_1^2,m_2^2,p,v) &=& -~ 2 \alpha 
I_1(\alpha+1,\beta,\lambda-1,m_1^2,m_2^2,p,v) \nonumber \\
&& -~ 2 \beta I_1(\alpha,\beta+1,\lambda-1,m_1^2,m_2^2,p,v) 
\end{eqnarray}
for $\lambda$~$\geq$~$2$, which allows one to reduce integrals with two source
propagators to a standard source free integral which has already been 
considered in the Gribov case in \cite{33}. For an odd number of source
propagators, this reduction will eventually lead to one source propagator but 
it can be treated using elementary contour integration in the time integral. 
So, for instance, we have, \cite{41}, 
\begin{equation}
I_1(1,1,1,m_1^2,m_2^2,p,v) ~=~ -~ \frac{i}{2} \int 
\frac{d^{d-1}{\mathbf k}}{(2\pi)^{d-1}} \frac{1}{[{\mathbf k}^2+m_1^2] 
[({\mathbf k}-{\mathbf p})^2+m_2^2]} ~.
\end{equation}
In other words in this case the treatment of the source propagator reduces the
dimensionality of the original integral by one. Finally, any remaining source
propagator free integral with propagator powers greater than unity can be
reduced by standard integration by parts identities. One final aspect of the
automatic calculation concerns the internal renormalization of the parameters
and sources. We follow the standard procedure of \cite{60} for this and compute
the static potential initially as a function of bare parameters. The
renormalized variables, and hence the underlying counterterms, are introduced
by replacing the bare parameters by the renormalized ones with the explicit
Landau gauge renormalization constants included. This also includes the 
renormalization associated with the source itself and that renormalization 
constant is derived in the same way as discussed in \cite{37} but for the
Landau gauge. For completeness, we note that the relevant $\MSbar$ Landau gauge
renormalization constants, in our conventions, are
\begin{eqnarray}
Z_A &=& 1 ~+~ \left[ \frac{13}{6} C_A - \frac{4}{3} T_F \Nf \right] 
\frac{a}{\epsilon} ~+~ O(a^2) \nonumber \\
Z_g &=& 1 ~+~ \left[ \frac{2}{3} T_F \Nf - \frac{11}{6} C_A \right]
\frac{a}{\epsilon} ~+~ O(a^2) \nonumber \\
Z_\gamma &=& 1 ~+~ \left[ \frac{1}{3} T_F \Nf - \frac{35}{48} C_A \right] 
\frac{a}{\epsilon} ~+~ O(a^2) \nonumber \\
Z_J &=& 1 ~+~ O(a^2) 
\end{eqnarray}
where $J^{a \, \mu}_{\mbox{\footnotesize{o}}}$~$=$~$\sqrt{Z_J} \, J^{a \, \mu}$
relates the bare and renormalized sources. Obviously a non-trivial check on the
renormalization procedure and the implementation of the explicit values for the
master Feynman integrals is that the ultimate static potential is finite and no
poles in $\epsilon$ remain when all the contributing diagrams are assembled. We
note now that this is indeed the case.

Having described the underlying tools for our calculation we now record that 
the one loop static potential for the Gribov-Zwanziger Lagrangian in the Landau
gauge is 
\begin{eqnarray}
\tilde{V}( \mathbf{p} ) 
&=& -~ \frac{C_F \mathbf{p}^2 g^2}{[(\mathbf{p}^2)^2+C_A \gamma^4]} 
\nonumber \\
&& +~ \left[ \frac{\pi \sqrt{C_A}}{768\gamma^2} 
- \frac{1}{768 \gamma^4} \tan^{-1} \left[ -~ 
\frac{\sqrt{4 C_A \gamma^4 - (\mathbf{p}^2)^2}}
{\mathbf{p}^2} \right] \sqrt{4 C_A \gamma^4 - (\mathbf{p}^2)^2}
\right. \nonumber \\
&& \left. ~~~~~+~ \frac{\sqrt{2}}{\gamma^2} \left[ 
\frac{\sqrt{C_A}}{768} \eta_1(\mathbf{p}^2) ~-~ 
\frac{\sqrt{C_A}}{192\sqrt{2}} \tan^{-1} 
\left[ \frac{\sqrt{C_A} \gamma^2}{\mathbf{p}^2} \right] \right] ~+~ 
\frac{231 \pi C_A^{3/2} \gamma^2}{128[(\mathbf{p}^2)^2+C_A \gamma^4]}
\right. \nonumber \\
&& \left. ~~~~~
+~ \frac{\sqrt{2}C_A \gamma^2}{[(\mathbf{p}^2)^2+C_A \gamma^4]} 
\left[ \frac{13 \sqrt{C_A}}{24\sqrt{2}} \tan^{-1} \left[ 
\frac{\sqrt{C_A} \gamma^2}{\mathbf{p}^2} \right] ~+~ \frac{79 \sqrt{C_A}}{128}
\eta_1(\mathbf{p}^2) \right] \right. \nonumber \\
&& \left. ~~~~~+~ \frac{2 \pi C_A^{3/2} \gamma^2}
{[(\mathbf{p}^2)^2-4C_A \gamma^4]} ~-~ 
\frac{1201 \pi C_A^{5/2} \gamma^6}{768[(\mathbf{p}^2)^2+C_A \gamma^4]^2} ~+~ 
\frac{\sqrt{2} C_A^{3/2} \gamma^2} {[(\mathbf{p}^2)^2+16 C_A \gamma^4]} 
\eta_1(\mathbf{p}^2) \right. \nonumber \\
&& \left. ~~~~~
-~ \frac{685 C_A^2 \gamma^4}{768[(\mathbf{p}^2)^2+C_A \gamma^4]^2} \tan^{-1} 
\left[ - \sqrt{ \frac{4 C_A \gamma^4 - (\mathbf{p}^2)^2 }{\mathbf{p}^2} } 
\right] \sqrt{ 4 C_A \gamma^4 - (\mathbf{p}^2)^2 }
\right. \nonumber \\
&& \left. ~~~~~-~ \frac{395\sqrt{2} C_A^{5/2} \gamma^6 \eta_1(\mathbf{p}^2)}
{768[(\mathbf{p}^2)^2+C_A \gamma^4]^2} ~+~ \frac{C_A}{\mathbf{p}^2} 
\left[ \frac{13}{96} \ln \left[ 1 + \frac{(\mathbf{p}^2)^2}{C_A\gamma^4} 
\right] - \frac{1}{192} \right]
\right. \nonumber \\
&& \left. ~~~~~+~ \frac{455 C_A}{384[(\mathbf{p}^2)^2+ C_A\gamma^4]} 
\tan^{-1} \left[ - \frac{\sqrt{4 C_A \gamma^4 - (\mathbf{p}^2)^2}} 
{\mathbf{p}^2} \right] \sqrt{ 4 C_A \gamma^4 - (\mathbf{p}^2)^2 }
\right. \nonumber \\
&& \left. ~~~~~+~ \frac{\pi C_A^{3/2} \gamma^2}{384(\mathbf{p}^2)^2} ~-~ 
\frac{C_A^{3/2} \gamma^2}{192(\mathbf{p}^2)^2} \tan^{-1} \left[ 
\frac{\sqrt{C_A}\gamma^2}{\mathbf{p}^2} \right] 
\right. \nonumber \\
&& \left. ~~~~~-~ \frac{C_A}{[(\mathbf{p}^2)^2-4 C_A \gamma^4]} 
\tan^{-1} \left[ \frac{\sqrt{4 C_A \gamma^4 - (\mathbf{p}^2)^2}}
{\mathbf{p}^2} \right] \sqrt{4 C_A \gamma^4 - (\mathbf{p}^2)^2} 
\right. \nonumber \\
&& \left. ~~~~~+~ 
\frac{C_A T_F \Nf \gamma^4 \mathbf{p}^2}{[(\mathbf{p}^2)^2+C_A \gamma^4]^2}
\left[ \frac{4}{3} \ln \left[ \frac{\mathbf{p}^2}{\mu^2} \right] 
- \frac{20}{9} \right] ~-~ 
\frac{T_F \Nf \mathbf{p}^2}{[(\mathbf{p}^2)^2+C_A \gamma^4]}
\left[ \frac{4}{3} \ln \left[ \frac{\mathbf{p}^2}{\mu^2} \right] 
- \frac{20}{9} \right] \right. \nonumber \\
&& \left. ~~~~~
+~ \frac{C_A^2 \gamma^4 \mathbf{p}^2}{[(\mathbf{p}^2)^2+C_A \gamma^4]^2} 
\left[ \frac{365}{72} - \frac{251}{192} \ln \left[ \frac{C_A \gamma^4}{\mu^4}
\right] - \frac{895\sqrt{2}}{3072} \eta_2(\mathbf{p}^2) 
- \frac{29}{96} \ln \left[ \frac{\mathbf{p}^2}{\mu^2} \right] \right]
\right. \nonumber \\
&& \left. ~~~~~+~ \frac{ C_A \mathbf{p}^2}{[(\mathbf{p}^2)^2+C_A \gamma^4]} 
\left[ \frac{125}{64} \ln \left[ \frac{C_A \gamma^4}{\mu^4} \right] ~-~ 
\frac{13}{48} \ln \left[ \frac{[C_A \gamma^4 + (\mathbf{p}^2)^2]}{\mu^4}
\right] ~-~ \frac{31}{9}
\right. \right. \nonumber \\
&& \left. \left. ~~~~~~~~~~~~~~~~~~~~~~~~~~~~~~~+~ \frac{167\sqrt{2}}{1536}
\eta_2(\mathbf{p}^2) ~+~ \frac{29}{96} \ln \left[ \frac{\mathbf{p}^2}{\mu^2} 
\right] 
\right] \right. \nonumber \\ 
&& \left. ~~~~~+~ \frac{\sqrt{2}C_A\mathbf{p}^2}
{4[(\mathbf{p}^2)^2+16 C_A\gamma^4]} \eta_2(\mathbf{p}^2) ~+~ 
\frac{\sqrt{2}\mathbf{p}^2}{3072\gamma^4} \eta_2(\mathbf{p}^2) 
\right] \frac{C_F g^4}{16\pi^2} ~+~ O(g^6) 
\label{stapot}
\end{eqnarray} 
where we use the notation that in four dimensions 
$p_\mu$~$=$~$(p_0,{\mathbf p})$ and we have introduced the intermediate
functions $\eta_1(\mathbf{p}^2)$ and $\eta_2(\mathbf{p}^2)$ for compactness
where
\begin{eqnarray}
\eta_1(\mathbf{p}^2) &=& 
-~ \ln \left[ 1 + \sqrt{ 1 
+ \frac{16 C_A \gamma^4}{(\mathbf{p}^2)^2} } \right] 
\sqrt{ - 1 + \sqrt{ 1 + \frac{16 C_A \gamma^4}{(\mathbf{p}^2)^2} } }
\nonumber \\
&& +~ \ln \left[ \frac{16 C_A \gamma^4}{(\mathbf{p}^2)^2} \right] 
\sqrt{ - 1 + \sqrt{ 1 + \frac{16 C_A \gamma^4}{(\mathbf{p}^2)^2} } }
\nonumber \\
&& -~ 2 \ln \left[ 
\sqrt{ 1 + \sqrt{ 1 + \frac{16 C_A \gamma^4}{(\mathbf{p}^2)^2} } } - \sqrt{2} 
\right] \sqrt{ - 1 + \sqrt{ 1 + \frac{16 C_A \gamma^4}{(\mathbf{p}^2)^2} } }
\nonumber \\
&& -~ 2 \tan^{-1} \left[ \frac{\sqrt{2}}{\sqrt{ - 1 + \sqrt{ 1 
+ \frac{16 C_A \gamma^4}{(\mathbf{p}^2)^2} } }} \right] 
\sqrt{ 1 + \sqrt{ 1 + \frac{16 C_A \gamma^4}{(\mathbf{p}^2)^2} } }
\end{eqnarray} 
and
\begin{eqnarray}
\eta_2(\mathbf{p}^2) &=& 
\ln \left[ \frac{16 C_A \gamma^4}{(\mathbf{p}^2)^2} \right] 
\sqrt{ 1 + \sqrt{1 + \frac{16 C_A \gamma^4}{(\mathbf{p}^2)^2} } } ~-~ 
\ln \left[ 1 + \sqrt{ 1 + \frac{16 C_A \gamma^4}{(\mathbf{p}^2)^2} } \right] 
\sqrt{ 1 + \sqrt{1 + \frac{16 C_A \gamma^4}{(\mathbf{p}^2)^2} } } \nonumber \\
&& -~ 2 \ln \left[ 
\sqrt{ 1 + \sqrt{1+\frac{16 C_A \gamma^4}{(\mathbf{p}^2)^2} } } 
- \sqrt{2} \right] 
\sqrt{ 1 + \sqrt{1 + \frac{16 C_A \gamma^4}{(\mathbf{p}^2)^2} } } \nonumber \\
&& +~ 2 \tan^{-1} \left[ \frac{\sqrt{2}}
{\sqrt{ - 1 + \sqrt{1+ \frac{16 C_A \gamma^4}{(\mathbf{p}^2)^2} } } } \right] 
\sqrt{ - 1 + \sqrt{1 + \frac{16 C_A \gamma^4}{(\mathbf{p}^2)^2} } } ~. 
\end{eqnarray} 
There is one main check on this result aside from the finiteness one above. 
This is that we recover the usual perturbative result in the 
$\gamma^2$~$\rightarrow$~$0$ limit. In other words we find
\begin{eqnarray}
\lim_{\gamma \rightarrow \, 0} \tilde{V}( \mathbf{p} ) &=& -~ 
\frac{4 \pi C_F \alpha_s(\mu)}{\mathbf{p}^2} \left[ 1 ~+~ \left[ \left[ 
\frac{31}{9} - \frac{11}{3} \ln \left[ \frac{\mathbf{p}^2}{\mu^2} \right] 
\right] C_A ~+~ \left[ \frac{4}{3} \ln \left[ \frac{\mathbf{p}^2}{\mu^2} 
\right] - \frac{20}{9} \right] T_F \Nf \right] a \right. \nonumber \\
&& \left. ~~~~~~~~~~~~~~~~~~~~+~ O(a^2) \right]
\label{potpert}
\end{eqnarray} 
where we note that (\ref{potpert}) agrees exactly with the earlier result of
\cite{36,37,38}. Whilst that was originally computed for the Feynman gauge 
taking the limit where the Gribov mass is removed recovers the perturbative 
result in the Landau gauge and its agreement with a gauge independent result is
another non-trivial check on our computational set-up.

We can now examine (\ref{stapot}) in the zero momentum limit by expanding the 
exact result in powers of ${\mathbf p}^2$ and find
\begin{equation}
\tilde{V} ( \mathbf{p} ) ~=~ -~ \frac{C_F \mathbf{p}^2 g^2}{C_A \gamma^4} ~-~ 
C_F \left[ \frac{\pi\sqrt{C_A}}{32\gamma^2} ~+~ \left( \frac{13}{72} 
- \frac{3}{8} \ln \left( \frac{C_A \gamma^4}{\mu^4} \right) \right) 
\frac{\mathbf{p}^2}{\gamma^4} \right] \frac{g^4}{16\pi^2} ~+~ 
O((\mathbf{p}^2)^2;g^6)
\label{potzero}
\end{equation}
where the order symbols denote the higher order terms in the momentum expansion
and two loop corrections separately. Thus (\ref{potzero}) would imply that
\begin{equation}
\tilde{V} (0) ~=~ -~ \frac{C_F\sqrt{C_A}g^4}{512\pi\gamma^2} ~+~
O(g^6) 
\label{potfrez}
\end{equation}
as a first approximation with no emergence of a dipole. To try and understand 
the implications of this freezing for the one loop coordinate space potential 
we first consider the leading term of (\ref{potzero}). For $\gamma$~$=$~$0$ the
exchange of Figure $1$ leads to the usual Coulomb potential but for non-zero 
$\gamma$ it is not clear what the $p^2$~$\rightarrow$~$0$ limit of the leading
term of (\ref{potzero}) implies for the corresponding 
$r$~$\rightarrow$~$\infty$ limit. If we write
\begin{equation}
\tilde{V} ( \mathbf{p} ) ~=~ \sum_{n=0}^\infty \tilde{V}_n (\mathbf{p}) 
\end{equation}
where the subscript $n$ labels the {\em loop} order, then the first term of 
(\ref{stapot}) gives 
\begin{equation}
\tilde{V}_0(\mathbf{p}) ~=~ -~ 
\frac{C_F\mathbf{p}^2 g^2}{[(\mathbf{p}^2)^2 + C_A \gamma^4]}
\end{equation}
whence the Fourier transform, (\ref{foudef}), leads to 
\begin{equation}
V_0(r) ~=~ -~ \frac{C_F g^2}{4\pi r} \exp \left[ - 
\frac{C_A^{\quarter} \gamma r}{\sqrt{2}} \right] \cos \left( 
\frac{C_A^{\quarter} \gamma r}{\sqrt{2}} \right) ~. 
\label{treepotr}
\end{equation}
The presence of the exponential factor is not unexpected since it is 
reminiscent of the Yukawa potential for a massive field. However, the Gribov
situation effectively corresponds to a width which is reflected in the
trignometric factor. Now examining the limit corresponding to
$p^2$~$\rightarrow$~$0$, which is $r$~$\rightarrow$~$\infty$ for {\em real}
$r$, we see that $V_0(r)$~$\rightarrow$~$0$. This is consistent on dimensional
grounds with naively taking the Fourier transform of the leading term of
(\ref{potzero}). As an aside we remark that similar potentials of this Friedel
form have recently emerged in leading order in models of plasmas and stellar 
nuclear reactions, \cite{61,62,63,64,65}. Briefly the underlying theory is 
based on a Higgs-like effective Lagrangian. By considering perturbations about 
a background then the gauge field fluctuations develop a Gribov type 
propagator. One claim is that a linearly rising potential will emerge in a
particular limit if the effective Higgs mass term has the wrong sign.

Turning to the one loop correction we do not have the luxury of being able to 
take the full Fourier transform of (\ref{stapot}) analytically. However, by the
same heuristic argument as leading order, the freezing to a finite value, 
(\ref{potfrez}), leads to a similar large distance limit for our potential. In 
other words $V(r)$~$\rightarrow$~$0$ at one loop and there is no linear growth.
However, on dimensional grounds such a term will clearly not be Coulombic in 
this limit. A $1/\mathbf{p}^2$ term would be required for that in the zero 
momentum limit. Indeed given this the next step would be to try and understand 
how the L\"{u}scher term, \cite{66}, emerges in the Gribov-Zwanziger formalism 
as $r$~$\rightarrow$~$\infty$. Within the present context it is not clear how 
such a term could appear. For instance, aside from the fact that the analysis 
of \cite{66} producing this term was based on chromoelectric flux tubes, the 
coefficient of the L\"{u}scher term is universal and independent of the 
coupling constant. One prospective way this could happen is that the 
coefficient of a $1/\mathbf{p}^2$ term, when it arises, is fixed by the gap 
equation satisfied by $\gamma$ with or without some renormalization group 
running of parameters to some non-perturbative fixed point. Moreover, the 
L\"{u}scher term appears together with a linearly rising potential term, 
\cite{66}, which would originate in momentum space from a dipole. However, a 
dipole behaviour in the zero momentum limit is clearly not evident in 
(\ref{stapot}). Although we have not taken the complete Fourier transform of 
(\ref{stapot}), we have examined it in detail for a confining behaviour to 
understand why a dipole is absent and there are some promising features. 

Clearly in (\ref{stapot}) there is an explicit $1/(\mathbf{p}^2)^2$ term which
gives a stronger singularity than that of the usual Coulomb case. Indeed
concentrating on this term for the moment and setting 
\begin{equation}
A_l ~=~ \frac{C_F C_A^{3/2}\gamma^2 g^4}{6144\pi}
\label{dip1}
\end{equation}
in (\ref{toypot}) we would have
\begin{equation}
V_l(r) ~=~ -~ \frac{C_F C_A^{3/2}\gamma^2 g^4}{49152\pi^2} r 
\end{equation} 
which would actually be a linearly {\em decreasing} potential. Ignoring this
point for the moment, there is, however, another dipole singularity in 
(\ref{stapot}) and this occurs in such a way that there is no overall 
singularity as $p^2$~$\rightarrow$~$0$. The specific term is 
\begin{equation}
-~ \frac{C_F C_A^{3/2} \gamma^2g^4}{3072\pi^2(\mathbf{p}^2)^2} \tan^{-1} 
\left[ \frac{\sqrt{C_A}\gamma^2}{\mathbf{p}^2} \right] 
\label{dip2}
\end{equation}
and in the zero momentum limit the inverse tan function tends to $\pi/2$ to 
cancel the previous pure dipole we considered. Thus, a linearly rising 
potential will not emerge. Actually for (\ref{dip2}) it is in fact possible to 
take the Fourier transform (\ref{foudef}) and verify that the overall absolute 
asymptotic behaviour as $r$~$\rightarrow$~$0$ is linear and precisely cancels 
that from the pure dipole term Fourier transform. However, we recall what was 
noted in \cite{33} concerning the overall sign of $\gamma^2$. In studying the
zero momentum limits of (\ref{dip1}) and (\ref{dip2}) we have tacitly assumed
that $\gamma^2$ is positive. The sign of $\gamma^2$, though, is not 
predetermined in (\ref{laggz}). It can only strictly be determined in relation 
to other evidence. Indeed it was suggested in \cite{33} that for consistency 
with other results, such as the power correction structure of the strong 
coupling constant in the next to high energy limit, that $\gamma^2$ was
negative and our sign convention needed to be modified by mapping 
$\gamma^2$~$\rightarrow$~$-$~$\gamma^2$ in situations where results depended 
explicitly on $\gamma^2$. It is worth noting that this is not cause for alarm 
since a similar situation always occurs with the choice of the sign of the 
coupling constant, $g$, in the covariant derivative. Its sign cannot be 
determined since in any computations it appears in the combination $g^2$. In
other words the choice of sign of $g$ is a convention as is that of $\gamma^2$
with the difference being here that one can actually make contact with other 
methods, such as the strong coupling constant power corrections, to determine 
it, \cite{33}. In our current set-up we had not made any a priori choice of 
sign and so there is a degree of freedom to make a specific choice. Therefore, 
if we return to (\ref{dip1}) and (\ref{dip2}), then since the former is an odd 
function and the latter is an even function of $\gamma^2$, flipping the sign of
$\gamma^2$ would mean that in fact the terms add and there is a net dipole.
This would lead to a linearly {\em rising} potential. Although this is 
encouraging and would very much be consistent with the expectation that the 
Gribov-Zwanziger Lagrangian describes a confined gluon since, for instance, it 
satisfies the Kugo-Ojima criterion, \cite{34,35}, it would on the other hand 
destroy gluon suppression, \cite{16}. This can be seen if one examines the 
explicit one loop corrections to the gluon propagator given in appendix B. More 
specifically examining the relevant pieces there, the net dipole arises purely 
from the $\xi^{ab}_\mu$ $2$-point function corrections in the one loop gluon 
propagator. Indeed this would be consistent with the Zwanziger localizing 
fields dominating the infrared as one moves towards the Gribov horizon. If such
a scenario is to be credible, however, one needs to check that at next loop 
order a triple pole in $\mathbf{p}^2$ does not arise in the static potential.
Such a term is possible on dimensional grounds, for instance. We stress, 
though, at this point that these are merely interim speculative remarks on the 
consequences of the sign choice of $\gamma^2$, which would require deeper 
consideration, and we will omit reference to this point hereafter.

Therefore, returning to our original $\gamma^2$ convention this dipole 
cancellation is not unexpected. Indeed if one examines the explicit exact one 
loop corrections to each of the transverse parts of the $2$-point functions in 
the $\{A^a_\mu, \xi^{ab}_\mu, \rho^{ab}_\mu\}$ sector, this dipole structure is
present in each case in individual terms but with no net dipole. Thus the 
apparent dipole behaviour is not truly present overall. However, on more 
general terms this lack of a linearly rising potential should not emerge in the
present set-up. Although we have followed an inherently perturbative approach 
the main lesson from Gribov's original article, which also underlies 
Zwanziger's construction, is that infrared properties emerge when the gap 
equation is realised. Then one is dealing with a gauge theory. For example, the
gap equation leads to Faddeev-Popov ghost enhancement and more recently 
produced a qualitative non-zero value of a renormalization group invariant 
coupling constant in the zero momentum limit, (\ref{alfre}). However, at no 
point have we implemented the gap equation within our static potential. (The 
simple replacement of $\gamma$ by a non-perturbative function of $a$ is not 
what we mean by this here.) Somehow, one would expect that a linearly rising 
potential can only emerge when the gap equation is used {\em explicitly}. In a 
later section we will present some considerations towards this point in the 
context of the static potential considered here. Though, recalling an earlier 
remark, at higher loop orders there would appear to be nothing in principle 
preventing pure terms such as $1/(\mathbf{p}^2)^3$ and higher emerging on 
dimensional grounds. These would have to have a compensating piece to exclude 
power terms in $r$ remaining in the coordinate space potential. Of course such 
terms would be irrelevant if the sign associated with it indicated a decreasing
contribution to the potential in contrast to an increasing part. Alternatively 
such terms might resum in such a way that there is matching to a linear type 
potential. One final point concerning the linear term relates to the soft BRST 
breaking of (\ref{laggz}) arising from the $\gamma^2$ term, \cite{12,13,18,30}.
There is an understanding that for gauge independent quantities results should 
depend on $\gamma^4$ and not on $\gamma^2$. However, on dimensional grounds a 
linear term in a potential must be accompanied by a factor of $\gamma^2$ and 
therefore it seems impossible to avoid having a $\gamma^2$ dependence in the 
ultimate static potential which is derived from the underlying gauge 
independent Wilson loop. Though, of course, any such scale can eventually in
principle be related back to $\Lambda_{\mbox{\footnotesize{$\MSbar$}}}$.

Next we make some specific remarks concerning (\ref{stapot}). Given that the 
gluon is now not a massless entity in (\ref{laggz}), we note that
(\ref{stapot}) has a pole and physical cut at $p^2$~$=$~$2\sqrt{C_A}\gamma^2$. 
This is a threshold type feature due to massless fields with a non-zero width. 
The associated denominator factor in the term with this cut arises from the 
Gram determinant of the integration by parts reduction formula for the one loop
Feynman integrals without source propagators. In \cite{13}, the glueball 
correlation function was analysed as a spectral density function and it was 
argued that there was a bound state of the same mass squared value, 
$2\sqrt{C_A}\gamma^2$, and it was identified as a potential glueball state. It 
is amusing that there is a threshold effect at the {\em same} mass value in the
static potential which is in some sense a scattering amplitude. It is not 
clear, though, if this corresponds to a {\em stable} bound state. However, it 
is worth recalling the computation of \cite{67} where the glueball spectrum was
determined by including a massive gluon in the evaluation of the leading order 
gluon scattering amplitude. This was then used to construct a potential within 
which bound states could be formed. However, the scattering amplitude with the 
massive gluons did not contain an explicit confining potential but this was 
added by hand prior to solving for the spectrum. Though strictly speaking the 
additional confining piece not only incorporated a linear rising part but also 
modelled string breaking. This final form of the potential energy appears to 
have endowed a stability on the glueball states. Therefore, it does not seem 
unreasonable to expect that bound state stability requires an explicit dipole 
term in (\ref{stapot}), with string breaking, for similar reasons. For 
completeness we note that in \cite{41} massive $W$ and $Z$ gauge bosons were 
included in the static potential formalism for a standard model study but there
the {\em real} mass did not lead to the same threshold singularity as 
(\ref{stapot}). Next we note that one can isolate the contribution to 
(\ref{stapot}) from purely single particle exchange diagrams corresponding to 
all the graphs of Figures $1$ and $2$, since it too has similar dipole 
behaviour. Indeed it could be the case that the single exchange graphs have a
net dipole term which is cancelled by a matching piece from the box and vertex
graphs. However, analysing the single exchange contribution in the same way as
(\ref{stapot}) the zero momentum limit is also non-singular. Though in 
comparison with (\ref{stapot}) there is an additional dipole-like term. As 
$p^2$~$\rightarrow$~$0$ these three terms conspire in a similar way as before 
to exclude an overall $1/(p^2)^2$ singularity. 

Although the dipole term does not emerge, we close this section by discussing
other aspects of (\ref{stapot}) in the context of \cite{36,37,38,39,40,41}. In
the perturbative approach one can define a strong coupling constant which 
matches the usual ultraviolet behaviour of a strong coupling constant in the
large momentum limit. It is based on the momentum space behaviour of the
potential and is defined by
\begin{equation} 
\tilde{V} ( \mathbf{p} ) ~=~ -~ \frac{4\pi C_F}{\mathbf{p}^2} 
\alpha_V(\mathbf{p})
\label{vdef}
\end{equation}
where the subscript ${}_V$ denotes the $V$-scheme, \cite{68,69}. One advantage
of this definition is that it is derived from a gauge independent quantity 
which is effectively the force between two coloured objects. The low momentum
behaviour can be studied. Clearly in the Gribov-Zwanziger case at both leading 
order and one loop $\alpha_V(0)$~$=$~$0$. Though for the former the behaviour
in the zero momentum limit is $O\left( (p^2)^2 \right)$ but $O(p^2)$ in the 
latter. If, however, an overall dipole piece dominated the infrared then
$\alpha_V(\mathbf{p})$ would be singular at low momentum when one factor of
$\mathbf{p}^2$ is cancelled from the coupling constant definition. This would 
be consistent with confinement in this scheme when the strength of a suitably 
defined coupling constant increases to produce infrared slavery. Though in the 
absence of quarks there would appear to be no string breaking related to 
saturation discussed, for example, in \cite{67}. Given these comments it might 
be an interesting exercise for lattice or Dyson Schwinger methods to be applied
to the static potential specifically in the Gribov-Zwanziger Landau gauge case.
This may be particularly apt given the current debate over whether the scaling 
or decoupling solution is the correct picture. Briefly, in essence we have 
concentrated on using the formalism associated with the scaling solution. In
the decoupling scenario, \cite{22,23,24,25,26,27,28,29}, it transpires that the 
gluon is not suppressed in the infrared, since the propagator freezes to a 
finite non-zero value but the form factor vanishes, and the Faddeev-Popov ghost
is not enhanced. It has a behaviour which is close to a $1/p^2$ dependence as 
$p^2$~$\rightarrow$~$0$ rather than the dipole of the Gribov scenario, 
\cite{22,23,24,25,26,27,28,29}.

\sect{Gap equation.}

In this section we focus on several aspects of the Gribov gap equation which
defines the Gribov parameter $\gamma$ using (\ref{laggz}) to correct the error
in the earlier two loop gap equation of \cite{32}. In our current notation the
gap equation is defined by the expectation value
\begin{equation}
f^{abc} \langle A^{a\,\mu} (x) \xi^{bc}_\mu(x) \rangle ~=~ 
\frac{i d\NA \gamma^2}{g^2} ~.
\end{equation}
Whilst the new features we present here are not immediately related to the 
static potential, we believe they are important and the computations do 
contribute to understanding further underlying properties of the 
Gribov-Zwanziger Lagrangian. Throughout our calculations $\gamma$ appears 
explicitly but is not an independent parameter of the theory. As emphasised in 
\cite{11,12,13,14,15,16,17,18}, the Lagrangian can only be interpreted as a 
gauge theory when the gap equation is satisfied explicitly. The original one 
loop expression was derived in \cite{10} in the $\MSbar$ scheme and it is 
straightforward to invert it to determine $\gamma$ as an explicit function of 
the coupling constant. For instance, we have, \cite{10}, 
\begin{equation} 
1 ~=~ C_A \left[ \frac{5}{8} ~-~ \frac{3}{8} \ln \left( 
\frac{C_A\gamma^4}{\mu^4} \right) \right] a ~+~ O(a^2) 
\label{gap1}
\end{equation} 
which gives 
\begin{equation}
\frac{C_A \gamma^4}{\mu^4} ~=~ \exp \left[ \frac{5}{3} ~-~ 
\frac{32\pi}{3C_A \alpha_s(\mu)}
\right]
\label{gapsoln1}
\end{equation}
or expressing it in numerical form we have
\begin{equation}
\frac{C_A \gamma^4}{\mu^4} ~=~ 5.294 \exp 
\left[ \,-~ \frac{33.510}{C_A \alpha_s(\mu)} \right] ~. 
\end{equation}
(In this section we revert to the $\MSbar$ scheme rather than continue with the
$V$-scheme.) This clearly demonstrates the non-perturbative behaviour of the 
Gribov mass parameter. However, as the two loop $\MSbar$ correction to the mass
gap is now available, \cite{32}, one interesting question is whether it is 
possible to rearrange the explicit expression to preserve the non-perturbative 
property emerging from the one loop result. Therefore, to do this we take 
\begin{equation} 
\frac{C_A \gamma^4}{\mu^4} ~=~ c_0 [ 1 ~+~ c_1 C_A \alpha_s(\mu) ] 
\exp \left[ \,-~ \frac{b_0}{C_A \alpha_s(\mu)} \right] ~.
\label{gapansz}
\end{equation}
as an ansatz for a corrected form of (\ref{gapsoln1}). Substituting this into 
the explicit gap equation, whose correct form now is, 
\begin{eqnarray} 
1 &=& C_A \left[ \frac{5}{8} - \frac{3}{8} \ln \left( 
\frac{C_A\gamma^4}{\mu^4} \right) \right] a \nonumber \\ 
&& +~ \left[ C_A^2 \left( \frac{3893}{1536} - \frac{22275}{4096} s_2
+ \frac{29}{128} \zeta(2)
- \frac{65}{48} \ln \left( \frac{C_A\gamma^4}{\mu^4} \right)
+ \frac{35}{128} \left( \ln \left( \frac{C_A\gamma^4}{\mu^4} \right)
\right)^2 \right. \right. \nonumber \\
&& \left. \left. ~~~~~~~~~~~~+~ \frac{411}{1024} \sqrt{5} \zeta(2) 
- \frac{1317\pi^2}{4096} \right) \right. \nonumber \\
&& \left. ~~~~~+~ C_A T_F \Nf \left( \frac{\pi^2}{8} - \frac{25}{24} - \zeta(2)
+ \frac{7}{12} \ln \left( \frac{C_A\gamma^4}{\mu^4} \right)
- \frac{1}{8} \left( \ln \left( \frac{C_A\gamma^4}{\mu^4} \right) \right)^2 
\right) \right] a^2 \nonumber \\
&& +~ O(a^3)
\label{gap2}
\end{eqnarray} 
where $s_2$~$=$~$(2\sqrt{3}/9) \mbox{Cl}_2(2\pi/3)$ and $\mbox{Cl}_2(x)$ is the
Clausen function, it ought to be possible to determine explicit expressions for
the unknown coefficients $b_0$ and $c_i$. If one cannot find a solution then an
alternative approach would be needed. Remarkably, it transpires that to two 
loop order one can achieve the inversion for arbitrary numbers of massless 
quarks and we find
\begin{eqnarray}
b_0 &=& \frac{32\pi\left[3C_A - \sqrt{79C_A^2-32C_A T_F \Nf}\right]}
{[35C_A-16T_F\Nf]} \nonumber \\ 
c_0 &=& \exp \left[ \frac{1}{[105C_A - 48 T_F \Nf]} 
\left[ 260 C_A - 112 T_F \Nf 
- \frac{[255C_A - 96 T_F \Nf] C_A}{\sqrt{79 C_A^2 - 32 C_A T_F \Nf}} \right]
\right] \nonumber \\ 
c_1 &=& \left[ 25855714080 \sqrt{5} \zeta(2) C_A^4 
             - 32766156288 \sqrt{5} \zeta(2) C_A^3 \Nf T_F 
             + 13817806848 \sqrt{5} \zeta(2) C_A^2 \Nf^2 T_F^2 
\right. \nonumber \\
&& \left. ~-~ 1939341312 \sqrt{5} \zeta(2) C_A \Nf^3 T_F^3 
- 20712880440 \pi^2 C_A^4 - 350326053000 s_2 C_A^4 
\right. \nonumber \\
&& \left. ~+~ 14594952960 \zeta(2) C_A^4 + 56581367360 C_A^4
+ 34301188224 \pi^2 C_A^3 \Nf T_F \right. \nonumber \\
&& \left. ~+~ 443957500800 s_2 C_A^3 \Nf T_F
- 82914840576 \zeta(2) C_A^3 \Nf T_F - 94986935296 C_A^3 \Nf T_F
\right. \nonumber \\
&& \left. ~-~ 21273919488 \pi^2 C_A^2 \Nf^2 T_F^2
             - 187221196800 s_2 C_A^2 \Nf^2 T_F^2
+ 89436192768 \zeta(2) C_A^2 \Nf^2 T_F^2 \right. \nonumber \\
&& \left. ~+~ 59735801856 C_A^2 \Nf^2 T_F^2 + 5856952320 \pi^2 C_A \Nf^3 T_F^3
             + 26276659200 s_2 C_A \Nf^3 T_F^3
\right. \nonumber \\
&& \left. ~-~ 35521560576 \zeta(2) C_A \Nf^3 T_F^3 
- 16679698432 C_A \Nf^3 T_F^3 - 603979776 \pi^2 \Nf^4 T_F^4 
\right. \nonumber \\
&& \left. ~+~ 4831838208 \zeta(2) \Nf^4 T_F^4 
+ 1744830464 \Nf^4 T_F^4 \right] \nonumber \\
&& \times ~ 
\frac{1}{36864 \pi [ 79 C_A - 32 T_F \Nf ]^{5/2} [ 35 C_A - 16 T_F \Nf ] 
\sqrt{C_A}} ~.  
\end{eqnarray} 
To assist with comparisons we have evaluated the coefficients numerically for
both $SU(2)$ and $SU(3)$ colour groups and for several values of $\Nf$ and 
these are given respectively in Tables $1$ and $2$. Finally, we note that to 
this order in the gap equation (\ref{gapansz}) can be rewritten as  
\begin{equation} 
\frac{C_A \gamma^4}{\mu^4} ~=~ c_0 \exp 
\left[ \,-~ \frac{b_0}{C_A \alpha_s(\mu)} ~+~ c_1 C_A \alpha_s(\mu) \right] 
\end{equation}
with the same values for $c_0$, $c_1$ and $b_0$.

\begin{table}[ht]
\begin{center}
\begin{tabular}{|c||c|c|c|}
\hline
$\Nf$ & $b_0$ & $c_0$ & $c_1$ \\
\hline 
$0$ & $16.913$ & $9.052$ & $-$ $0.133$ \\ 
$2$ & $18.383$ & $8.994$ & $-$ $0.180$ \\ 
$3$ & $19.303$ & $8.963$ & $-$ $0.210$ \\ 
\hline 
\end{tabular}
\end{center}
\begin{center}
{Table 1. Numerical values of parameters for $SU(2)$.}
\end{center}
\end{table}

{\begin{table}[ht]
\begin{center}
\begin{tabular}{|c||c|c|c|}
\hline
$\Nf$ & $b_0$ & $c_0$ & $c_1$ \\
\hline 
$0$ & $16.913$ & $9.052$ & $-$ $0.133$ \\ 
$2$ & $17.864$ & $9.014$ & $-$ $0.163$ \\ 
$3$ & $18.383$ & $8.994$ & $-$ $0.180$ \\ 
\hline 
\end{tabular}
\end{center}
\begin{center}
{Table 2. Numerical values of parameters for $SU(3)$.}
\end{center}
\end{table}} 

Although the perturbative correction to the one loop form appears to be 
converging the actual convergence cannot fully be commented on until a three
loop gap equation is available. One indication of the issues related to this
is the change in value of $c_0$ between loop orders. In one sense this is not
unexpected since the ansatz we have taken has a relatively simple form and may 
not be the best approximation to the full relation for $\gamma$ as an explicit 
function of the coupling constant which is not known. To illustrate the
subtleties of solving the gap equation at two loops, we had tried to solve 
(\ref{gap2}) numerically as a quadratic in the coupling constant. However, this
was problematic since for certain values of the parameters, it resulted in 
solutions where there was an imaginary part. This was in contradiction with the
understanding that $\gamma$ and the coupling constant are real parameters.
Moreover, if the solution to the quadratic had turned out to be real, it was
not clear whether this would persist at higher loop orders when one would in
principle have to solve cubic and quartic equations. The potential
proliferation of roots would at some point be problematic and render such an
approach meaningless. Therefore, we took the ansatz approach discussed above.
As an aside we note that with the new two loop $\MSbar$ gap equation,
(\ref{gap2}), we have verified that the Kugo-Ojima confinement criterion, 
\cite{34,35}, still holds and that both the Faddeev-Popov and $\omega^{ab}_\mu$
propagators enhance in the zero momentum limit at two loops, as before. 

We turn now to another aspect of the gap equation and that is to do with its
form in another renormalization scheme. Whilst the $\MSbar$ scheme is standard
in (high order) loop calculations it is not always the most appropriate since
it is not a scheme motivated from a physical point of view. One such scheme,
however, is the MOM scheme, \cite{70,71}, which has also been used, for 
example, in lattice calculations of the gluon $2$-point function and the 
evaluation of the renormalization group invariant effective strong coupling 
constant, \cite{72,73,74}. The latter is derived from the gluon ghost vertex, 
though the triple gluon vertex has also been considered. As the MOM scheme 
seems to be the one most studied in that context it would seem appropriate to 
develop the Gribov gap equation in the MOM scheme to see if there is any scheme
dependence. Therefore, we repeat the computation of section $3$ but for the MOM
scheme which is not a trivial exercise.

As a starting point we note that the exact expressions for the one loop 
$2$-point functions are required as a function of $\gamma$. These are necessary
since the MOM scheme is a mass dependent scheme and therefore the {\em full} 
expressions are needed to determine the correct finite parts for the MOM 
renormalization constants. However, as in the $\gamma$~$=$~$0$ case the MOM 
renormalization has to be carried out in such a way that the underlying 
Slavnov-Taylor identities are not violated. For the Gribov-Zwanziger Lagrangian 
specifically, this means the renormalization constants for $A^a_\mu$, $c^a$, 
$\rho^{ab}_\mu$, $\xi^{ab}_\mu$, $\omega^{ab}_\mu$ and $\gamma$, which are 
$Z_A$, $Z_c$, $Z_\rho$, $Z_\xi$, $Z_\omega$ and $Z_\gamma$, must satisfy, 
\cite{18,30,31}, 
\begin{equation}
Z_c ~=~ Z_\rho ~=~ Z_\xi ~=~ Z_\omega ~=~ \frac{1}{Z_g \sqrt{Z_A}} ~~~,~~~ 
Z_\gamma ~=~ \left( Z_A Z_c \right)^{-1/4} 
\label{rencondef}
\end{equation}
where $Z_g$ is the coupling constant renormalization constant. Given this we
have proceeded as follows. First, we determined the wave function 
renormalization constants for $A^a_\mu$, $\rho^{ab}_\mu$ and $\xi^{ab}_\mu$ by 
rendering the respective $2$-point functions exactly equal to unity when the 
external momentum satisfies $p^2$~$=$~$\mu^2$ where $\mu$ is the mass scale of 
the renormalization point. We record the explicit forms for $Z_A$, $Z_\rho$ and
$Z_\xi$ in the MOM scheme are
\begin{eqnarray}
Z_A^{\mbox{MOM}} &=& 1 ~+~ \left[ \frac{13}{6} C_A - \frac{4}{3} T_F \Nf
\right] \frac{a}{\epsilon} - \frac{20}{9} T_F \Nf a \nonumber \\
&& +~ \left[ \frac{1939}{576} ~+~ \frac{7}{64} \ln 
\left[ \frac{[C_A\gamma^4+\mu^4]}{\mu^4} \right] ~-~ \frac{135}{128} 
\ln \left[ \frac{C_A\gamma^4}{\mu^4} \right] ~-~ \frac{241\sqrt{2}}{768} 
\eta_2(\mu^2) \right. \nonumber \\
&& \left. ~~~~~+~ \frac{25\mu^2\sqrt{4C_A \gamma^4 - \mu^4}}{768C_A\gamma^4} 
\tan^{-1} \left[ -~ \frac{\sqrt{4C_A \gamma^4 -\mu^4}}{\mu^2} \right] ~-~
\frac{131\pi\mu^2}{768\sqrt{C_A}\gamma^2} \right. \nonumber \\
&& \left. ~~~~~-~ \frac{37\pi\sqrt{C_A}\gamma^2}{128\mu^2} ~-~ 
\frac{3C_A\gamma^4\sqrt{4C_A \gamma^4 - \mu^4}}{8\mu^6} 
\tan^{-1} \left[ -~ \frac{\sqrt{4C_A \gamma^4 -\mu^4}}{\mu^2} \right] 
\right. \nonumber \\ 
&& \left. ~~~~~+~ \left[ \frac{35}{64} - \frac{47}{192} \ln 
\left[ 1 + \frac{\mu^4}{C_A\gamma^4} \right] \right] \frac{C_A\gamma^4}{\mu^4}
~-~ \frac{59\pi C_A^{3/2} \gamma^6}{128\mu^6}
\right. \nonumber \\ 
&& \left. +~ \frac{11 C_A^{3/2} \gamma^6}{64\mu^6} \tan^{-1} \left[
\frac{\sqrt{C_A}\gamma^2}{\mu^2} \right] ~-~ 
\frac{7\sqrt{4C_A \gamma^4 - \mu^4}}{192\mu^2} 
\tan^{-1} \left[ -~ \frac{\sqrt{4C_A \gamma^4 -\mu^4}}{\mu^2} \right] 
\right. \nonumber \\ 
&& \left. +~ \left[ \frac{1}{96} \ln \left[ \frac{C_A\gamma^4}{\mu^4} 
\right] ~-~ \frac{1}{48} \ln \left[ 
\frac{[C_A\gamma^4+\mu^4]}{\mu^4} \right] ~-~ \frac{3\sqrt{2}}{1024} 
\eta_2(\mu^2) \right] \frac{\mu^4}{C_A\gamma^4} \right. \nonumber \\
&& \left. +~ \left[ \frac{19}{192} \tan^{-1} \left[ 
\frac{\sqrt{C_A}\gamma^2}{\mu^2} \right] ~-~ \frac{35\sqrt{2}}{384} 
\eta_1(\mu^2) \right] \frac{\mu^2}{\sqrt{C_A}\gamma^2} 
\right. \nonumber \\
&& \left. -~ \left[ \frac{25\sqrt{2}}{64} \eta_1(\mu^2) ~+~ 
\frac{23}{48} \tan^{-1} \left[ \frac{\sqrt{C_A}\gamma^2}{\mu^2} \right] \right] 
\frac{\sqrt{C_A}\gamma^2}{\mu^2} \right] C_A a ~+~ O(a^2) 
\end{eqnarray} 
and
\begin{eqnarray}
Z_\rho^{\mbox{MOM}} &=& Z_\xi^{\mbox{MOM}} \nonumber \\
&=& 1 ~+~ \frac{3C_A}{4} \frac{a}{\epsilon} ~+~ 
\left[ \frac{5}{4} - \frac{3}{8} \ln \left[ \frac{[C_A\gamma^4+\mu^4]}{\mu^4} 
\right] ~-~ \frac{C_A\gamma^4}{8\mu^4} \ln \left[
\frac{C_A\gamma^4}{[C_A\gamma^4+\mu^4]} \right] \right. \nonumber \\
&& \left. -\, \frac{3\pi\sqrt{C_A}\gamma^2}{8\mu^2} + \left[ 
\frac{3\sqrt{C_A}\gamma^2}{4\mu^2} - \frac{\mu^2}{4\sqrt{C_A}\gamma^2} \!
\right] \! \tan^{-1} \!\! \left[ \frac{\sqrt{C_A}\gamma^2}{\mu^2} \right] 
\right] C_A a \,+ O(a^2) \,.
\end{eqnarray}
We have checked that the latter is consistent by renormalizing the 
Faddeev-Popov ghost $2$-point function in the same way and verified explicitly 
that the first of the Slavnov-Taylor identities holds at one loop. Equipped 
with these renormalization constants then that for $\gamma$ is now already 
fixed in the MOM scheme through the second identity of (\ref{rencondef}). In 
order to verify that this is not inconsistent we have repeated the one loop gap
equation computation but using these MOM renormalization constants. This is 
achieved in the same way as \cite{32}, by closing the legs on the mixed 
propagator which gives the one loop contribution to the gap equation. As this 
is a vacuum diagram there is no external momentum to set to a mass shell value 
to extract the MOM scheme expression for $Z_\gamma$ which is why we have 
proceeded with the $2$-point functions first. Therefore, the renormalization 
constants which have already been determined will remove the divergences and 
their finite parts will influence the final form of the gap equation. 
Ultimately we find the relatively simple expression\footnote{This corrects a 
sign error in the expression given in \cite{75}.} 
\begin{eqnarray}
1 &=& \left[ \frac{3}{8} \ln \left[ \frac{[C_A\gamma^4+\mu^4]}{C_A\gamma^4} 
\right] ~-~ \frac{5}{8} ~+~ \frac{C_A\gamma^4}{8\mu^4} 
\ln \left[ \frac{C_A\gamma^4}{[C_A\gamma^4+\mu^4]} \right] ~+~ 
\frac{3\pi\sqrt{C_A}\gamma^2}{8\mu^2} \right. \nonumber \\
&& \left. ~-~ \left[ \frac{3\sqrt{C_A}\gamma^2}{4\mu^2}
- \frac{\mu^2}{4\sqrt{C_A}\gamma^2} \right] \tan^{-1} \left[
\frac{\sqrt{C_A}\gamma^2}{\mu^2} \right] \right] C_A a ~+~ O(a^2) \nonumber \\ 
& \equiv & \mbox{gap}(\gamma,\mu,\mbox{MOM}) C_A a ~+~ O(a^2)
\label{gap1mom}
\end{eqnarray}
where for later purposes we have introduced a shorthand definition of the right
hand side of the gap equation. Given that we are following the standard way to 
proceed in the MOM scheme there could be a doubt as to whether this is 
ultimately correct. However, in the Gribov-Zwanziger context the main check on 
this is whether the Kugo-Ojima criterion of Faddeev-Popov ghost enhancement,
\cite{34,35}, emerges in the $c^a$ $2$-point function in the zero momentum 
limit when the MOM gap equation of (\ref{gap1mom}) is satisfied. We note that 
not unexpectedly it does so. As a final point we have returned to the 
renormalization group invariant effective coupling constant defined from the 
gluon ghost vertex and reconsidered it in the infrared limit similar to section
3. Using the MOM versions of the respective form factors, 
$D^{\mbox{\footnotesize{MOM}}}_A(p^2)$ and
$D^{\mbox{\footnotesize{MOM}}}_c(p^2)$ then we define the MOM effective 
renormalization group invariant coupling constant as
\begin{equation}
\left. \frac{}{} \alpha_{\mbox{\footnotesize{eff}}} (p^2) 
\right|_{\mbox{\footnotesize{MOM}}} ~=~ \alpha_s(\mu) 
D^{\mbox{\footnotesize{MOM}}}_A(p^2) \left( 
D^{\mbox{\footnotesize{MOM}}}_c(p^2) \right)^2 ~.
\label{rgieffcc}
\end{equation}
Equipped with this we find that the coupling constant freezing calculation 
becomes 
\begin{equation}
\left. \frac{}{} \alpha_{\mbox{\footnotesize{eff}}} (0) 
\right|_{\mbox{\footnotesize{MOM}}} ~=~ \lim_{p^2 \rightarrow 0} \left[ 
\frac{ \alpha_s(\mu) \left[ 1 - C_A \left( \mbox{gap}(\gamma,\mu,\mbox{MOM}) 
+ \frac{5}{8} - \frac{11}{16} \right) a \right] (p^2)^2 }
{ C_A \gamma^4 \left[ 1 - C_A \left( \mbox{gap}(\gamma,\mu,\mbox{MOM})
- \frac{\pi p^2}{8 \sqrt{C_A} \gamma^2} \right) a \right]^2 } \right]
\end{equation}
where we have displayed the rationals in the numerator explicitly in order to
compare with the analogous $\MSbar$ computation of \cite{33}. Hence, 
\begin{equation}
\left. \frac{}{} \alpha_{\mbox{\footnotesize{eff}}} (0) 
\right|_{\mbox{\footnotesize{MOM}}} ~=~ \frac{16}{\pi C_A} ~.
\label{ccfre}
\end{equation}
This is the same value as the $\MSbar$ scheme, \cite{33}, as expected and can 
be regarded as a robust check on the finite parts of the MOM renormalization 
constants we derived. One final comment on the MOM gap equation in the context 
of the earlier discussion is that it is evident that inverting (\ref{gap1mom})
is not as straightforward as that for the $\MSbar$ scheme and we have not 
proceeded. This is partly because the two loop expression is not available in 
order to do a proper comparison but also because if it was, it is clear that 
nothing substantial would emerge from a complicated explicit expression. Indeed
a two loop calculation of the MOM gap equation would require the finite part of
the $2$-point functions {\em exactly} which would be a huge computation. To
elaborate on the complexity of this problem in order to obtain, for instance, 
the wave function renormalization constants there are well over $1000$ two loop 
Feynman diagrams correcting all the $A^a_\mu$, $\rho^{ab}_\mu$ and 
$\xi^{ab}_\mu$ $2$-point functions. Aside from this, it is not clear if all the
master scalar two loop self-energy Feynman integrals have been evaluated for 
all the necessary massive propagator configurations. For instance, the exact 
expression for the scalar master two loop self-energy corrections with purely 
$3$-point vertices with all masses equal is not known. Even it were available 
one would still require the expression for different masses in order to 
determine the expressions for the Gribov mass combinations of $\pm i \sqrt{C_A}
\gamma^2$. Indeed to appreciate how complicated this problem would be, one can 
examine the structure for the simpler two loop sunset self-energy correction 
for arbitrary masses, discussed in \cite{76}. The arbitrary mass expression in 
this simplest case involves Lauricella functions where the arguments are a 
function of the masses and the external momentum. In an MOM renormalization 
such functions, and the as yet undetermined master integrals for the two loop 
topologies with four and five propagators, would have to be evaluated exactly 
at the point where $p^2$~$=$~$\mu^2$. Therefore, given these considerations it 
is inconceivable that a two loop MOM gap equation is viable in the immediate 
future, even if such a quantity were of interest. Finally, we note that the 
different values for the renormalization group invariant and $V$-scheme 
coupling constants at zero momentum are not inconsistent. In this region there 
is no unique way of defining the strong coupling constant.

\sect{$\xi^{ab}_\mu$ and $\rho^{ab}_\mu$ enhancement.} 

Whilst our potential has been derived from the usual static potential formalism
at one loop, it is clear that it lacks a linearly rising part. This is due to
the cancellation of similar $1/(p^2)^2$ type terms as $p^2$~$\rightarrow$~$0$.
However, from experience of the Gribov-Zwanziger Lagrangian any infrared
behaviour which is in qualitative agreement with expectations from other work
has always required the explicit use of the gap equation satisfied by 
$\gamma$. For example, the enhancement of the Faddeev-Popov ghost is due to 
the gap equation, \cite{10}, and we recall that the theory can only be regarded
as a gauge theory when the gap equation is satisfied. Indeed the ghost 
enhancement is also a key component of the Kugo-Ojima confinement criterion, 
\cite{34,35}, in Yang-Mills theory. Its role in the Gribov-Zwanziger path 
integral has recently been clarified in \cite{77} and is also used as a
boundary condition in Dyson Schwinger studies, \cite{29}. Whilst the 
Faddeev-Popov ghosts and localizing ghosts $\omega^{ab}_\mu$ satisfy the 
enhancement, they cannot play a direct role in actually confining a gluon since
they are Grassmann fields. Hence they cannot be directly exchanged in 
gluon-gluon interactions as is evident from the graphs of Figures $1$ and $2$. 
Instead one clearly requires a commuting field to enhance to be at least in a 
situation where a $1/(p^2)^2$ singularity could be exchanged. Unlike the 
original expectations of Mandelstam and others, \cite{6,7,8,9}, this does not 
appear to be the gluon in the Gribov-Zwanziger set-up. However, as recently 
pointed out by Zwanziger there appears to be a clue in the enhancement of the 
propagators of the bosonic ghosts from a Schwinger Dyson analysis, \cite{50}. 
Therefore, our aim in this section is to first demonstrate that that 
enhancement can also actually be accessed in perturbation theory and then 
discuss its implications for the static potential. Aside from the fields $c^a$ 
and $\omega^{ab}_\mu$ the explicit forms of the $2$-point functions for 
$\xi^{ab}_\mu$ and $\rho^{ab}_\mu$ have the potential for enhancement. More 
concretely the explicit one loop correction to the colour channel of the 
Lagrangian kinetic term for each of $\xi^{ab}_\mu$ and $\rho^{ab}_\mu$, given
in appendix B, are exactly equivalent to that of the Faddeev-Popov ghost 
$2$-point function. To illustrate this for completeness here we recall the 
$p^2$~$\rightarrow$~$0$ behaviour of the $2$-point functions are,
\begin{eqnarray}
\langle \xi^{ab}_\mu(-p) \xi^{cd}_\nu(p) \rangle^{-1} &=& -~ \left[
\delta^{ac} \delta^{bd} \left[ 1 ~-~ C_A \left( \frac{5}{8} ~-~ \frac{3}{8} \ln
\left( \frac{C_A\gamma^4}{\mu^4} \right) \right) a \right] p^2
~+~ \frac{7}{144} f^{ace} f^{bde} p^2 a \right. \nonumber \\
&& \left. ~~~~+~ \frac{11}{288} f^{abe} f^{cde} p^2 a ~+~ \frac{7}{24} 
d_A^{abcd} \frac{p^2}{C_A} a ~+~ O(a^2) \right] P_{\mu\nu}(p) \nonumber \\
&& -~ \left[
\delta^{ac} \delta^{bd} \left[ 1 ~-~ C_A \left( \frac{5}{8} ~-~ \frac{3}{8} \ln
\left( \frac{C_A\gamma^4}{\mu^4} \right) \right) a \right] p^2
~+~ \frac{5}{48} f^{ace} f^{bde} p^2 a \right. \nonumber \\
&& \left. ~~~~-~ \frac{5}{96} f^{abe} f^{cde} p^2 a ~+~ \frac{5}{8} 
d_A^{abcd} \frac{p^2}{C_A} a ~+~ O(a^2) \right] L_{\mu\nu}(p) \nonumber \\
&& +~ O\left((p^2)^2\right)
\label{xi2pt}
\end{eqnarray}
and
\begin{equation}
\langle \rho^{ab}_\mu(-p) \rho^{cd}_\nu(p) \rangle^{-1} ~=~ -~ \left[
\delta^{ac} \delta^{bd} \left[ 1 ~-~ C_A \left( \frac{5}{8} ~-~ \frac{3}{8} \ln
\left( \frac{C_A\gamma^4}{\mu^4} \right) \right) a \right] p^2 \right] 
\eta_{\mu\nu} ~+~ O\left((p^2)^2\right) 
\label{rho2pt}
\end{equation}
where to avoid confusion with the propagator we have formally indicated the 
inverse. The first terms on the right hand side of (\ref{xi2pt}) and 
(\ref{rho2pt}) clearly correspond to the one loop gap equation, (\ref{gap1}), 
which would imply enhancement similar to the Grassmann ghost fields and the 
emergence of an infrared dipole form. The inversion of the $\rho^{ab}_\mu$ 
$2$-point function to deduce the propagator is similar to that of the 
Faddeev-Popov ghost and $\omega^{ab}_\mu$ and so $\rho^{ab}_\mu$ will have an 
enhanced infrared propagator too. However, a $\xi^{ab}_\mu$ enhancement is not 
as straightforward to observe as those fields due to the extra terms in 
(\ref{xi2pt}) and the group structure.

To examine how $\xi^{ab}_\mu$ enhancement emerges, we first recall the 
situation in the Faddeev-Popov case. There one first computes the $2$-point
function in the zero momentum limit and applies (\ref{gap1}). This produces a 
leading term of $O\left( (p^2)^2 \right)$ which one then inverts to discover
the dipole infrared behaviour of the Faddeev-Popov ghost. Turning to the
$\xi^{ab}_\mu$ case the algorithm is the same but not as straightforward due to
the mixing in the quadratic part of the $\{ A^a_\mu, \xi^{ab}_\mu \}$ sector of
the Lagrangian. So the analogous inversion has to involve the {\em full} 
$2$~$\times$~$2$ mixing matrix. In section $3$ this was carried out formally at
one loop but without first enforcing the gap equation. Given the potential for 
$\xi^{ab}_\mu$ enhancement of \cite{50} in a Dyson Schwinger analysis, we 
reconsider the formal matrix inversion in a more general way. Specifically we
will focus on the $\{ A^a_\mu, \xi^{ab}_\mu \}$ sector and in particular the
transverse piece. We do this because at one loop $V$~$=$~$0$ and so the
transverse part of (\ref{twoptdef}) becomes block diagonal. It is the upper
$2$~$\times$~$2$ matrix which is of interest. If we now define this 
$2$~$\times$~$2$ matrix by $\Lambda_2^{\{ab|cd\}}$, where we can drop the 
Lorentz indices, then formally
\begin{equation}
\Lambda_2^{\{ab|cd\}} ~=~ \left(
\begin{array}{cc}
{\cal X} \delta^{ac} & {\cal U} f^{acd} \\
{\cal U} f^{cab} & {\cal Q}^{abcd}_\xi \\
\end{array}
\right) 
\label{matdef}
\end{equation} 
where we use the more general decomposition  
\begin{equation}
{\cal Q}^{abcd}_\xi ~=~ {\cal Q}_\xi \delta^{ac} \delta^{bd} ~+~ {\cal W}_\xi
f^{ace} f^{bde} ~+~ {\cal R}_\xi f^{abe} f^{cde} ~+~ 
{\cal S}_\xi d_A^{abcd} ~+~ {\cal P}_\xi \delta^{ab} \delta^{cd} ~+~ 
{\cal T}_\xi \delta^{ad} \delta^{bc}  
\end{equation}
and the quantities ${\cal X}$, ${\cal U}$, ${\cal Q}_\xi$, ${\cal W}_\xi$, 
${\cal R}_\xi$ and ${\cal S}_\xi$ are the formal $2$-point functions
{\em including} the part from the quadratic part of the Lagrangian. We have 
used similar notation to sections $2$ and $3$ but in calligraphic font to 
indicate that these are not solely the one loop corrections. We have also 
included two extra terms, ${\cal P}_\xi$ and ${\cal T}_\xi$, to complete the 
basis. Although these are zero up to and including one loop they are required
here since we will be multiplying $\Lambda_2^{\{ab|cd\}}$ by its {\em full}
inverse rather than drop the $O(a^2)$ part as was carried out in deriving 
(\ref{traprop}). Here there will be extra group structures when, for example, 
two $d_A^{abcd}$ tensors are partially contracted. Whilst we are aiming at 
being general our matrix can only really be regarded as one loop in one sense
since the colour group structure in the final element may not be the most 
general. Our choice there is motivated by what actually emerges from the 
explicit one loop computations. For instance, if quarks are present in higher 
loop diagrams then the tensor $d_F^{abcd}$ could also be present due to 
light-by-light subgraphs or some peculiar contraction of this with other 
tensors. For a comprehensive discussion of potential high rank (adjoint) 
tensors, see \cite{78}. At present we make no assumptions about the behaviour 
of (\ref{matdef}) in the $p^2$~$\rightarrow$~$0$ limit but note that like 
$Q_\rho$ in (\ref{rho2pt}), ${\cal Q}_\xi$ will be the key function in driving 
any enhancement. Also the loop order will play a key role in the inversion and 
we note that
\begin{equation}
{\cal X} ~=~ {\cal U} ~=~ {\cal Q}_\xi ~=~ O(1) ~~,~~
{\cal W}_\xi ~=~ {\cal R}_\xi ~=~ {\cal S}_\xi ~=~ O(a) ~~,~~
{\cal P}_\xi ~=~ {\cal T}_\xi ~=~ O(a^2) ~.
\end{equation}
Given $\Lambda_2^{\{ab|cd\}}$ we define the general inverse 
$\Pi_2^{\{cd|pq\}}$, which will be the matrix of propagators, in the same 
formal way by 
\begin{equation}
\Pi_2^{\{cd|pq\}} ~=~ \left(
\begin{array}{cc}
{\cal A} \delta^{cp} & {\cal B} f^{cpq} \\
{\cal B} f^{pcd} & {\cal D}^{cdpq}_\xi \\
\end{array}
\right) 
\end{equation} 
where now 
\begin{equation}
{\cal D}^{cdpq}_\xi ~=~ {\cal D}_\xi \delta^{cp} \delta^{dq} ~+~ {\cal J}_\xi 
f^{cpe} f^{dqe} ~+~ {\cal K}_\xi f^{cde} f^{pqe} ~+~ 
{\cal L}_\xi d_A^{cdpq} ~+~ {\cal M}_\xi \delta^{cd} \delta^{pq} ~+~ 
{\cal N}_\xi \delta^{cq} \delta^{dp}  
\end{equation} 
similar to (\ref{ffdef}) but allowing for the extra colour tensors to have a
basis. For clarity we note that the two matrices must satisfy the standard 
inversion on the smaller subspace given by
\begin{equation}
\Lambda_2^{\{ab|cd\}} \Pi_2^{\{cd|pq\}} ~=~ \left(
\begin{array}{cc}
\delta^{cp} & 0 \\
0 & \delta^{cp} \delta^{dq} \\
\end{array}
\right) 
\end{equation} 
where the right hand side is effectively the unit matrix. Multiplying out the 
matrices explicitly leads to the formal linear equations satisfied by the 
$2$-point functions and propagators. We have 
\begin{eqnarray}
1 &=& {\cal AX} ~+~ C_A {\cal UB} ~~~,~~~ 0 ~=~ {\cal XB} ~+~ 
\left( {\cal D}_\xi - {\cal N}_\xi + C_A {\cal K}_\xi 
+ \frac{1}{2} C_A {\cal J}_\xi \right) {\cal U} \nonumber \\
0 &=& {\cal A} {\cal U} ~+~ \left( {\cal Q}_\xi + C_A {\cal R}_\xi 
+ \frac{1}{2} C_A {\cal W}_\xi - {\cal T}_\xi \right) {\cal B} \nonumber \\
1 &=& {\cal Q}_\xi {\cal D}_\xi ~+~ b_2 {\cal L}_\xi {\cal W}_\xi ~+~ 
b_2 {\cal S}_\xi {\cal J}_\xi ~+~ a_2 {\cal S}_\xi {\cal L}_\xi ~+~
{\cal T}_\xi {\cal N}_\xi \nonumber \\ 
0 &=& \left( {\cal Q}_\xi +  C_A {\cal W}_\xi + \frac{5}{6} C_A^2 {\cal S}_\xi 
+ \NA {\cal P}_\xi + {\cal T}_\xi \right) {\cal M}_\xi ~+~ 
\left( C_A {\cal J}_\xi + \frac{5}{6} C_A^2 {\cal L}_\xi + {\cal D}_\xi 
+ {\cal N}_\xi \right) {\cal P}_\xi \nonumber \\
&& +~ b_1 {\cal W}_\xi {\cal L}_\xi ~+~ b_1 {\cal S}_\xi {\cal J}_\xi ~+~ 
a_1 {\cal S}_\xi {\cal L}_\xi \nonumber \\ 
0 &=& b_2 {\cal L}_\xi {\cal W}_\xi ~+~ b_2 {\cal S}_\xi {\cal J}_\xi ~+~
a_2 {\cal S}_\xi {\cal L}_\xi ~+~ {\cal Q}_\xi {\cal N}_\xi ~+~ 
{\cal T}_\xi {\cal D}_\xi \nonumber \\ 
0 &=& {\cal Q}_\xi {\cal L}_\xi ~+~ {\cal W}_\xi {\cal J}_\xi ~+~ 
{\cal S}_\xi {\cal D}_\xi ~+~ b_4 {\cal W}_\xi {\cal L}_\xi ~+~ 
b_4 {\cal S}_\xi {\cal J}_\xi ~+~ a_4 {\cal S}_\xi {\cal L}_\xi ~+~ 
{\cal S}_\xi {\cal N}_\xi ~+~ {\cal T}_\xi {\cal L}_\xi \nonumber \\
0 &=& {\cal W}_\xi {\cal D}_\xi ~+~ {\cal Q}_\xi {\cal J}_\xi ~+~ 
\frac{1}{6} C_A {\cal W}_\xi {\cal J}_\xi ~+~ 
2 b_3 {\cal S}_\xi {\cal J}_\xi ~+~ 2 a_3 {\cal S}_\xi {\cal L}_\xi ~+~ 
2 b_3 {\cal W}_\xi {\cal L}_\xi ~+~ {\cal W}_\xi {\cal N}_\xi ~+~ 
{\cal T}_\xi {\cal J}_\xi \nonumber \\
0 &=& {\cal UB} ~+~ {\cal Q}_\xi {\cal K}_\xi ~+~ \frac{1}{6} C_A {\cal W}_\xi 
{\cal J}_\xi ~+~ \frac{1}{2} C_A {\cal W}_\xi {\cal K}_\xi ~+~ 
\frac{1}{2} C_A {\cal R}_\xi {\cal J}_\xi ~+~ 
C_A {\cal R}_\xi {\cal K}_\xi ~-~ b_3 {\cal W}_\xi {\cal L}_\xi \nonumber \\
&& -~ {\cal W}_\xi {\cal N}_\xi ~+~ {\cal R}_\xi {\cal D}_\xi ~-~ 
{\cal R}_\xi {\cal N}_\xi ~-~ b_3 {\cal S}_\xi {\cal J}_\xi ~-~ 
a_3 {\cal S}_\xi {\cal L}_\xi ~-~ {\cal T}_\xi {\cal J}_\xi ~-~ 
{\cal T}_\xi {\cal K}_\xi  
\label{propsdef}
\end{eqnarray} 
where the coefficients $a_i$ and $b_i$ derive from the group decompositions
defined and discussed in appendix A where their explicit forms are given for an 
arbitrary colour group. It is clear we have nine equations for nine unknowns. 
So it is a straightforward exercise to determine the general solution. 
First, we record that 
\begin{equation}
{\cal A} ~=~ \frac{[{\cal Q}_\xi + C_A {\cal R}_\xi + \half C_A {\cal W}_\xi]}
{[ ({\cal Q}_\xi + C_A {\cal R}_\xi + \half C_A {\cal W}_\xi) {\cal X} 
- C_A {\cal U}^2]} ~~~,~~~ {\cal B} ~=~ -~ \frac{{\cal U}}
{[ ({\cal Q}_\xi + C_A {\cal R}_\xi + \half C_A {\cal W}_\xi) {\cal X} 
- C_A {\cal U}^2]} 
\label{propgen1}
\end{equation}
where we have assumed ${\cal P}_\xi$~$=$~${\cal T}_\xi$~$=$~$0$ initially in
accordance to what we found at one loop. The explicit forms of the form factors
for $\xi^{ab}_\mu$ propagator are cumbersome. So for these cases we record the 
$SU(3)$ expressions where the values for $a_i$ and $b_i$ of appendix A have 
been used. We have 
\begin{eqnarray}
{\cal D}_\xi &=& \frac{1}{2{\cal Q}_\xi} \left[ 3 ( 3 {\cal S}_\xi 
- 2 {\cal W}_\xi ) ( {\cal S}_\xi + 2 {\cal W}_\xi ) ( {\cal S}_\xi 
+ {\cal W}_\xi ) + 8 ( 7 {\cal S}_\xi + 3 {\cal W}_\xi ) {\cal Q}_\xi^2 \right.
\nonumber \\
&& \left. ~~~~~+~ 16 {\cal Q}_\xi^3 + 2 ( 27 {\cal S}_\xi^3 + 20 {\cal S}_\xi 
{\cal W}_\xi - 8 {\cal W}_\xi^2 ) {\cal Q}_\xi \right] \nonumber \\
&& ~ \times \left[ 2 {\cal Q}_\xi + 3 {\cal S}_\xi + 3 {\cal W}_\xi 
\right]^{-1} \left[ 2 {\cal Q}_\xi + 3 {\cal S}_\xi - 2 {\cal W}_\xi 
\right]^{-1} \left[ 2 {\cal Q}_\xi + {\cal S}_\xi + 2 {\cal W}_\xi \right]^{-1}
\nonumber \\ 
{\cal J}_\xi &=& -~ \frac{4 {\cal W}_\xi}{[ 2 {\cal Q}_\xi + 3 {\cal S}_\xi
+ 3 {\cal W}_\xi ] [ 2 {\cal Q}_\xi + 3 {\cal S}_\xi - 2 {\cal W}_\xi ]} 
\nonumber \\ 
{\cal K}_\xi &=& \frac{1}{{\cal Q}_\xi} \left[ ( 4 ( 3 {\cal S}_\xi 
- {\cal W}_\xi ) {\cal Q}_\xi + 3 ( 3 {\cal S}_\xi - 2 {\cal W}_\xi ) 
( {\cal S}_\xi + {\cal W}_\xi ) ) ( 2 {\cal U}^2 - {\cal W}_\xi {\cal X} 
- 2 {\cal R}_\xi {\cal X} ) \right. \nonumber \\
&& \left. ~~~~-~ 8 ( {\cal R}_\xi {\cal X} - {\cal U} )^2 {\cal Q}_\xi^2 
\right] \nonumber \\
&& ~ \times \left[ 2 {\cal Q}_\xi {\cal X} + 6 {\cal R}_\xi {\cal X} 
- 6 {\cal U}^2 + 3 {\cal W}_\xi {\cal X} \right]^{-1} \left[ 2 {\cal Q}_\xi 
+ 3 {\cal S}_\xi + 3 {\cal W}_\xi \right]^{-1} \left[ 2 {\cal Q}_\xi 
+ 3 {\cal S}_\xi - 2 {\cal W}_\xi \right]^{-1} \nonumber \\ 
{\cal L}_\xi &=& -~ 4 \left[ 2 {\cal Q}_\xi {\cal S}_\xi + ( 3 {\cal S}_\xi 
+ 2 {\cal W}_\xi ) ( {\cal S}_\xi - {\cal W}_\xi ) \right] \nonumber \\
&& ~~ \times \left[ 2 {\cal Q}_\xi + 3 {\cal S}_\xi + 3 {\cal W}_\xi 
\right]^{-1} \left[ 2 {\cal Q}_\xi + 3 {\cal S}_\xi - 2 {\cal W}_\xi 
\right]^{-1} \left[ 2 {\cal Q}_\xi + {\cal S}_\xi + 2 {\cal W}_\xi 
\right]^{-1} \nonumber \\
{\cal M}_\xi &=& 6 \left[ 21 {\cal S}_\xi^3 + {\cal S}_\xi^2 {\cal W}_\xi 
- 12 {\cal S}_\xi {\cal W}_\xi^2 - 4 {\cal W}_\xi^3 + 2 ( 7 {\cal S}_\xi 
+ 4 {\cal W}_\xi ) {\cal Q}_\xi {\cal S}_\xi \right] 
\left[ 2 {\cal Q}_\xi + 15 {\cal S}_\xi + 6 {\cal W}_\xi \right]^{-1} 
\nonumber \\
&& \times \left[ 2 {\cal Q}_\xi + 3 {\cal S}_\xi + 3 {\cal W}_\xi \right]^{-1} 
\left[ 2 {\cal Q}_\xi + 3 {\cal S}_\xi - 2 {\cal W}_\xi \right]^{-1} 
\left[ 2 {\cal Q}_\xi + {\cal S}_\xi + 2 {\cal W}_\xi \right]^{-1} \nonumber \\
{\cal N}_\xi &=& -~ \frac{3}{2{\cal Q}_\xi} \left[ ( 3 {\cal S}_\xi 
- 2 {\cal W}_\xi ) ( {\cal S}_\xi + 2 {\cal W}_\xi ) ( {\cal S}_\xi 
+ {\cal W}_\xi ) + 2 ( {\cal S}_\xi + 4 {\cal W}_\xi ) {\cal Q}_\xi 
{\cal S}_\xi \right] \nonumber \\
&& ~~~ \times \left[ 2 {\cal Q}_\xi + 3 {\cal S}_\xi + 3 {\cal W}_\xi 
\right]^{-1} \left[ 2 {\cal Q}_\xi + 3 {\cal S}_\xi - 2 {\cal W}_\xi 
\right]^{-1} \left[ 2 {\cal Q}_\xi + {\cal S}_\xi + 2 {\cal W}_\xi 
\right]^{-1} ~. 
\label{propgen2}
\end{eqnarray}
We have checked that our full arbitrary group solution correctly reproduces the
one loop propagator corrections of (\ref{traprop}). Indeed the gluon propagator
remains suppressed in the zero momentum limit at one loop. Aside from 
indicating how involved the final expression for the $\xi^{ab}_\mu$ propagator 
is, the main point of (\ref{propgen2}) is to illustrate which of the form 
factors enhance. From (\ref{xi2pt}) the initial terms of the small momentum 
expansion of ${\cal Q}_\xi$ represents the gap equation. Therefore, in this 
limit when the gap equation is realised ${\cal Q}_\xi$ is effectively 
$O\left( (p^2)^2 \right)$ which corresponds to the same situation with the 
Faddeev-Popov ghost, $\omega^{ab}_\mu$ and $\rho^{ab}_\mu$ fields giving rise 
to the enhancement of these fields. From (\ref{propgen2}) the situation is 
similar since several of the amplitudes have an overall factor of 
${\cal Q}_\xi$. Hence these colour channels will enhance whilst the others will
not in the zero momentum limit. More specifically the leading order behaviour 
of the transverse part of both bosonic ghost propagators in the 
$p^2$~$\rightarrow$~$0$ limit for an arbitrary group is 
\begin{eqnarray}
\langle \xi^{ab}_\mu(p) \xi^{cd}_\nu(-p) \rangle & \sim & \left[ 
\frac{4 \gamma^2}{\pi \sqrt{C_A} (p^2)^2 a} \left[ \delta^{ad} \delta^{bc}
- \delta^{ac} \delta^{bd} \right] + 
\frac{8 \gamma^2}{\pi C_A^{3/2} (p^2)^2 a} f^{abe} f^{cde} \right]
P_{\mu\nu}(p) \nonumber \\ 
\langle \rho^{ab}_\mu(p) \rho^{cd}_\nu(-p) \rangle & \sim & -~ 
\frac{8 \gamma^2}{\pi \sqrt{C_A} (p^2)^2 a} \delta^{ac} \delta^{bd} 
P_{\mu\nu}(p) ~. 
\label{xirhoenh}
\end{eqnarray} 
So one colour channel in addition to those of the original $\xi^{ab}_\mu$
propagator enhances. Although we have used the numerical values deriving from 
the one loop corrections to ${\cal Q}_\xi$ this enhanced behaviour is more 
general. If instead we examine the leading ${\cal Q}_\xi$ behaviour of 
(\ref{propgen2}) but for a general colour group then we find 
\begin{equation}
{\cal D}_\xi ~\sim~ \frac{1}{2{\cal Q}_\xi} ~~,~~ 
{\cal K}_\xi ~\sim~ -~ \frac{1}{C_A{\cal Q}_\xi} ~~,~~ 
{\cal N}_\xi ~\sim~ -~ \frac{1}{2{\cal Q}_\xi} 
\end{equation}
with the other form factors being non-singular in the Laurent expansion in
${\cal Q}_\xi$. This implies that the leading ${\cal Q}_\xi$ behaviour of the 
transverse part of the $\xi^{ab}_\mu$ propagator is 
\begin{equation}
\langle \xi^{ab}_\mu(p) \xi^{cd}_\nu(-p) \rangle ~ \sim ~ 
\frac{1}{2{\cal Q}_\xi} \left[ \delta^{ac} \delta^{bd} ~-~ 
\delta^{ad} \delta^{bc} ~-~ \frac{2}{C_A} f^{abe} f^{cde} \right] 
P_{\mu\nu}(p) ~. 
\label{xirhoenhgen}
\end{equation}
So, for instance, when the gap equation is realised in the {\em three}
dimensional Gribov-Zwanziger Lagrangian, then the corresponding $\xi^{ab}_\mu$ 
propagator will also be enhanced together with the Grassmann fields with the
{\em same} colour group structure as four dimensions, (\ref{xirhoenh}). 
Interestingly for $\xi^{ab}_\mu$ the colour channel which dominates in the 
infrared is that which is antisymmetric in the colour indices of the field 
itself. In other words the colour projected field $f^{abc} \xi^{bc}_\mu$ does 
not enhance at one loop given the relative coefficients of (\ref{xirhoenhgen}). 

\begin{figure}[ht]
\hspace{2.65cm}
\epsfig{file=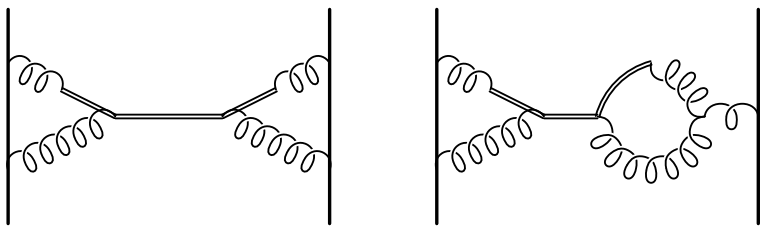,height=3cm}
\vspace{0.5cm}
\caption{Several two loop topologies with $\xi^{ab}_\mu$ exchange.}
\end{figure}

Having established that $\xi^{ab}_\mu$ and $\rho^{ab}_\mu$ both enhance we can
consider the implications of this for the static potential. These are better
candidates for a confinement mechanism since they are bosonic fields and 
therefore can be exchanged between coloured static sources. However, the same 
points that were noted in \cite{50} equally apply here in that $\xi^{ab}_\mu$ 
and $\rho^{ab}_\mu$ do not couple {\em directly} to a (static) field such as a 
quark or gluon. In the context of the static potential, the colour source only 
couples to the gluon field, $A^a_\mu$, in (\ref{potdefz}). However, one can 
have a single $\xi^{ab}_\mu$ exchange {\em indirectly}, for example, through
various topologies such as those illustrated in Figure $4$. The first graph 
involves source $\xi^{ab}_\mu$ vertex corrections whereas the second graph is 
a correction to the mixed $2$-point function. If we consider the first in an 
extended computation one can replace the single $\xi^{ab}_\mu$ by the enhanced
propagator whilst the second graph could be ignored since that is part of our 
earlier matrix inversion. Given (\ref{xirhoenh}) this exchange will actually 
give an $O(a^2)$ contribution to the static potential and not an $O(a^3)$ one. 
Although this is clearly contrary to a perturbative approach it is a 
qualitative indication that a reordering of the series will emerge even in a 
Dyson Schwinger approach as already discussed in \cite{50}. Moreover, a similar
reordering to produce the renormalization group invariant effective coupling 
constant freezing occured in (\ref{alfre}). In the context of freezing it is 
worth noting that as the Lagrangian channel of both $\xi^{ab}_\mu$ and
$\rho^{ab}_\mu$ have enhancement, one could define an effective coupling 
constant based on either the gluon $\xi^{ab}_\mu$ vertex or gluon 
$\rho^{ab}_\mu$ vertex rather than the ghost gluon vertex which has a freezing 
value the same as (\ref{ccfre}) in either the $\MSbar$ or MOM schemes. Briefly 
instead of using either the ghost or $\omega^{ab}_\mu$ $2$-point function in 
the denominator of the definition, one uses instead the first part of 
(\ref{xi2pt}). However, this is offered as an observation since the 
renormalization group invariance of such a construction is clearly not 
established and the $\xi^{ab}_\mu$ gluon vertex is certainly not on a par with 
the ghost gluon vertex in the context of non-renormalizability.

If we take the point of view that our reasoning is at a qualitative level, we 
can then examine some possible implications of $\xi^{ab}_\mu$ enhancement 
on the static potential. We have computed the eight Feynman diagrams 
contributing to the topology illustrated in the left hand diagram of Figure $4$
for the static potential formalism. With the $\xi^{ab}_\mu$ propagator of 
(\ref{propdef}) we find in the $p^2$~$\rightarrow$~$0$ limit that
\begin{equation}
\tilde{V}^{\mbox{\footnotesize{Fig 4a}}}( \mathbf{p} ) 
~=~ \left[ \frac{5\pi C_A^{3/2} (\mathbf{p}^2)^2}{4608\gamma^6} ~+~ 
O \left( (\mathbf{p}^2)^3 \right) \right] \frac{g^6}{(16\pi^2)^2} 
\label{pot4a}
\end{equation} 
where, since we are working at two loops, we have put the sources in the
adjoint representation. Useful in handling the group theory aspects of this 
calculation was the ${\tt color.h}$ {\sc Form} routine from \cite{57} based on 
\cite{78}. However, with the enhanced propagator of (\ref{xirhoenh}) we find in
the same limit
\begin{equation}
\left. \tilde{V}^{\mbox{\footnotesize{Fig 4a}}}( \mathbf{p} ) 
\right|^{\mbox{\footnotesize{$SU(3)$}}}_{\mbox{\footnotesize{enhance}}} ~=~ 
\left[ \frac{3\mathbf{p}^2}{14\gamma^4} ~+~ O \left( (\mathbf{p}^2)^2 
\right) \right] \frac{g^4}{16\pi^2} 
\label{pot4aen}
\end{equation} 
where we record the $SU(3)$ result given that the arbitrary group expression
would be too cumbersome. The reason for this is that this leading order term
does not derive from the enhanced part of (\ref{xirhoenh}) but instead the
$O( 1/p^2 )$ correction. In other words the dominant part of the exchange 
diagram in the zero momentum limit does not depend on the enhancement. This is 
because the one loop vertex subgraphs corrections are proportional to a 
structure function. More specifically the colour dependence of the vertex is
$f^{abc} T^c$ where the group generator carries the indices of the source legs.
Closing to form the Wilson loop leads to a trace over these indices when the
other vertex is included. Since the colour channel of the enhanced part of the 
$\xi^{ab}_\mu$ propagator vanishes identically when the indices of this 
structure function are contracted then that contribution is absent. Indeed we
have evaluated the one loop source $\xi^{ab}_\mu$ vertex function for the
static potential momentum configuration exactly and verified that this colour
structure is the source of the cancellation. Moreover, in this set-up the 
vertex function in principle involves two contributions deriving from the two 
possible vectors, $v_\mu$ and $p_\mu$, which the vertex can be decomposed into 
since it has one free Lorentz index. It transpires that to this order the form 
factor of that of the $p_\mu$ term is zero leaving only that for $v_\mu$. 
Hence, we do not need to consider the longitudinal part of the $\xi^{ab}_\mu$ 
propagator since we are in a static situation where $vp$~$=$~$0$. Whilst there 
is the possibility of enhancement from the $\rho^{ab}_\mu$ propagator through 
similar topologies to those illustrated on the left in Figure $4$, each of the 
contributing vertex corrections vanish identically since the $P_{\mu\nu}(k)$ is
contracted with $k^\mu$ where $k$ is the loop momentum. This occurs because the
gluon $\xi^{ab}_\mu$ $\rho^{cd}_\nu$ vertex, which we have not neglected, is 
essentially proportional to the Landau gauge fixing condition. So in fact 
(\ref{pot4aen}) represents the inclusion of both enhanced propagators of 
(\ref{xirhoenh}). Therefore, we appear to be forced to conclude that the single
$\xi^{ab}_\mu$ exchange process of Figure $4$ cannot lead to the dipole 
behaviour underpinning the linearly rising potential if the one loop 
enhancement of (\ref{xirhoenh}) is accepted. 

Indeed considering the higher loop corrections to the source $\xi^{ab}_\mu$ 
vertex would not appear to remedy the situation. For instance, one would have 
to have a colour structure for the vertex which does not involve $f^{abc} T^c$
if the colour structure of (\ref{xirhoenh}) was preserved beyond one loop.
Another candidate would be $d^{abc} T^c$ where $d^{abc}$ is the totally 
symmetric rank three tensor but it clearly gives zero when contracted with 
(\ref{xirhoenh}). Whilst one might contrive something to circumvent this then 
the coefficient of whatever this colour tensor is would have to be constant in 
the zero momentum limit. It is not clear at which loop order this could emerge.
Alternatively when the higher loop corrections to the propagators are computed 
then it may be the case that other colour channels aside from ${\cal Q}_\xi$ 
are enhanced. In such a case the colour tensor of (\ref{xirhoenh}) should be 
different allowing the enhanced piece to dominate. Clearly such considerations 
are beyond the scope of the current article but suggest that using resummation
methods such as the Schwinger Dyson technique could probe this in more detail. 
However, it is worth noting the situation with regard to the original 
observation of bosonic ghost enhancement of Zwanziger, \cite{50}. In \cite{50} 
an over-enhancement was obtained for various colour channels. In order to 
obtain a linearly rising potential it was argued, \cite{50}, that the quark 
gluon vertex develops a Lorentz tensor coupling involving 
$\sigma^{\mu\nu}$~$=$~$[\gamma^\mu,\gamma^\nu]$ together with one momentum 
vector to reduce the overall exchange to a dipole. Since we have considered in 
essence static gluons we do not have the same freedom in Lorentz space to 
accommodate a tensor coupling. However, provided the colour structure did not 
make the problem trivial then an over-enhancement would clearly reduce the 
power of momentum in (\ref{pot4aen}) at least by another power. Indeed given 
this it could be the case that the higher loop corrections to (\ref{xi2pt}) 
might also lead to an over-enhanced $\xi^{ab}_\mu$ propagator deriving from 
colour channels other than the propagator one with another colour structure to
(\ref{xirhoenh}). Though we do note that $\rho^{ab}_\mu$ enhancement is 
preserved at {\em two} loops in $\MSbar$ using (\ref{gap2}) similar to $c^a$ 
and $\omega^{ab}_\mu$. To verify this enhancement we evaluated the $212$ two 
loop Feynman diagrams of the $\rho^{ab}_\mu$ $2$-point function in the zero 
momentum limit using the vacuum bubble expansion. We recall one advantage of 
the enhancement feature is that it avoids the proliferation of higher order 
powers in $1/p^2$ which could emerge if one calculated the static potential 
order by order in perturbation theory as indicated earlier. The use of the gap 
equation in being central to this appears to be unavoidably essential to any 
analysis of studying the zero momentum limit. If anything these remarks might 
only serve to indicate how delicate it is to determine the zero momentum 
behaviour of the localizing bosonic ghost propagators via the inversion of the 
matrix of $2$-point functions. Having said this it is important to state that 
we are not ruling out the existence of a linearly rising potential in the 
Gribov-Zwanziger Lagrangian. There are other avenues one could consider aside 
from the single exchange of Figure $4$. For instance, the enhanced 
$\xi^{ab}_\mu$ propagator can appear inside loop diagrams and hence one would 
have to use a Schwinger Dyson style of analysis to study the implications for 
the overall topologies in the infrared. The enhanced propagator would only be 
significant at low virtual loop momenta. Alternatively the dominant topologies 
in the infrared could be something such as ladder graphs instead of the simple 
single $\xi^{ab}_\mu$ graphs analysed here. Again that would require techniques
beyond those discussed here. 

\sect{Power corrections.} 

Next, we return to our earlier comments concerning the appearance of power type
corrections of the form $\gamma^2/\mathbf{p}^2$ at higher loops in the static 
potential. However, we restrict them to our one loop static potential. First, 
one widely used tool to probe towards the infrared in QCD is the operator 
product expansion. In essence it provides a tool to include corrections to 
perturbative expressions where the corrections involve the vacuum expectation 
values of gauge invariant operators, such as $G_{\mu\nu}^a G^{a \, \mu\nu}$. In
the conventional perturbative vacuum the expectation value of such operators is
zero but in the true non-perturbative vacuum they acquire a non-zero value. 
Therefore such dimensionful quantities provide the mass scale required to have 
an expansion in inverse powers of the key momentum in the operator product 
expansion of the particular Green's function of interest. One situation where 
this formalism is applied is to the static potential. See, for example, 
\cite{79,80}. For instance, in \cite{79} the zero momentum exchange of a single
gluon with a one loop self-energy correction in gluon scattering is examined 
and a $1/(p^2)^3$ correction is produced where the dimensionality is made 
consistent by the presence of the dimension four quantity 
$\langle G_{\mu\nu}^a G^{a \, \mu\nu} \rangle$. However, clearly there is 
nothing to prevent higher order powers of $1/p^2$ appearing in such analyses 
and so it is worthwhile considering (\ref{stapot}) in a similar power series 
expansion. Therefore, from (\ref{stapot}) we have
\begin{eqnarray}
\tilde{V}( \mathbf{p} ) &=& -~ 
\frac{4 \pi C_F \alpha_s(\mu)}{\mathbf{p}^2} \nonumber \\
&& ~~ \times \left[ \left[ 1 - \frac{C_A\gamma^4}{(\mathbf{p}^2)^2} 
+ O \left( \frac{\gamma^8}{(\mathbf{p}^2)^4} \right) \right] \right.
\nonumber \\
&& \left. ~~~~~~+ \left[ \left[ \frac{31}{9} - \frac{11}{3} \ln \left[ 
\frac{\mathbf{p}^2}{\mu^2} \right] \right] C_A ~+~ \left[ 
\frac{4}{3} \ln \left[ \frac{\mathbf{p}^2}{\mu^2} \right] 
- \frac{20}{9} \right] T_F \Nf ~-~ 
\frac{2\pi C_A^{3/2}\gamma^2}{\mathbf{p}^2} \right. \right. \nonumber \\
&& \left. \left. ~~~~~~~~~~~+ \left[
\left[ \frac{79}{12} \ln \left[ \frac{\mathbf{p}^2}{\mu^2} \right] 
+ \frac{9}{8} \ln \left[ \frac{C_A\gamma^4}{(\mathbf{p}^2)^2} \right] 
- \frac{1315}{72} \right] C_A^2 \right. \right. \right. \nonumber \\
&& \left. \left. \left. ~~~~~~~~~~~~~~~+ \left[ \frac{40}{9} - \frac{8}{3} 
\ln \left[ \frac{\mathbf{p}^2}{\mu^2} \right] \right] T_F \Nf \right] 
\frac{C_A\gamma^4}{(\mathbf{p}^2)^2} ~+~ O \left( 
\frac{\gamma^6}{(\mathbf{p}^2)^3} \right) \right] a \right. \nonumber \\
&& \left. ~~~~~~+~ O(a^2) \right] ~.
\label{potpow}
\end{eqnarray} 
Clearly the leading term in this $\gamma^2/\mathbf{p}^2$ expansion is the 
perturbative result of \cite{36,37,38}. Moreover, at leading order in the 
perturbative expansion the next to leading power correction is 
$O(\gamma^4/(\mathbf{p}^2)^3)$ which follows trivially since the full term is
a function of $\gamma^4$ and not $\gamma^2$ which the one loop correction 
clearly is. Therefore, the leading tree part of the potential mimicks the
correction observed in \cite{79} although clearly here the quantity used to
ensure the dimensions balance is $\gamma^4$ and not 
$\langle G_{\mu\nu}^a G^{a \, \mu\nu} \rangle$. Though it ought to be stressed
here as was emphasised in \cite{79}, that this is a short distance 
approximation to the potential which has not fully been accounted for on the
lattice. The reasoning is that the confinement force is in principle accessible
at both low and higher energy scales. A recent exposition on this point has 
been provided in \cite{81}. Though we believe one needs to be careful in this 
power correction approximation since mathematically a short distance expansion 
of the full potential can only give an insight into the $r$ dependence for a 
limited range of $r$. So before considering the loop correction, the tree part 
of the potential considered as a power series in $\gamma$ gives an interesting 
insight into applying the Fourier transform to coordinate space. As the exact 
transform is known in (\ref{treepotr}), it can be expanded in powers of 
$\gamma r$ which is also the combination of variables which appear and not 
their square or higher powers. We find 
\begin{equation}
V_0(r) ~=~ -~ \frac{C_F g^2}{4\pi r} \left[ 1 ~-~ 
\frac{C_A^{\quarter} \gamma r}{\sqrt{2}} ~+~ 
\frac{C_A^{\threequarters} \gamma^3 r^3}{6\sqrt{2}} ~-~ 
\frac{C_A \gamma^4 r^4}{24} ~+~ O \left( \gamma^5 r^5 \right) \right] ~. 
\end{equation} 
The absence of a term linear in $r$ is consistent with there being no dipole
in the momentum space tree potential. However, the one-to-one power series
matching clearly breaks down with the appearance of a quadratic correction
which would ordinarily be associated with a momentum space term of 
$1/(\mathbf{p}^2)^{5/2}$ on dimensional grounds. 

In the one loop correction of (\ref{potpow}) one observes that a dipole 
correction appears at next to leading order in the power expansion rather than 
the triple pole at leading order. Parenthetically recalling our comments
concerning the sign of $\gamma^2$, then this term would then lead to an 
effective string tension, $\sigma^{\mbox{\footnotesize{eff}}}$, of 
\begin{equation} 
\sigma^{\mbox{\footnotesize{eff}}} ~=~ 
\frac{C_F C_A^{3/2} \gamma^2 g^4}{64\pi^2} 
\end{equation}
in the conventional definition of the potential. However, as is evident from
our full expression there is actually no net explicit dipole term in 
(\ref{stapot}) whose small momentum or large distance potential effectively
becomes Coulomb-like. If there was a pure dipole present in addition to the
remaining $\gamma$ dependent part, then performing a power series expansion
of the full expression would mean it would be difficult to isolate its 
contribution {\em uniquely}. Moreover, it would be difficult to interpret such
a correction in an operator power series context since it would involve the
square root of the gluon condensate or the vacuum expectation value of a
dimension two object. In the Gribov-Zwanziger context such an operator could
be $f^{abc} A^{a \, \mu} \xi^{ab}_\mu$ which from the equation of motion
\begin{equation}
\xi^{ab}_\mu ~=~ i \gamma^2 f^{abc} \frac{1}{\partial^\nu D_\nu} A^c_\mu
\end{equation}
would effectively equate to the presence of the Gribov horizon condition 
operator. Though one other application of (\ref{stapot}) could be to use it as 
a testbed for examining renormalon style corrections, like \cite{80}, since the
gap equation solution (\ref{gapsoln1}) is clearly non-perturbative. For
instance, in the $V$-scheme we would have a power correction beyond the
perturbative contribution of (\ref{potpert}) in (\ref{vdef}) which is
\begin{equation} 
\alpha_V(\mathbf{p}) ~=~ 
\alpha_V^{\mbox{\footnotesize{pert}}} (\mathbf{p}) ~-~
\frac{C_A^{3/2} \gamma^2 \alpha_s^2(\mu)}{2\mathbf{p}^2} ~+~ 
O \left( \frac{\gamma^4}{(\mathbf{p}^2)^2} \right) ~. 
\end{equation}
where, \cite{36,37,38},
\begin{eqnarray}
\alpha_V^{\mbox{\footnotesize{pert}}} (\mathbf{p}) &=& \alpha_s(\mu) 
\left[ 1 ~+~ \left[ \left[ \frac{31}{9} - \frac{11}{3} \ln \left[ 
\frac{\mathbf{p}^2}{\mu^2} \right] \right] C_A ~+~ \left[ \frac{4}{3} \ln 
\left[ \frac{\mathbf{p}^2}{\mu^2} \right] - \frac{20}{9} \right] T_F \Nf 
\right] a(\mu) \right. \nonumber \\
&& \left. ~~~~~~~~+~ O(a^2) \right] ~.
\label{vpert}
\end{eqnarray} 
A formally similar power correction was observed in the effective coupling of
(\ref{rgieffcc}) in \cite{33} where the correction had the same sign. In order
to compare and for completeness we record that the power correction to this
renormalization group invariant coupling in the same notation as (\ref{vpert}),
\cite{33}, is
\begin{equation}
\alpha_{\mbox{\footnotesize{eff}}} (\mathbf{p}) ~=~ 
\alpha_{\mbox{\footnotesize{eff}}}^{\mbox{\footnotesize{pert}}} (\mathbf{p}) ~-~
\frac{9C_A^{3/2} \gamma^2 \alpha_s^2(\mu)}{16\mathbf{p}^2} ~+~ 
O \left( \frac{\gamma^4}{(\mathbf{p}^2)^2} \right) 
\end{equation}
where 
\begin{eqnarray}
\alpha_{\mbox{\footnotesize{eff}}}^{\mbox{\footnotesize{pert}}} (\mathbf{p}) 
&=& \alpha_s(\mu) \left[ 1 ~+~ \left[ \left[ \frac{169}{36} - \frac{11}{3} \ln 
\left[ \frac{\mathbf{p}^2}{\mu^2} \right] \right] C_A ~+~ \left[ \frac{4}{3} 
\ln \left[ \frac{\mathbf{p}^2}{\mu^2} \right] - \frac{20}{9} \right] T_F \Nf 
\right] a(\mu) \right. \nonumber \\
&& \left. ~~~~~~~~+~ O(a^2) \right] ~.
\end{eqnarray} 

\begin{figure}[ht]
\hspace{5cm}
\epsfig{file=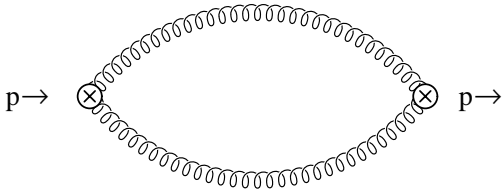,height=2.5cm}
\vspace{0.5cm}
\caption{Leading contribution to operator correlation function.}
\end{figure}

Next, we comment on one aspect concerning potentials and that is whether it is
possible to observe {\em stable} bound states. If there were such states then 
these could possibly be identified with glueballs. However, since we have 
demonstrated that the static potential does not have a linear confining part at
one loop but a form not dissimilar to that of $V_0(r)$, we expect any states 
formed in, say, a Schr\"{o}dinger equation analysis of (\ref{stapot}) to be 
unstable. For instance, in \cite{67} a confining piece had to be included in 
that glueball analysis which assumed a gluon mass. For a similar calculation in
the Gribov-Zwanziger context one would have to implement the gap equation for 
the Gribov mass in some fashion in addition. Aside from this one way of 
possibly quantifying properties of bound state masses is by considering the 
correlation of operators with the same quantum numbers as the bound states 
along the lines of \cite{13}. For glueballs the obvious candidate operators are
those involving the field strength given by
\begin{equation}
{\cal O}_S ~=~ -~ \frac{1}{4} G_{\mu\nu}^a G^{a \, \mu\nu} ~~~,~~~
{\cal O}_T^{\mu\nu} ~=~ -~ \frac{1}{4} \left( \frac{\eta^{\mu\nu}}{d}
G_{\sigma\rho}^a G^{a \, \sigma\rho} ~-~ G^{a \, \mu\sigma} G^a_{\nu\sigma} 
\right) 
\end{equation}
since they are gluonic and gauge invariant, where the subscripts $S$ and $T$
denote Lorentz scalar and tensor operators respectively and the second operator
is symmetric and traceless. It is based on a similar operator considered in 
\cite{82}. Indeed Zwanziger has examined the first operator in the context of 
(\ref{laggz}) by considering the leading term of the correlator of 
${\cal O}_S$, \cite{13}. In \cite{13} it was suggested that there was evidence 
for a state with mass squared $2 \sqrt{C_A} \gamma^2$, in our conventions, by 
rewriting the correlation function in a spectral representation using tools 
such as Schwinger and Feynman parameters and searching for physical cuts. 
However, a full explicit expression for the correlator as a function of the 
momentum and $\gamma$ at leading order was not given. Therefore, to partly 
address this we have computed the leading order term of the correlator of 
${\cal O}_S$ {\em exactly}. The Feynman diagram is illustrated in Figure $5$ 
where the crossed circled denotes the location of the operator insertion with 
momentum $p$ flowing through it. Defining

\begin{equation}
\Pi_S(p^2) ~=~ (4\pi)^2 i \int d^4 x \, e^{ipx} \langle 0 | {\cal O}_S(x) 
{\cal O}_S(0) | 0 \rangle 
\end{equation}
we have
\begin{eqnarray}
\Pi_S(p^2) &=& \! \! \left[ \frac{(p^2)^2}{4\epsilon} ~-~ 
\frac{3 C_A \gamma^4}{\epsilon} ~+~ \frac{C_A \gamma^4 
\sqrt{4C_A \gamma^4 - (p^2)^2}}{4 p^2} \tan^{-1} \left[ 
\frac{\sqrt{4 C_A \gamma^4 - (p^2)^2}}{p^2} \right] \right. \nonumber \\
&& \left. -~ \frac{\pi C_A^{3/2}\gamma^6}{4 p^2} ~+~ \frac{p^2}{8} 
\sqrt{4C_A \gamma^4 - (p^2)^2} \tan^{-1} \left[ 
\frac{\sqrt{4 C_A \gamma^4 - (p^2)^2}}{p^2} \right] \right. \nonumber \\
&& \left. +~ \left[ \frac{3}{2} \ln \left[ \frac{C_A \gamma^4}{\mu^4} \right]
- \frac{3}{2} + \frac{3\sqrt{2}}{16} \eta_2(p^2) \right] C_A \gamma^4 
~+~ \left[ \frac{3\pi}{8} + \frac{\sqrt{2}}{8} \eta_1(p^2)  
\right] \sqrt{C_A} \gamma^2 p^2 \right. \nonumber \\
&& \left. +~ \left[ \frac{1}{4} ~-~ \frac{1}{8} \ln \left[ 
\frac{C_A \gamma^4}{\mu^2} \right] - \frac{\sqrt{2}}{32} \eta_2(p^2) \right] 
(p^2)^2 \right] \NA ~+~ O(a) 
\label{opcor}
\end{eqnarray}
where the divergent contact terms are included for completeness. These are 
absorbed by the canonical contact renormalization for operator correlation 
functions but we note that as there is a Gribov parameter present this contact 
renormalization actually has a mixing aspect which is not unexpected. Given 
this one can regard the finite part as the leading piece of the correlation 
function since the first appearance of an explicit coupling constant is at next
order, ignoring the coupling constant implicit in $\gamma$, (\ref{gapsoln1}). 
We have checked that there are no poles in the expression at obvious places 
such as $\sqrt{C_A} \gamma^2$, $2\sqrt{C_A} \gamma^2$ or $4\sqrt{C_A} \gamma^2$
and hence regard this correlation function to be regular as a function of 
$p^2$. However, one can see that there is a cut at the same value as observed
in \cite{13} which is $p^2$~$=$~$2 \sqrt{C_A} \gamma^2$ and this is at the same
point as the cut in (\ref{stapot}). Equally it is elementary to verify 
Zwanziger's other observation in \cite{13} that there are unphysical cuts at 
$p^2$~$=$~$\pm$~$4i \sqrt{C_A} \gamma^2$ in our conventions similar to
(\ref{stapot}). One final point concerning the structure of $\Pi_S(p^2)$ in
relation to both glueball states and the physical cut structure and that is 
that the same conclusion would be obtained if other operators with similar 
properties to ${\cal O}_S$ were considered. For instance, the correlation 
functions of both the operators $\mbox{Tr} \left[ \left( D_\mu G_{\nu\sigma}
\right) \left( D^\mu G^{\nu\sigma} \right) \right]$ and $\mbox{Tr} \left[ 
\left( D_\mu D_\nu G_{\sigma\rho} \right) \left( D^\mu D^\nu G^{\sigma\rho} 
\right) \right]$ with themselves, where $G_{\mu\nu}$~$=$~$G^a_{\mu\nu} T^a$, 
will produce the same physical and unphysical cuts as (\ref{opcor}). These 
Lorentz scalar operators are both gauge invariant and have the same number of 
leading gluon legs. 

\begin{figure}[ht]
\hspace{5cm}
\epsfig{file=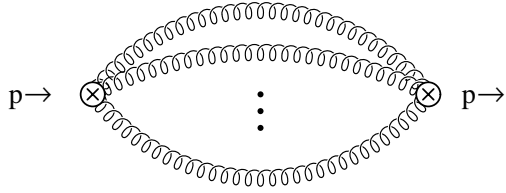,height=2.5cm}
\vspace{0.5cm}
\caption{Higher order graphs contributing to operator correlation functions.}
\end{figure}

For higher loop graphs contributing to the gluonic operator correlation 
functions it is straightforward to determine the location of the physical cuts.
Such graphs are illustrated in Figure $6$ where the dots denote more and more 
gluon propagators. When there are an odd number of gluons then there are 
unphysical cuts but no physical ones. For an even number of gluons, say $2n$, 
then there are both types of cuts with the physical ones being at 
$p^2$~$=$~$2 n^2 \sqrt{C_A} \gamma^2$. The pattern of unphysical cuts cannot be
written in as compact a formula. As noted by Zwanziger, \cite{13}, for the two 
gluon case the cuts additional to the physical ones are at 
$p^2$~$=$~$\pm$~$4 i \sqrt{C_A} \gamma^2$. It is straightforward to record 
those for the lowest order cases. For the three gluon situation the only cuts 
are at $p^2$~$=$~$\pm$~$i \sqrt{C_A} \gamma^2$,
$p^2$~$=$~$\pm$~$9 i \sqrt{C_A} \gamma^2$ and 
$p^2$~$=$~$(\pm$~$4$~$\pm$~$3 i) \sqrt{C_A} \gamma^2$ where all possible sign 
combinations are taken in the final off axis cut. For four gluons, the
unphysical cuts are now at $p^2$~$=$~$\pm$~$4i \sqrt{C_A} \gamma^2$,
$p^2$~$=$~$\pm$~$16 i \sqrt{C_A} \gamma^2$ and 
$p^2$~$=$~$2(\pm$~$4$~$\pm$~$3 i) \sqrt{C_A} \gamma^2$. For other Green's 
functions, such as the higher loop corrections to (\ref{stapot}), we expect the
physical (and unphysical) cut structure to be the same. Returning to Figure $6$
and, for instance, ignoring the presence of internal vertices for the moment, 
the next physical cut in (\ref{opcor}) will be at 
$p^2$~$=$~$8\sqrt{C_A} \gamma^2$. If instead of (\ref{opcor}) one considers the
correlation function of a gauge invariant operator involving three gluons in 
the leading leg term, such as 
$f^{abc} G^a_{\mu\nu} G^{b\,\nu}_{~~\,\sigma} G^{c\,\sigma\mu}$ which was 
considered in \cite{83}, then the first cut would be at 
$p^2$~$=$~$8\sqrt{C_A} \gamma^2$ with the subsequent one at 
$p^2$~$=$~$18\sqrt{C_A} \gamma^2$. In \cite{83} QCD sum rules were used to 
estimate the masses of various glueball states. For instance, the two gluon 
scalar glueball mass was $M_{2g}$~$=$~$(1.50 \pm 0.19)\mbox{GeV}$ and that for 
the three gluon bound state was $M_{3g}$~$=$~$3.1\mbox{GeV}$ with no errors 
quoted, \cite{83}. Intriguingly the mass ratio is 
$M_{3g}/M_{2g}$~$\approx$~$2$. If we take the ratio of the leading physical 
cuts for the appropriate gauge invariant gluonic operator correlation functions
then we have the {\em same} ratio deriving from the expression of the ratio in 
the crude form $\sqrt{8/2}$. By contrast, however, if one was working with 
gluons which had an ordinary massive propagator instead of the Gribov width,
then one would have a canonical cut structure in the corresponding Feynman 
diagrams. Overlooking the obvious loss of gauge invariance for the moment, then
the corresponding ratio for $M_{3g}/M_{2g}$ would be $3/2$. This seems to be
too far away to accommodate the mass ratio of \cite{83}. Returning to the 
Gribov-Zwanziger case, an estimate was also given in \cite{83} for the tensor 
$2^{++}$ state which was $M_{2^{++}}$~$=$~$(2.0 \pm 0.1)\mbox{GeV}$. This gives
the ratio $M_{2^{++}}/M_{2g}$~$\approx$~$4/3$. Such a ratio can be accommodated
in terms of ratios of physical cuts, such as $\sqrt{32/18}$ in crude form. 
Similarly, $M_{3g}/M_{2^{++}}$ can be accommodated by the ratio $\sqrt{18/8}$. 
However, we acknowledge that the justification of these latter ratios from a 
simple diagram argument similar to the earlier one seems to be tenuous. This is
partly because (\ref{stapot}) can only really be regarded as being relevant for
spinless zero angular momentum bound states such as the lowest lying two and 
three gluon states. For states such as $2^{++}$ one requires more detailed 
structure beyond the simple radial dependence provided in (\ref{stapot}), such 
as that considered in depth in \cite{67}. It would be interesting, though, to 
see if the ratio of $4/3$ could be justified from that point of view in an 
extended potential incorporating spin and angular momentum. However, even 
though we are primarily concentrating on Yang-Mills here the issue of 
degeneracy will arise at some point for the higher states. Whilst not as 
involved a problem as when quarks are present, which would clearly extend the 
number of potential bound states, resolving any mixing will require a much more
detailed analysis of a potential derived from (\ref{laggz}). Another test of 
the ratio hypothesis would be the mass of a four gluon state which ought to be 
roughly three times that of the lowest state if our reasoning is sound. 
However, such a state has not been extensively studied, for example on the 
lattice, as far as we are aware. Finally, in this simple cut analysis we need 
to temper our remarks with the fact that with the inclusion of higher loops it 
is not inconceivable that the cut locations will be shifted by corrections. So 
our method of estimating mass ratios should only regarded as a rough guide. 

Whilst trying to understand the relative sizes of various glueball states is an
interesting exercise in itself, ultimately one has to eventually estimate one 
of the masses which is a non-trivial task. We draw attention to several 
possibilities in the context of (\ref{laggz}). In each case one has to somehow 
fix a mass scale in relation to a measured quantity from other methods such as 
the lattice in order to obtain a numerical value for $\sqrt{C_A}\gamma^2$ which
is the fundamental mass parameter in (\ref{laggz}). One approach to extract a 
glueball mass would be to consider the power corrections to (\ref{opcor}) and 
apply sum rule technology akin to that used in \cite{82,83} where the gluon 
condensate is the quantity used to fix a scale. In \cite{82} a tachyonic gluon 
mass was introduced to account for apparent discrepancies with experimental 
data. Performing the detailed analyses for both these approaches is clearly 
outside the scope of the current article. For example, a proper sum rule 
analysis would first require a reworking of the original operator product 
expansion but using the Lagrangian of (\ref{laggz}) rather than the usual QCD 
Lagrangian. However, we have computed the leading power correction to 
(\ref{opcor}) in the context of the gluonia channels of $\Pi_S(q^2)$ considered
in \cite{82}. Therefore, using the same moment definition and notation as 
\cite{82} we have 
\begin{equation}
\Pi_S(M^2) ~=~ ( \mbox{parton model} ) \left[ 1 ~-~ 
\frac{6C_A \gamma^4}{M^4} ~+~ O \left( \frac{\gamma^8}{M^8} \right) \right]
\end{equation}
where 
\begin{equation}
\Pi_S(M^2) ~\equiv~ \frac{(p^2)^n}{(n-1)!} \left( \frac{d~}{dp^2} \right)^n
\Pi_S(p^2)
\end{equation}
and $M^2$~$=$~$p^2/n$ is finite in the limit of large $p^2$ and $n$. As a check
on our derivation of this power correction using {\sc Form}, \cite{57}, we have
replaced the gluon propagator of (\ref{laggz}) by the tachyonic gluon mass 
propagator of \cite{82} and correctly reproduced the power correction 
{\em quadratic} in this mass recorded in \cite{82}. As we are dealing with a 
gauge invariant operator correlation function in the Gribov-Zwanziger case the 
first power correction is quartic with the correction numerator being related 
on dimensional grounds to the gluon condensate. Although in the static 
potential case the leading order function was a function of $\gamma^4$, the 
loop correction produced $O(\gamma^2)$ terms in the power series expansion. 
Such a scenario could emerge for $\Pi_S(p^2)$ when the $O(a)$ correction is 
computed. Though this is currently beyond the scope of this article as well as 
using sum rules to estimate scalar or tensor glueball masses akin to \cite{83}.
Finally, for the correlation function of the tensor operator 
${\cal O}_T^{\mu\nu}$, similar power corrections should emerge in each of the 
scalar amplitudes of its Lorentz decomposition. We record the full expressions 
for the tensor case for completeness, and for comparison with \cite{82}, in 
appendix C as they are similar to (\ref{opcor}). However, there is nothing 
formally different from the scalar case above having, for example, the same 
physical cut.

Another approach would be to make direct contact with a non-zero vacuum
expectation value, whose numerical value is known, by explicit computation 
using (\ref{laggz}) itself. This is possible since now the gluon propagator is 
not massless, (\ref{propdef}), and so vacuum expectation values of gluonic
operators are non-zero. Such an approach has already been used in 
\cite{84,85,86} but where the gluon is assumed to have a canonical mass term. 
Ignoring the first of these three papers, since it appears to have a result 
inconsistent with the latter two, an estimate for the gluon mass was deduced by
comparing the leading order gluon condensate value with the quark current 
correlator in the high energy limit at the same order. However, we can also 
consider the dimension two condensate based on the operator $\frac{1}{2} 
( A^a_\mu )^2$ which has been measured on the lattice in \cite{72,73}. At 
leading order we have
\begin{equation}
\left\langle \frac{1}{2} \left( A^a_\mu \right)^2 \right\rangle ~=~ 
\frac{3\NA\sqrt{C_A}\gamma^2}{64\pi} ~+~ O(a) ~.
\end{equation}
So, for instance, at leading order we have the formal relation for a glueball 
mass, if we regard our lowest cut as a glueball mass, of
$2\sqrt{C_A}\gamma^2$~$=$~$(16\pi/3) \langle \half ( A^a_\mu )^2 \rangle$. 
Taking the lattice estimate of $\langle ( A^a_\mu )^2 
\rangle$~$=$~$3.1\mbox{GeV}^2$, from \cite{73}, for example, then this rough
way of estimating would give an unrealistic glueball mass of $5.1\mbox{GeV}$.
Aside from this one ought not to overlook the gap equation, (\ref{gap1}),
which is assumed to be valid at this order of approximation. If one uses 
representative numerical values for $\sqrt{C_A}\gamma^2$ derived in this way
and a reasonable estimate for $\Lambda_{\mbox{\footnotesize{$\MSbar$}}}$ of, 
say, $\Lambda_{\mbox{\footnotesize{$\MSbar$}}}$~$=$~$300\mbox{MeV}$ then one 
would obtain a {\em negative} value of $\alpha_s$ which is clearly 
unacceptable. Though this is partly due to the Landau pole problem of the 
running coupling constant. (We find similar conclusions when the gluon
condensate is used similar to \cite{85}.) Of course, this simple exercise has 
been recorded to merely illustrate some of the potential difficulties in 
estimating a value for the underlying mass parameter of (\ref{laggz}). Indeed 
we have only considered leading order and including higher order corrections 
will in principle change the estimates. However, the full vacuum expectation 
value should have a non-perturbative piece which would need to be properly 
incorporated into any deeper analysis as well as dealing with infrared issues 
which are known to exist in the case of the gluon condensate. The lack of 
consistency with the Gribov gap equation to the order we considered, clearly 
indicates that the approximations assumed here could not be validated in a 
consistent way. By contrast there is no equivalent gap equation constraint on 
the fundamental gluon mass used in the analyses of \cite{84,85,86}. Also, it 
may be the case that the Gribov gap equation would require a more complete 
function of $a$ beyond the perturbative approximation. Therefore, whilst 
carrying out this rough leading order analysis has led to a null conclusion, we
feel it is useful to include it as a moderating point of view since it does 
perhaps illustrate the difficulty in producing a reasonable estimate for the 
underlying mass parameter of (\ref{laggz}). Indeed it may also be indicative 
that the use of condensates to fix a mass scale may not be compatible when a 
Gribov mass is present. 

\sect{Discussion.}

The main result of this article is the explicit construction of the one loop
static potential using the Gribov-Zwanziger Lagrangian. The original motivation
was to study a gauge invariant object related to the non-perturbative structure
of a formulation of a non-abelian gauge theory which has properties suggesting 
it describes a confined gluon. Indeed one aim was to see what functions of 
momentum appeared in the explicit final expression. Whilst a linear potential 
clearly does not emerge in the final result, there is an intriguing hint with 
the presence of a pure dipole term but which has a compensating term in the 
zero momentum limit for positive $\gamma^2$. Ultimately we have to conclude 
that the usefulness of this approach to the static potential is that it will in
principle closely match the full potential if one adds successive loop 
corrections to (\ref{stapot}). So it would appear that in the present context a
linearly rising potential for a significant range of $r$ arises out of a truly 
non-perturbative effect. Indeed it is hard to see how the compensating term 
could be split from the pure dipole in higher loop corrections to leave a net 
dipole as well as no net higher order momentum poles. Therefore, some 
non-perturbative feature must be taken into account. 

In the context of this 
field theoretic approach in the Gribov set-up the most promising candidate for 
this is the (non-perturbative) Gribov gap equation. To this end we revisited 
the behaviour of the bosonic $\rho^{ab}_\mu$ and $\xi^{ab}_\mu$ localizing 
ghosts which dominate in the infrared and are responsible in essence for 
implementing the horizon condition in (\ref{laggz}). We have been able to 
derive enhancement for both bosonic localizing ghosts, together with the colour
structure but it differs from that derived in Zwanziger's Schwinger Dyson 
analysis, \cite{50}. Moreover, we do not find the over-enhancement discovered
in \cite{50}. We have discussed various possibilities that might lead to
similar behaviour but this would at least require a higher loop computation. 
One lesson appears to be that the inversion of the matrix of $2$-point 
functions to obtain the propagators is intricately tied to the colour group 
structure of the corrections. For instance, to extend the perturbative analysis
to the two loop level is in principle possible in the zero momentum limit but 
requires the computation of over $1000$ Feynman diagrams since one has to 
consider the full matrix of $2$-point functions in the $\{A^a_\mu, 
\xi^{ab}_\mu\}$ sector. This would at least give some indication of the zero 
momentum behaviour of the colour channels not present in the original 
Lagrangian and whether the gap equation emerges in the corrections similar to 
(\ref{xi2pt}). It could be the case that additional enhancement arises in other
channels or an over-enhancement akin to that of \cite{50}. The aim of such an 
analysis would be to see if a dipole dominated the exchange between the static 
colour sources of the formalism but in such a way that higher order poles were 
excluded. 

Further, one underlying feature of the static potential formalism is 
the absence of a direct coupling of the source to either $\rho^{ab}_\mu$ or 
$\xi^{ab}_\mu$ which prevents the direct single $\rho^{ab}_\mu$ or 
$\xi^{ab}_\mu$ field exchange with an enhanced propagator being the simple 
explanation for a linear potential, provided such a vertex did not depend on 
the exchange momentum. Though one would naively believe $\xi^{ab}_\mu$ has to 
play some role in this way since from its equation of motion it is a non-local 
projection of the gluon. Perhaps the original Wilson loop static potential 
formalism would need to be reconsidered in the Gribov-Zwanziger case due to the
restriction of the path integral to the Gribov region of configuration space. 
From another point of view, although the Gribov-Zwanziger Lagrangian is 
consistent with the Kugo-Ojima confinement criterion, \cite{34,35}, that is a 
necessary but not sufficient condition for confinement which would imply that 
an additional feature might be necessary in (\ref{laggz}) to obtain a confining
potential. For instance, the Gribov construction is founded on the 
{\em infinitesimal} behaviour of the gauge fixing condition, \cite{10}, and the
less local aspects of that construction might need to be included now. 

Throughout the article we have concentrated on what is referred to now as the
conformal or scaling solution. In recent years there has been interest in an
alternative point of view which is called the decoupling solution, 
\cite{22,23,24,25,26,27,28,29}. Essentially this scenario differs from the 
gluon suppression and ghost enhancement properties of the conformal solution, 
in having no Faddeev-Popov ghost enhancement and the gluon propagator freezes 
to a finite non-zero value excluding suppression. The evidence for this 
behaviour derives from both lattice gauge theory and Schwinger Dyson analyses,
\cite{22,23,24,25,26,27,28,29}. As yet there is no definitive concensus as to 
which of the conformal or decoupling solutions is the correct picture of 
infrared Landau gauge Yang-Mills theory. 

One proposal to explain the lack of 
suppression and enhancement in the Gribov-Zwanziger Lagrangian is the 
condensation of the BRST invariant operator 
$\bar{\phi}^{ab}_\mu \phi^{ab\,\mu}$~$-$~$\bar{\omega}^{ab}_\mu 
\omega^{ab\,\mu}$, \cite{52,53}, using the notation of the original localizing 
fields. With a non-zero value for the vacuum expectation value of this operator
the observed behaviour of the gluon and Faddeev-Popov ghost propagators could 
be accommodated. However, to include such an operator one applies the local 
composite operator formalism developed in \cite{87,88,89}. Briefly the method 
introduces a new source coupled to the dimension two operator and then 
constructs the effective potential of the operator. Studying the minima of this
effective potential one observes that the perturbative vacuum solution is 
unstable in favour of a vacuum solution where the operator condenses. Hence a 
new dynamically generated mass is introduced which modifies the propagators of 
the appropriate fields. In the application to the Gribov-Zwanziger Lagrangian, 
\cite{52,53}, both sets of localizing ghosts acquire a mass and induce extra 
mass dependence in the gluon and Faddeev-Popov ghost propagators. Moreover, the
corresponding gap equation is insufficient to produce a Faddeev-Popov 
enhancement. 

This raises several points for our static potential calculation. 
First, we have taken the point of view that the $\rho^{ab}_\mu$, $\xi^{ab}_\mu$
and $\omega^{ab}_\mu$ localizing fields are purely {\em internal} fields with 
no coupling to external sources in the way the gluon does. Therefore, it seems 
unclear how to incorporate all the dynamically modified propagators at the 
outset since including them would require coupling an {\em operator} of 
internal fields to an external source. Moreover, to have a homogeneous 
renormalization group interpretation in the local composite operator formalism 
one has to allow for the generation of the square of this external source. 
However, even if one relaxed such assumptions or used the modified propagators 
directly in the analogous computation of (\ref{stapot}) then it seems difficult
to see how a linearly rising term could emerge in the corresponding static 
potential due to the absence of enhancement. 

This leads on to the other point 
of view we examined and that was the enhancement of $\rho^{ab}_\mu$ and
$\xi^{ab}_\mu$. It now appears evident, at least in the conformal solution, 
that the enhancement of the Faddeev-Popov ghost, $\rho^{ab}_\mu$, 
$\xi^{ab}_\mu$ and $\omega^{ab}_\mu$ are on the same footing and inextricably 
linked. Therefore, if there is a loss of enhancement in the decoupling solution
for the Faddeev-Popov ghost it would seem inevitable that this would be the 
case for the localizing fields too. Whilst we analysed the simple two loop 
contribution of Figure $4$ to the static potential with enhanced $\xi^{ab}_\mu$
and $\rho^{ab}_\mu$ propagators, it was insufficient with the enhancement to 
produce a dipole behaviour at one loop primarily due to group theory
considerations. If there was an non-enhanced $\xi^{ab}_\mu$ propagator in the 
decoupling case then there would appear to be no emergence of anything like a
dipole behaviour in the zero momentum limit. However, the behaviour of the 
$\rho^{ab}_\mu$ and $\xi^{ab}_\mu$ propagators in the infrared for the 
decoupling solution has not been established yet. If the decoupling solution is
established as the correct description of the infrared properties of Landau 
gauge Yang-Mills it would be interesting to see how the linearly rising 
potential emerges in the field theory context. For instance, in \cite{19} the 
Wilson loop was considered in the context of (\ref{laggz}) and it was shown 
that the string tension of a linearly rising potential depended on the zero 
momentum value of a particular combination of the gluon and Faddeev-Popov ghost
propagator form factors. Briefly, if the gluon propagator froze to a zero or
non-zero value but the Faddeev-Popov ghost was {\em enhanced} then a linear 
potential emerged. If the argument of \cite{19} remains valid in the decoupling
scenario then the lack of Faddeev-Popov enhancement would seem to exclude a 
linear potential in that case. 
 
\vspace{1cm} 
\noindent
{\bf Acknowledgement.} The author thanks Dr F.R. Ford, Prof. S.J. Hands, Dr 
P.E.L. Rakow, Prof. S.P. Sorella and Prof. D. Zwanziger for useful discussions. 
Support from IHES, Paris, France, where part of this work was carried out, is 
also gratefully acknowledged.

\appendix

\sect{Group theory.}

In this appendix we discuss the decomposition of products of group generators 
into the basis of Casimirs. For a comprehensive review of such Casimirs we 
refer the reader to \cite{78}. In section $6$ we introduced the two sets of 
coefficients $\{a_i\}$ and $\{b_i\}$ where 
\begin{equation}
d_A^{abpq} d_A^{cdpq} ~=~ a_1 \delta^{ab} \delta^{cd} ~+~
a_2 \left( \delta^{ac} \delta^{bd} ~+~ \delta^{ad} \delta^{bc} \right) ~+~
a_3 \left( f^{ace} f^{bde} ~+~ f^{ade} f^{bce} \right) ~+~ 
a_4 d_A^{abcd}
\label{dec1}
\end{equation} 
and 
\begin{equation}
f^{ape} f^{bqe} d_A^{cdpq} ~=~ b_1 \delta^{ab} \delta^{cd} ~+~
b_2 \left( \delta^{ac} \delta^{bd} ~+~ \delta^{ad} \delta^{bc} \right) ~+~
b_3 \left( f^{ace} f^{bde} ~+~ f^{ade} f^{bce} \right) ~+~ 
b_4 d_A^{abcd} ~.
\label{dec2}
\end{equation} 
These coefficients can be computed by the projection method. By this we
mean that each tensor on the right hand side is used in sequence to multiply
each equation. Using the properties of the Lie algebra this leads to a set
of linear equations for the $a_i$ and $b_i$ which can be solved by simple
matrix inversion. It transpires that the matrix which needs to be inverted is
the same for both (\ref{dec1}) and (\ref{dec2}) and is 
\begin{equation}
M ~=~ \left(
\begin{array}{cccc}
\NA^2 & 2 \NA & 2 C_A \NA & \frac{5}{6} C_A^2 \NA \\ 
2 \NA & 2 \NA ( \NA + 1 ) & -~ 2 C_A \NA & \frac{5}{3} C_A^2 \NA \\ 
2 C_A \NA & -~ 2 C_A \NA & 3 C_A^2 \NA & 0 \\ 
\frac{5}{6} C_A^2 \NA & \frac{5}{3} C_A^2 \NA & 0 & d_A^{abcd} d_A^{abcd} \\ 
\end{array}
\right) ~.
\end{equation} 
To determine $a_i$ and $b_i$ the inverse of $M$ multiplies the respective
vectors
\begin{equation}
\left( 
\begin{array}{c}
\frac{25}{36} C_A^4 \NA \\ 2 d_A^{abcd} d_A^{abcd} \\ \frac{2}{3} C_A
d_A^{abcd} d_A^{abcd} \\ d_A^{abcd} d_A^{cdpq} d_A^{abpq} \\ 
\end{array}
\right) 
~~~~~~ \mbox{and} ~~~~~~
\left( 
\begin{array}{c}
\frac{5}{6} C_A^3 \NA \\ 0 \\ 2 d_A^{abcd} d_A^{abcd} \\ \frac{1}{3} C_A
d_A^{abcd} d_A^{abcd} \\
\end{array}
\right) ~.
\end{equation}
Hence, we find 
\begin{eqnarray}
a_1 &=& -~ \left[ 540 C_A^2 \NA (\NA-3) d_A^{abcd} d_A^{cdpq} d_A^{abpq}
+ 144 (2\NA+19) \left( d_A^{abcd} d_A^{abcd} \right)^2 \right. \nonumber \\
&& \left. ~~~~-~ 150 C_A^4 \NA (3\NA+11) d_A^{abcd} d_A^{abcd} 
+ 625 C_A^8 \NA^2 \right] \nonumber \\
&& ~~~ \times ~ 
\frac{1}{54\NA(\NA-3) [ 12 (\NA + 2) d_A^{efgh} d_A^{efgh} - 25 C_A^4 \NA]}
\nonumber \\
a_2 &=& \left[ 144 (11\NA - 8) \left( d_A^{abcd} d_A^{abcd} \right)^2 \right.
- 1080 C_A^2 \NA (\NA - 3) d_A^{abcd} d_A^{cdpq} d_A^{abpq} \nonumber \\
&& \left. ~+~ 625 C_A^8 \NA^2 - 3000 C_A^4 \NA d_A^{abcd} d_A^{abcd} \right] 
\nonumber \\
&& \times ~ 
\frac{1}{108\NA(\NA-3) [ 12 (\NA + 2) d_A^{efgh} d_A^{efgh} - 25 C_A^4 \NA]}
\nonumber \\
a_3 &=& \frac{[ 12 (\NA + 2) d_A^{abcd} d_A^{abcd} - 25 C_A^4 \NA]}
{54 C_A \NA (\NA-3)} \nonumber \\
a_4 &=& \frac{[ 216 (\NA + 2) d_A^{abcd} d_A^{cdpq} d_A^{abpq} - 125 C_A^6 \NA
- 360 C_A^2 d_A^{abcd} d_A^{abcd}]}{18[ 12 (\NA + 2) d_A^{efgh} d_A^{efgh} 
- 25 C_A^4 \NA]}
\end{eqnarray}
and
\begin{equation}
b_1 ~=~ -~ 2 b_2 ~=~ \frac{[5C_A^4 \NA - 12 d_A^{abcd} d_A^{abcd}]}
{9 C_A \NA (\NA - 3)} ~~,~~ 
b_3 ~=~ \frac{[6 (\NA-1) d_A^{abcd} d_A^{abcd} - 5 C_A^4 \NA]}
{9 C_A^2 \NA (\NA - 3)} ~~,~~ b_4 ~=~ \frac{C_A}{3} ~.
\end{equation} 
These expressions have been derived by making use of the {\tt color.h} package
of \cite{57,78}. Both sets of coefficients can be evaluated explicitly for 
$SU(N_c)$ and for completeness we note the values of the various Casimirs in
this instance are
\begin{equation}
\NA ~=~ N_c^2 ~-~ 1 ~~~,~~~ C_A ~=~ N_c ~~~,~~~ 
\frac{d_A^{abcd} d_A^{abcd}}{\NA} ~=~ \frac{N_c^2[N_c^2+36]}{24} 
\label{cassun}
\end{equation}
where the last expression is given in \cite{90}. The set $\{a_i\}$ also require 
$d_A^{abcd} d_A^{cdpq} d_A^{abpq}$ which we have evaluated directly using the 
$SU(N_c)$ identity of \cite{91}
\begin{equation}
f^{abe} f^{cde} ~=~ \frac{2}{N_c} \left[ \delta^{ac} \delta^{bd} - \delta^{ad}
\delta^{bc} \right] ~+~ d^{ace} d^{bde} ~-~ d^{ade} d^{bce}
\end{equation}
and relations for products of the structure functions and the totally symmetric
rank $3$ tensor $d^{abc}$, \cite{91}. As an intermediate step we have 
\begin{equation}
d_A^{abcd} ~=~ \frac{2}{3} \left[ \delta^{ab} \delta^{cd} ~+~ 
\delta^{ac} \delta^{bd} ~+~ \delta^{ad} \delta^{bc} \right] ~+~
\frac{N_c}{12} \left[ d^{abe} d^{cde} ~+~ d^{ace} d^{bde} ~+~ d^{ade} d^{bce} 
\right] 
\end{equation} 
for $SU(N_c)$ which we have checked correctly reproduces the final expression 
of (\ref{cassun}). Hence, we found
\begin{equation}
\frac{d_A^{abcd} d_A^{cdpq} d_A^{abpq}}{\NA} ~=~ 
\frac{N_c^2 \left[ N_c^4 + 135 N_c^2 + 324 \right]}{216} ~.
\end{equation}
We have checked the consistency of this expression by explicitly evaluating the
left hand side for both colour groups $SU(2)$ and $SU(3)$. To do this we used 
the explicit respective $3$~$\times$~$3$ and $8$~$\times$~$8$ matrix 
representations of the adjoint group generators using {\sc Form}. Although it
is not required here, as a corollary we have also determined the $SU(N_c)$ 
value for $d_A^{abcdef} d_A^{abcdef}$ which is introduced in \cite{78}. We find
\begin{equation}
\frac{d_A^{abcdef} d_A^{abcdef}}{\NA} ~=~ 
\frac{N_c^2 \left[ N_c^4 + 666 N_c^2 + 1800 \right]}{1920} ~.
\end{equation}
Equipped with the $SU(N_c)$ values of these Casimirs we have 
\begin{equation}
a_1 ~=~ \frac{7N_c^2}{12} ~~,~~ a_2 ~=~ -~ \frac{N_c^2}{24} ~~,~~ 
a_3 ~=~ \frac{N_c(N_c^2-9)}{108} ~~,~~ a_4 ~=~ \frac{(N_c^2+9)}{9} 
\end{equation} 
and
\begin{equation}
b_1 ~=~ -~ 2 b_2 ~=~ \frac{N_c}{2} ~~,~~ 
b_3 ~=~ \frac{(N_c^2+18)}{36} ~~,~~ b_4 ~=~ \frac{N_c}{3} ~. 
\end{equation} 
It is worth noting that some of the numerator factors as well as the 
denominator factors of $a_i$, which are $\NA$, $(\NA^2-3)$ and 
$[12 (\NA + 2) d_A^{abcd} d_A^{abcd} - 25 C_A^4 \NA]$, have zeroes at 
$N_c$~$=$~$1$, $2$ and $3$. Therefore, one cannot directly evaluate $a_i$ for
the latter two values of $N_c$ but must derive the $N_c$ dependent expressions 
first before determining numerical values.

\sect{Transverse parts.}

In this appendix we collect the explicit one loop $\MSbar$ expressions for the 
transverse parts of the $2$-point functions of (\ref{twoptdef}). We have  
\begin{eqnarray}
X &=& \left[ -~ \frac{37\pi}{128} \sqrt{C_A^3} \gamma^2 ~-~ 
\left[ \frac{23}{48} \tan^{-1} \left[ \frac{\sqrt{C_A}\gamma^2}{p^2} \right]
+ \frac{25\sqrt{2}}{64} \eta_1(p^2) \right] \sqrt{C_A^3} \gamma^2 \right. 
\nonumber \\
&& \left. ~-~ \frac{7 C_A \sqrt{4C_A \gamma^4 - (p^2)^2}}{192} \tan^{-1} 
\left[ -~ \frac{\sqrt{4C_A \gamma^4 -(p^2)^2}}{p^2} \right] \right. 
\nonumber \\
&& \left. ~-~ \frac{3C_A^2\gamma^4}{8(p^2)^2} \sqrt{4C_A\gamma^4-(p^2)^2}
\tan^{-1} \left[ - \frac{\sqrt{4C_A\gamma^4-(p^2)^2}}{p^2} \right] 
\right. \nonumber \\
&& \left. ~-~ \frac{59\pi}{128} \frac{\sqrt{C_A^5}\gamma^6}{(p^2)^2} ~+~ 
\frac{11}{64} \frac{\sqrt{C_A^5}\gamma^6}{(p^2)^2} \tan^{-1} \left[ 
\frac{\sqrt{C_A}\gamma^2}{p^2} \right] \right. \nonumber \\
&& \left. ~+~ \left[ \frac{35}{64} - \frac{47}{192} \ln \left[ 1 +
\frac{(p^2)^2}{C_A\gamma^4} \right] \right] \frac{C_A^2\gamma^4}{p^2} ~+~ 
\left[ \frac{4}{3} \ln \left[ \frac{p^2}{\mu^2} \right] - \frac{20}{9} \right] 
T_F \Nf p^2 \right. \nonumber \\
&& \left. ~+~ \left[ \frac{1939}{576} - \frac{135}{128} \ln \left[
\frac{C_A\gamma^4}{\mu^4} \right] + \frac{7}{64} \ln \left[
\frac{[(p^2)^2+C_A\gamma^4]}{\mu^4} \right] \right. \right. \nonumber \\
&& \left. \left. ~~~~~~~-~ \frac{53}{192} \ln \left[ \frac{p^2}{\mu^2} \right] 
- \frac{241\sqrt{2}}{768} \eta_2(p^2) \right] C_A p^2 \right. \nonumber \\
&& \left. ~+~ \frac{25(p^2)^2}{768\gamma^4} \sqrt{4C_A \gamma^4 - (p^2)^2} 
\tan^{-1} \left[ -~ \frac{\sqrt{4C_A \gamma^4 -(p^2)^2}}{p^2} \right] ~-~
\frac{131\pi}{768} \frac{\sqrt{C_A}(p^2)^2}{\gamma^2} \right. \nonumber \\ 
&& \left. ~+~ \left[ \frac{19}{192} \tan^{-1} \left[ 
\frac{\sqrt{C_A}\gamma^2}{p^2} \right] - \frac{35\sqrt{2}}{384} \eta_1(p^2) 
\right] \frac{\sqrt{C_A}(p^2)^2}{\gamma^2} \right. \nonumber \\ 
&& \left. ~+~ \left[ \frac{1}{96} \ln \left[ \frac{C_A\gamma^4}{\mu^4} 
\right] - \frac{1}{48} \ln \left[ \frac{[(p^2)^2+C_A\gamma^4]}{\mu^4} \right]
\right. \right. \nonumber \\
&& \left. \left. ~~~~~~~+~ \frac{1}{48} \ln \left[ \frac{p^2}{\mu^2} \right] 
- \frac{3\sqrt{2}}{1024} \eta_2(p^2) \right] \frac{(p^2)^3}{C_A\gamma^4}
\right] a ~+~ O(a^2)  
\\
U &=& i \left[ C_A \left[ \frac{1}{64} \ln \left[ 1 
+ \frac{(p^2)^2}{C_A\gamma^4} \right] - \frac{31}{64} \right] ~-~
\frac{C_A^2\gamma^4}{96(p^2)^2} \ln \left[ 1 + \frac{(p^2)}{C_A\gamma^4} 
\right] ~+~ \frac{179\pi\sqrt{C_A^3}\gamma^2}{384p^2} 
\right. \nonumber \\
&& \left. ~~~-~ \frac{11\sqrt{C_A^3}\gamma^2}{192p^2} \tan^{-1} \left[
\frac{\sqrt{C_A}\gamma^2}{p^2} \right] \right. \nonumber \\
&& \left. ~~~+~ \left[ \frac{7 C_A}{16p^2} - \frac{7 p^2}{64\gamma^4} \right] 
\sqrt{4C_A \gamma^4 - (p^2)^2} 
\tan^{-1} \left[ -~ \frac{\sqrt{4C_A \gamma^4 -(p^2)^2}}{p^2} \right] ~-~
\frac{39\pi \sqrt{C_A} p^2}{128\gamma^2} \right. \nonumber \\
&& \left. ~~~-~ \left[ \frac{1}{24} \tan^{-1} \left[ 
\frac{\sqrt{C_A}\gamma^2}{p^2} \right] + \frac{7\sqrt{2}}{64} \eta_1(p^2)
\right] \frac{\sqrt{C_A}p^2}{\gamma^2} \right. \nonumber \\ 
&& \left. ~~~+~ \left[ \frac{5}{192} \ln \left[ 
\frac{[(p^2)^2+C_A\gamma^4]}{\mu^4} \right] - \frac{3}{128} \ln \left[
\frac{C_A\gamma^4}{\mu^4} \right] - \frac{1}{192} \ln \left[ \frac{p^2}{\mu^2} 
\right] + \frac{5\sqrt{2}}{256} \eta_2(p^2) \right] \frac{(p^2)^2}{\gamma^4} 
\right. \nonumber \\
&& \left. ~~~-~ \frac{\pi (p^2)^3}{256\sqrt{C_A}\gamma^6} ~+~ \left[ 
\frac{1}{64} \tan^{-1} \left[ \frac{\sqrt{C_A}\gamma^2}{p^2} \right] 
- \frac{\sqrt{2}}{512} \eta_1(p^2) \right] \frac{(p^2)^3}{\sqrt{C_A}\gamma^6} 
\right] \gamma^2 a ~+~ O(a^2) 
\\
V &=& W_\rho ~=~ R_\rho ~=~ S_\rho ~=~ O(a^2)
\\
Q_\xi &=& Q_\rho ~=~ \! \! \left[ \frac{3 \sqrt{C_A^3} \gamma^2}{4} \tan^{-1} 
\left[ \frac{\sqrt{C_A}\gamma^2}{p^2} \right] ~-~ 
\frac{3 \pi \sqrt{C_A^3} \gamma^2}{8} ~+~ 
\frac{C_A^2\gamma^4}{8p^2} \ln \left[ 1 + \frac{(p^2)^2}{C_A\gamma^4} \right] 
\right. \nonumber \\
&& \left. ~~~~~~~~~~+~ \frac{5 C_A p^2}{4} ~-~ \frac{3 C_A p^2}{8} \ln 
\left[ \frac{C_A \gamma^4}{\mu^4} \right] ~-~ 
\frac{\sqrt{C_A}(p^2)^2}{4\gamma^2} \tan^{-1} \left[ 
\frac{\sqrt{C_A}\gamma^2}{p^2} \right] \right] a \nonumber \\
&& ~~~~~~~~~+~ O(a^2) \\
W_\xi &=& \left[ \frac{\pi}{24} \sqrt{C_A} \gamma^2 ~-~ 
\frac{\sqrt{C_A}\gamma^2}{9} \tan^{-1} \left[ \frac{\sqrt{C_A}\gamma^2}{p^2} 
\right] \right. \nonumber \\
&& \left. ~-~ \frac{\sqrt{4C_A \gamma^4 - (p^2)^2}}{72} \tan^{-1} 
\left[ -~ \frac{\sqrt{4C_A \gamma^4 -(p^2)^2}}{p^2} \right] \right. 
\nonumber \\
&& \left. ~-~ \frac{\pi}{192} \frac{\sqrt{C_A^3}\gamma^6}{(p^2)^2} ~+~ 
\frac{1}{96} \frac{\sqrt{C_A^3}\gamma^6}{(p^2)^2} \tan^{-1} \left[ 
\frac{\sqrt{C_A}\gamma^2}{p^2} \right] ~+~ \left[ \frac{1}{96} 
- \frac{1}{36} \ln \left[ 1 + \frac{(p^2)^2}{C_A\gamma^4} \right] \right] 
\frac{C_A\gamma^4}{p^2} \right. \nonumber \\
&& \left. ~+~ \left[ -~ \frac{1}{96} - \frac{5}{96} \ln \left[
\frac{C_A\gamma^4}{\mu^4} \right] + \frac{5}{96} \ln \left[
\frac{[(p^2)^2+C_A\gamma^4]}{\mu^4} \right] - \frac{\sqrt{2}}{72} \eta_2(p^2) 
\right] p^2 \right. \nonumber \\
&& \left. ~+~ \frac{(p^2)^2}{288C_A\gamma^4} \sqrt{4C_A \gamma^4 - (p^2)^2} 
\tan^{-1} \left[ -~ \frac{\sqrt{4C_A \gamma^4 -(p^2)^2}}{p^2} \right] ~-~
\frac{\pi}{144} \frac{(p^2)^2}{\sqrt{C_A}\gamma^2} \right. \nonumber \\ 
&& \left. ~+~ \left[ \frac{13}{288} \tan^{-1} \left[ 
\frac{\sqrt{C_A}\gamma^2}{p^2} \right] - \frac{\sqrt{2}}{288} \eta_1(p^2) 
\right] \frac{(p^2)^2}{\sqrt{C_A}\gamma^2} \right. \nonumber \\
&& \left. ~+~ \left[ \frac{1}{576} \ln \left[ \frac{C_A\gamma^4}{\mu^4} \right]
- \frac{1}{288} \ln \left[ \frac{[(p^2)^2+C_A\gamma^4]}{\mu^4} \right]
+ \frac{1}{288} \ln \left[ \frac{p^2}{\mu^2} \right]
\right] \frac{(p^2)^3}{C_A\gamma^4} \right] a \nonumber \\
&& +~ O(a^2)  
\\
R_\xi &=& \left[ \frac{5\pi}{96} \sqrt{C_A} \gamma^2 ~-~ 
\frac{11\sqrt{C_A}\gamma^2}{72} \tan^{-1} \left[ \frac{\sqrt{C_A}\gamma^2}{p^2} 
\right] \right. \nonumber \\
&& \left. ~-~ \frac{\sqrt{4C_A \gamma^4 - (p^2)^2}}{72} \tan^{-1} 
\left[ -~ \frac{\sqrt{4C_A \gamma^4 -(p^2)^2}}{p^2} \right] \right. 
\nonumber \\
&& \left. ~+~ \frac{C_A\gamma^4\sqrt{4C_A \gamma^4 - (p^2)^2}}{12(p^2)^2} 
\tan^{-1} \left[ -~ \frac{\sqrt{4C_A \gamma^4 -(p^2)^2}}{p^2} \right] ~+~ 
\frac{13\pi}{192} \frac{\sqrt{C_A^3}\gamma^6}{(p^2)^2} 
\right. \nonumber \\
&& \left. ~+~ \frac{1}{32} \frac{\sqrt{C_A^3}\gamma^6}{(p^2)^2} \tan^{-1} 
\left[ \frac{\sqrt{C_A}\gamma^2}{p^2} \right] ~-~ \left[ \frac{5}{96} 
+ \frac{17}{288} \ln \left[ 1 + \frac{(p^2)^2}{C_A\gamma^4} \right] \right] 
\frac{C_A\gamma^4}{p^2} \right. \nonumber \\
&& \left. ~+~ \left[ -~ \frac{1}{96} - \frac{1}{32} \ln \left[
\frac{C_A\gamma^4}{\mu^4} \right] + \frac{1}{32} \ln \left[
\frac{[(p^2)^2+C_A\gamma^4]}{\mu^4} \right] - \frac{\sqrt{2}}{72} \eta_2(p^2) 
\right] p^2 \right. \nonumber \\
&& \left. ~-~ \frac{(p^2)^2}{576C_A\gamma^4} \sqrt{4C_A \gamma^4 - (p^2)^2} 
\tan^{-1} \left[ -~ \frac{\sqrt{4C_A \gamma^4 -(p^2)^2}}{p^2} \right] ~+~
\frac{5\pi}{576} \frac{(p^2)^2}{\sqrt{C_A}\gamma^2} \right. \nonumber \\ 
&& \left. ~+~ \left[ -~ \frac{5}{288} \tan^{-1} \left[ 
\frac{\sqrt{C_A}\gamma^2}{p^2} \right] + \frac{\sqrt{2}}{576} \eta_1(p^2) 
\right] \frac{(p^2)^2}{\sqrt{C_A}\gamma^2} \right. \nonumber \\
&& \left. ~+~ \left[ -~ \frac{1}{288} \ln \left[ \frac{C_A\gamma^4}{\mu^4} 
\right] + \frac{1}{144} \ln \left[ \frac{[(p^2)^2+C_A\gamma^4]}{\mu^4} \right]
\right. \right. \nonumber \\
&& \left. \left. ~~~~~~~-~ \frac{1}{144} \ln \left[ \frac{p^2}{\mu^2} \right] 
- \frac{\sqrt{2}}{768} \eta_2(p^2) \right] \frac{(p^2)^3}{C_A\gamma^4} \right] 
a ~+~ O(a^2)  
\end{eqnarray} 
and 
\begin{eqnarray}
S_\xi &=& \left[ \frac{\pi\gamma^2}{4\sqrt{C_A}} ~-~ 
\frac{2\gamma^2}{3\sqrt{C_A}} \tan^{-1} \left[ \frac{\sqrt{C_A}\gamma^2}{p^2} 
\right] ~-~ \frac{\sqrt{4C_A \gamma^4 - (p^2)^2}}{12C_A} \tan^{-1} 
\left[ -~ \frac{\sqrt{4C_A \gamma^4 -(p^2)^2}}{p^2} \right] \right. 
\nonumber \\
&& \left. ~-~ \frac{\pi}{32} \frac{\sqrt{C_A}\gamma^6}{(p^2)^2} ~+~ 
\frac{1}{16} \frac{\sqrt{C_A}\gamma^6}{(p^2)^2} \tan^{-1} \left[ 
\frac{\sqrt{C_A}\gamma^2}{p^2} \right] ~+~ \left[ \frac{1}{16} 
- \frac{1}{6} \ln \left[ 1 + \frac{(p^2)^2}{C_A\gamma^4} \right] \right] 
\frac{\gamma^4}{p^2} \right. \nonumber \\
&& \left. ~+~ \left[ -~ \frac{1}{16} - \frac{5}{16} \ln \left[
\frac{C_A\gamma^4}{\mu^4} \right] + \frac{5}{16} \ln \left[
\frac{[(p^2)^2+C_A\gamma^4]}{\mu^4} \right] - \frac{\sqrt{2}}{12} \eta_2(p^2) 
\right] \frac{p^2}{C_A} \right. \nonumber \\
&& \left. ~+~ \frac{(p^2)^2}{48C_A^2\gamma^4} \sqrt{4C_A \gamma^4 - (p^2)^2} 
\tan^{-1} \left[ -~ \frac{\sqrt{4C_A \gamma^4 -(p^2)^2}}{p^2} \right] ~-~
\frac{\pi}{24} \frac{(p^2)^2}{\sqrt{C_A^3}\gamma^2} \right. \nonumber \\ 
&& \left. ~+~ \left[ \frac{13}{48} \tan^{-1} \left[ 
\frac{\sqrt{C_A}\gamma^2}{p^2} \right] - \frac{\sqrt{2}}{48} \eta_1(p^2) 
\right] \frac{(p^2)^2}{\sqrt{C_A^3}\gamma^2} \right. \nonumber \\
&& \left. ~+~ \left[ \frac{1}{96} \ln \left[ \frac{C_A\gamma^4}{\mu^4} 
\right] - \frac{1}{48} \ln \left[ \frac{[(p^2)^2+C_A\gamma^4]}{\mu^4} \right]
+ \frac{1}{48} \ln \left[ \frac{p^2}{\mu^2} \right] 
\right] \frac{(p^2)^3}{C_A^2\gamma^4} \right] a ~+~ O(a^2) \,.
\end{eqnarray} 
Taking the zero momentum limit of each expression we find 
\begin{eqnarray}
X &=& \left[ \left[ \frac{101}{72} - \frac{121}{128} \ln \left[ 
\frac{C_A\gamma^4}{\mu^4} \right] - \frac{53}{192} \ln \left[ \frac{p^2}{\mu^2} 
\right] \right] C_A p^2 ~-~ \frac{69\pi}{128} \sqrt{C_A^3} \gamma^2 
\right. \nonumber \\
&& \left. ~+~ \left[ \frac{4}{3} \ln \left[ \frac{p^2}{\mu^2} \right] 
- \frac{20}{9} \right] T_F \Nf p^2 ~+~ O\left((p^2)^2\right) \right] a ~+~ 
O(a^2) \nonumber \\
U &=& i \left[ -~ \frac{31\pi}{192} \sqrt{C_A} p^2 ~+~ O\left((p^2)^2\right) 
\right] \gamma^2 a ~+~ O(a^2) \nonumber \\
V &=& W_\rho ~=~ R_\rho ~=~ S_\rho ~=~ O(a^2) \nonumber \\
Q_\xi &=& Q_\rho ~=~ \left[ \left[ \frac{5}{8} ~-~ \frac{3}{8} \ln \left[ 
\frac{C_A\gamma^4}{\mu^4} \right] \right] C_A p^2 ~+~ O\left((p^2)^2\right) 
\right] a ~+~ O(a^2) \nonumber \\
W_\xi &=& \left[ -~ \frac{7p^2}{144} ~+~ O\left((p^2)^2\right) \right] a ~+~ 
O(a^2) ~~~,~~~
R_\xi ~=~ \left[ -~ \frac{11p^2}{288} ~+~ O\left((p^2)^2\right) \right] a ~+~ 
O(a^2) \nonumber \\
S_\xi &=& \left[ -~ \frac{7p^2}{24C_A} ~+~ O\left((p^2)^2\right) \right] a ~+~ 
O(a^2) ~. 
\end{eqnarray}

\sect{Longitudinal parts.}

In this appendix we provide the one loop $\MSbar$ expressions for the
longitudinal parts of the $2$-point functions of (\ref{twoptdef}). We find 
\begin{eqnarray}
X^L &=& \left[ -~ \frac{69}{64} \sqrt{C_A^3} \gamma^2 \tan^{-1} \left[ 
\frac{\sqrt{C_A}\gamma^2}{p^2} \right]
~+~ \frac{9 C_A \sqrt{4C_A \gamma^4 - (p^2)^2}}{64} \tan^{-1} 
\left[ -~ \frac{\sqrt{4C_A \gamma^4 -(p^2)^2}}{p^2} \right] \right. 
\nonumber \\
&& \left. ~+~ \frac{9C_A^2\gamma^4}{8(p^2)^2} \sqrt{4C_A\gamma^4-(p^2)^2}
\tan^{-1} \left[ - \frac{\sqrt{4C_A\gamma^4-(p^2)^2}}{p^2} \right] ~+~ 
\frac{177\pi}{128} \frac{\sqrt{C_A^5}\gamma^6}{(p^2)^2} \right. \nonumber \\
&& \left. ~-~ \frac{33}{64} \frac{\sqrt{C_A^5}\gamma^6}{(p^2)^2} \tan^{-1} 
\left[ \frac{\sqrt{C_A}\gamma^2}{p^2} \right] ~+~ \left[ -~ \frac{105}{64} 
+ \frac{9}{64} \ln \left[ 1 + \frac{(p^2)^2}{C_A\gamma^4} \right] \right] 
\frac{C_A^2\gamma^4}{p^2} \right. \nonumber \\
&& \left. ~+~ \left[ \frac{9}{128} \ln \left[
\frac{C_A\gamma^4}{\mu^4} \right] + \frac{27}{64} \ln \left[
\frac{[(p^2)^2+C_A\gamma^4]}{\mu^4} \right] - \frac{63}{64} \ln \left[ 
\frac{p^2}{\mu^2} \right] \right] C_A p^2 \right] a ~+~ O(a^2)  
\\
U^L &=& i \left[ C_A \left[ \frac{3}{32} - \frac{15}{64} \ln \left[ 1 
+ \frac{(p^2)^2}{C_A\gamma^4} \right] \right] ~-~
\frac{C_A^2\gamma^4}{32(p^2)^2} \ln \left[ 1 + \frac{(p^2)^2}{C_A\gamma^4} 
\right] ~+~ \frac{\pi\sqrt{C_A^3}\gamma^2}{32p^2} 
\right. \nonumber \\
&& \left. ~~~+~ \frac{5\sqrt{C_A^3}\gamma^2}{16p^2} \tan^{-1} \left[
\frac{\sqrt{C_A}\gamma^2}{p^2} \right] \,+\, 
\frac{3 C_A \sqrt{4C_A \gamma^4 - (p^2)^2}}{16p^2} \tan^{-1} 
\left[ -~ \frac{\sqrt{4C_A \gamma^4 -(p^2)^2}}{p^2} \right] \right. 
\nonumber \\
&& \left. ~~~-~ \frac{3 p^2 \sqrt{4C_A \gamma^4 - (p^2)^2}}{64\gamma^4} 
\tan^{-1} \left[ -~ \frac{\sqrt{4C_A \gamma^4 -(p^2)^2}}{p^2} \right] ~-~ 
\frac{\sqrt{C_A}p^2}{4\gamma^2} \tan^{-1} \left[ \frac{\sqrt{C_A}\gamma^2}{p^2}
\right] \right. \nonumber \\ 
&& \left. ~~~+~ \left[ \frac{1}{64} \ln \left[ 
\frac{[(p^2)^2+C_A\gamma^4]}{\mu^4} \right] - \frac{3}{128} \ln \left[
\frac{C_A\gamma^4}{\mu^4} \right] + \frac{1}{64} \ln \left[ \frac{p^2}{\mu^2} 
\right] \right] \frac{(p^2)^2}{\gamma^4} \right] \gamma^2 a \nonumber \\
&& +~ O(a^2) 
\\
V^L &=& i \left[ C_A \left[ \frac{3}{32} \ln \left[ 1 
+ \frac{(p^2)^2}{C_A\gamma^4} \right] - \frac{1}{16} \right] ~+~
\frac{\pi\sqrt{C_A^3}\gamma^2}{32p^2} 
\right. \nonumber \\
&& \left. ~~~-~ \frac{\sqrt{C_A^3}\gamma^2}{16p^2} \tan^{-1} \left[
\frac{\sqrt{C_A}\gamma^2}{p^2} \right] ~+~ \frac{3\sqrt{C_A}p^2}{16\gamma^2} 
\tan^{-1} \left[ \frac{\sqrt{C_A}\gamma^2}{p^2} \right] \right. \nonumber \\ 
&& \left. ~~~+~ \left[ -~ \frac{1}{32} \ln \left[ 
\frac{[(p^2)^2+C_A\gamma^4]}{\mu^4} \right] + \frac{1}{16} \ln \left[ 
\frac{p^2}{\mu^2} \right] \right] \frac{(p^2)^2}{\gamma^4} \right] 
\gamma^2 a ~+~ O(a^2) 
\\
W^L_\rho &=& R^L_\rho ~=~ S^L_\rho ~=~ O(a^2)
\\
Q^L_\xi &=& Q^L_\rho ~=~ \left[ \frac{3 \sqrt{C_A^3} \gamma^2}{4} 
\tan^{-1} \left[ \frac{\sqrt{C_A}\gamma^2}{p^2} \right] ~-~ 
\frac{3 \pi \sqrt{C_A^3} \gamma^2}{8} ~+~ 
\frac{C_A^2\gamma^4}{8p^2} \ln \left[ 1 + \frac{(p^2)^2}{C_A\gamma^4} 
\right] \right. \nonumber \\
&& \left. ~~~~~~~~~~~+~ \frac{5 C_A p^2}{4} ~-~ \frac{3 C_A p^2}{8} \ln \left[ 
\frac{C_A \gamma^4}{\mu^4} \right] ~-~ \frac{\sqrt{C_A}(p^2)^2}{4\gamma^2} 
\tan^{-1} \left[ \frac{\sqrt{C_A}\gamma^2}{p^2} \right] \right] a \nonumber \\
&& ~~~~~~~~~~+~ O(a^2) 
\\
W^L_\xi &=& \left[ \frac{5\pi}{48} \sqrt{C_A} \gamma^2 ~-~ 
\frac{5\sqrt{C_A}\gamma^2}{24} \tan^{-1} \left[ \frac{\sqrt{C_A}\gamma^2}{p^2} 
\right] ~+~ \frac{\pi}{64} \frac{\sqrt{C_A^3}\gamma^6}{(p^2)^2} \right.
\nonumber \\
&& \left. ~-~ \frac{1}{32} \frac{\sqrt{C_A^3}\gamma^6}{(p^2)^2} \tan^{-1} 
\left[ \frac{\sqrt{C_A}\gamma^2}{p^2} \right] ~-~ \frac{1}{32} 
\frac{C_A\gamma^4}{p^2} \right. \nonumber \\
&& \left. ~+~ \left[ \frac{1}{32} - \frac{5}{32} \ln \left[
\frac{C_A\gamma^4}{\mu^4} \right] + \frac{5}{32} \ln \left[
\frac{[(p^2)^2+C_A\gamma^4]}{\mu^4} \right] - \frac{\sqrt{2}}{24} \eta_2(p^2) 
\right] p^2 \right. \nonumber \\
&& \left. ~-~ \frac{5\pi}{96} \frac{(p^2)^2}{\sqrt{C_A}\gamma^2} ~+~ \left[ 
\frac{5}{32} \tan^{-1} \left[ \frac{\sqrt{C_A}\gamma^2}{p^2} \right] 
- \frac{\sqrt{2}}{48} \eta_1(p^2) \right] \frac{(p^2)^2}{\sqrt{C_A}\gamma^2} 
\right. \nonumber \\
&& \left. ~+~ \left[ \frac{1}{192} \ln \left[ \frac{C_A\gamma^4}{\mu^4} \right]
- \frac{1}{96} \ln \left[ \frac{[(p^2)^2+C_A\gamma^4]}{\mu^4} \right]
+ \frac{1}{96} \ln \left[ \frac{p^2}{\mu^2} \right] + \frac{\sqrt{2}}{384}
\eta_2(p^2) \right] \frac{(p^2)^3}{C_A\gamma^4} \right] a \nonumber \\
&& +~ O(a^2)  
\\
R^L_\xi &=& \left[ -~ \frac{5\pi}{96} \sqrt{C_A} \gamma^2 ~+~ 
\frac{5\sqrt{C_A}\gamma^2}{12} \tan^{-1} \left[ \frac{\sqrt{C_A}\gamma^2}{p^2} 
\right] \right. \nonumber \\
&& \left. ~+~ \left[ \frac{1}{8} - \frac{C_A\gamma^4}{4(p^2)^2} \right]
\sqrt{4C_A \gamma^4 - (p^2)^2} \tan^{-1} 
\left[ -~ \frac{\sqrt{4C_A \gamma^4 -(p^2)^2}}{p^2} \right] ~-~ 
\frac{13\pi}{64} \frac{\sqrt{C_A^3}\gamma^6}{(p^2)^2} 
\right. \nonumber \\
&& \left. ~-~ \frac{3}{32} \frac{\sqrt{C_A^3}\gamma^6}{(p^2)^2} \tan^{-1} 
\left[ \frac{\sqrt{C_A}\gamma^2}{p^2} \right] ~+~ \left[ \frac{5}{32} 
+ \frac{5}{32} \ln \left[ 1 + \frac{(p^2)^2}{C_A\gamma^4} \right] \right] 
\frac{C_A\gamma^4}{p^2} \right. \nonumber \\
&& \left. ~+~ \left[ \frac{1}{32} + \frac{5}{32} \ln \left[
\frac{C_A\gamma^4}{\mu^4} \right] - \frac{5}{32} \ln \left[
\frac{[(p^2)^2+C_A\gamma^4]}{\mu^4} \right] + \frac{\sqrt{2}}{48} \eta_2(p^2) 
\right] p^2 \right. \nonumber \\
&& \left. ~-~ \frac{(p^2)^2}{64C_A\gamma^4} \sqrt{4C_A \gamma^4 - (p^2)^2} 
\tan^{-1} \left[ -~ \frac{\sqrt{4C_A \gamma^4 -(p^2)^2}}{p^2} \right] ~+~
\frac{5\pi}{192} \frac{(p^2)^2}{\sqrt{C_A}\gamma^2} \right. \nonumber \\ 
&& \left. ~+~ \left[ -~ \frac{5}{32} \tan^{-1} \left[ 
\frac{\sqrt{C_A}\gamma^2}{p^2} \right] + \frac{\sqrt{2}}{96} \eta_1(p^2) 
\right] \frac{(p^2)^2}{\sqrt{C_A}\gamma^2} \right. \nonumber \\
&& \left. ~+~ \left[ -~ \frac{1}{96} \ln \left[ \frac{C_A\gamma^4}{\mu^4} 
\right] + \frac{1}{48} \ln \left[ \frac{[(p^2)^2+C_A\gamma^4]}{\mu^4} \right]
\right. \right. \nonumber \\
&& \left. \left. ~~~~~~~-~ \frac{1}{48} \ln \left[ \frac{p^2}{\mu^2} \right] 
- \frac{\sqrt{2}}{768} \eta_2(p^2) \right] \frac{(p^2)^3}{C_A\gamma^4} \right] 
a ~+~ O(a^2)  
\end{eqnarray} 
and 
\begin{eqnarray}
S^L_\xi &=& \left[ \frac{5\pi\gamma^2}{8\sqrt{C_A}} ~-~ 
\frac{5\gamma^2}{4\sqrt{C_A}} \tan^{-1} \left[ \frac{\sqrt{C_A}\gamma^2}{p^2} 
\right] \right. \nonumber \\
&& \left. ~+~ \frac{3\pi}{32} \frac{\sqrt{C_A}\gamma^6}{(p^2)^2} ~-~ 
\frac{3}{16} \frac{\sqrt{C_A}\gamma^6}{(p^2)^2} \tan^{-1} \left[ 
\frac{\sqrt{C_A}\gamma^2}{p^2} \right] ~-~ \frac{3}{16} \frac{\gamma^4}{p^2} 
\right. \nonumber \\
&& \left. ~+~ \left[ \frac{3}{16} - \frac{15}{16} \ln \left[
\frac{C_A\gamma^4}{\mu^4} \right] + \frac{15}{16} \ln \left[
\frac{[(p^2)^2+C_A\gamma^4]}{\mu^4} \right] - \frac{\sqrt{2}}{4} \eta_2(p^2) 
\right] \frac{p^2}{C_A} \right. \nonumber \\
&& \left. ~-~ \frac{5\pi}{16} \frac{(p^2)^2}{\sqrt{C_A^3}\gamma^2} ~+~ \left[ 
\frac{15}{16} \tan^{-1} \left[ \frac{\sqrt{C_A}\gamma^2}{p^2} \right] 
- \frac{\sqrt{2}}{8} \eta_1(p^2) \right] \frac{(p^2)^2}{\sqrt{C_A^3}\gamma^2} 
\right. \nonumber \\
&& \left. ~+~ \left[ \frac{1}{32} \ln \left[ \frac{C_A\gamma^4}{\mu^4} 
\right] - \frac{1}{16} \ln \left[ \frac{[(p^2)^2+C_A\gamma^4]}{\mu^4} \right]
+ \frac{1}{16} \ln \left[ \frac{p^2}{\mu^2} \right] ~+~ \frac{\sqrt{2}}{64} 
\eta_2(p^2) \right] \frac{(p^2)^3}{C_A^2\gamma^4} \right] a \nonumber \\
&& +~ O(a^2) ~.
\end{eqnarray} 
As in the previous appendix, taking the zero momentum limit of each expression 
we find 
\begin{eqnarray}
X^L &=& \left[ \left[ \frac{35}{32} + \frac{63}{128} \ln \left[ 
\frac{C_A\gamma^4}{\mu^4} \right] - \frac{63}{64} \ln \left[ \frac{p^2}{\mu^2} 
\right] \right] C_A p^2 \,-\, \frac{69\pi}{128} \sqrt{C_A^3} \gamma^2 \,+\, 
O\left((p^2)^2\right) \right] a ~+~ O(a^2) \nonumber \\
U^L &=& i \left[ -~ \frac{7\pi}{128} \sqrt{C_A} p^2 ~+~ O\left((p^2)^2\right) 
\right] \gamma^2 a ~+~ O(a^2) \nonumber \\
V^L &=& i \left[ \frac{3\pi}{32} \sqrt{C_A} p^2 ~+~ O\left((p^2)^2\right) 
\right] \gamma^2 a ~+~ O(a^2) \nonumber \\
W^L_\rho &=& R^L_\rho ~=~ S^L_\rho ~=~ O(a^2) \nonumber \\
Q^L_\xi &=& Q^L_\rho ~=~ \left[ \left[ \frac{5}{8} ~-~ \frac{3}{8} \ln \left[ 
\frac{C_A\gamma^4}{\mu^4} \right] \right] C_A p^2 ~+~ O\left((p^2)^2\right) 
\right] a ~+~ O(a^2) \nonumber \\
W^L_\xi &=& \left[ -~ \frac{5p^2}{48} ~+~ O\left((p^2)^2\right) \right] a ~+~ 
O(a^2) ~~~,~~~
R^L_\xi ~=~ \left[ \frac{5p^2}{96} ~+~ O\left((p^2)^2\right) \right] a ~+~ 
O(a^2) \nonumber \\
S^L_\xi &=& \left[ -~ \frac{5p^2}{8C_A} ~+~ O\left((p^2)^2\right) \right] a ~+~ 
O(a^2) ~. 
\end{eqnarray}
Interestingly the $O(\gamma^2)$ part of $X^L$ is equivalent to the 
$O(\gamma^2)$ term of $X$ in the same limit. 

\sect{Tensor operator correlation function.}

In this appendix we record the explicit form of the correlation function of the
Lorentz tensor operator ${\cal O}_T^{\mu\nu}$ which is defined by 
\begin{equation}
\Pi^{\{\mu\nu|\sigma\rho\}}_T(p^2) ~=~ (4\pi)^2 i \int d^4 x \, e^{ipx} 
\langle 0 | {\cal O}^{\mu\nu}_T(x) {\cal O}^{\sigma\rho}_T(0) | 0 \rangle ~. 
\end{equation}
We decompose this into a similar basis of Lorentz tensors to that used in
\cite{82} with the only difference being that we work completely in 
$d$-dimensions and not four dimensions, as we use dimensional regularization.
We have 
\begin{eqnarray}
\Pi^{\{\mu\nu|\sigma\rho\}}_T(p^2) &=& 
\left[ \eta_{\mu\sigma} \eta_{\nu\rho} + \eta_{\mu\rho} \eta_{\nu\sigma} 
- \frac{2}{d} \eta_{\mu\nu} \eta_{\sigma\rho} \right] \Pi^T_{1}(p^2) 
\nonumber \\
&& +~ \left[ \frac{4}{d^2} \eta_{\mu\nu} \eta_{\sigma\rho} 
+ \left( \eta_{\mu\sigma} \frac{p_\nu p_\rho}{p^2} 
+ \eta_{\nu\rho} \frac{p_\mu p_\sigma}{p^2} 
+ \eta_{\mu\rho} \frac{p_\nu p_\sigma}{p^2} 
+ \eta_{\nu\sigma} \frac{p_\mu p_\rho}{p^2} \right) \right. \nonumber \\
&& \left. ~~~~-~ \frac{4}{d} \left( \eta_{\mu\nu} \frac{p_\sigma p_\rho}{p^2}
+ \eta_{\sigma\rho} \frac{p_\mu p_\nu}{p^2} \right) \right] \Pi^T_{2}(p^2) 
\nonumber \\
&& +~ \left[ \frac{1}{d^2} \eta_{\mu\nu} \eta_{\sigma\rho} 
- \frac{1}{d} \left( \eta_{\mu\nu} \frac{p_\sigma p_\rho}{p^2}
+ \eta_{\sigma\rho} \frac{p_\mu p_\nu}{p^2} \right) 
+ \frac{p_\mu p_\nu p_\sigma p_\rho}{(p^2)^2} \right] \Pi^T_{3}(p^2) ~. 
\label{tenscordef}
\end{eqnarray}
To determine the scalar amplitudes $\Pi^T_i(p^2)$ we use a projection method.
This involves multiplying the original correlator by a Lorentz tensor which is
a linear combination of the three basis Lorentz tensors defining the 
decomposition, (\ref{tenscordef}). The construction of each of the three
projection tensors is similar to that used in appendix A for the decomposition
of colour group tensors. We first construct a matrix where the entries are 
determined by multiplying (\ref{tenscordef}) by each basis tensor in turn. This
produces the $d$-dependent matrix 
\begin{equation}
P_T ~=~ \left(
\begin{array}{ccc}
2 (d-1)(d+2) & \frac{4}{d} (d-1)(d+2) & \frac{2}{d} (d-1) \\
\frac{4}{d} (d-1)(d+2) & \frac{4}{d^2} (d-1) (d^2+4d-4) & 
\frac{4}{d^2} (d-1)^2 \\
\frac{2}{d} (d-1) & \frac{4}{d^2} (d-1)^2 & \frac{(d-1)^2}{d^2} \\ 
\end{array}
\right) ~.
\end{equation}
The coefficients of the basis tensors in each of the three projectors are then 
determined from the inverse which is 
\begin{equation}
P^{-1}_T ~=~ \frac{1}{(d^2-1)(d-2)} 
\left(
\begin{array}{ccc}
\frac{1}{2} (d-1) & - \frac{1}{2} (d-1) & (d-2) \\
- \frac{1}{2} (d-1) & \frac{1}{4} (d^2+d-4) & - (d^2-4) \\
(d-2) & - (d^2-4) & (d^2-4) (d+4) \\ 
\end{array}
\right) ~. 
\end{equation}
Equipped with this we find the three scalar amplitudes are
\begin{eqnarray}
\Pi^T_1(p^2) &=& \! \! \left[ \frac{(p^2)^2}{160\epsilon} ~-~ 
\frac{C_A \gamma^4}{12\epsilon} ~+~ \frac{C_A^2 \gamma^8 
\sqrt{4C_A \gamma^4 - (p^2)^2}}{120 (p^2)^3} \tan^{-1} \left[ 
\frac{\sqrt{4 C_A \gamma^4 - (p^2)^2}}{p^2} \right] \right. \nonumber \\
&& \left. -~ \frac{\pi C_A^{5/2} \gamma^{10}}{120(p^2)^3} ~+~ 
\frac{C_A^2 \gamma^8}{120(p^2)^2} ~+~ \frac{C_A \gamma^4 
\sqrt{4C_A \gamma^4 - (p^2)^2}}{60 p^2} \tan^{-1} \left[ 
\frac{\sqrt{4 C_A \gamma^4 - (p^2)^2}}{p^2} \right] \right. \nonumber \\
&& \left. -~ \frac{\pi C_A^{3/2}\gamma^6}{64 p^2} ~+~ \frac{p^2}{320} 
\sqrt{4C_A \gamma^4 - (p^2)^2} \tan^{-1} \left[ 
\frac{\sqrt{4 C_A \gamma^4 - (p^2)^2}}{p^2} \right] \right. \nonumber \\
&& \left. +~ \left[ \frac{1}{24} \ln \left[ \frac{C_A \gamma^4}{\mu^4} 
\right] ~-~ \frac{71}{1440} ~+~ \frac{3\sqrt{2}}{640} \eta_2(p^2) \right] C_A 
\gamma^4 \right. \nonumber \\
&& \left. +~ \left[ \frac{\pi}{128} ~+~ \frac{3\sqrt{2}}{1280} \eta_1(p^2)  
\right] \sqrt{C_A} \gamma^2 p^2 \right. \nonumber \\
&& \left. +~ \left[ \frac{9}{1600} ~-~ \frac{1}{320} \ln \left[ 
\frac{C_A \gamma^4}{\mu^2} \right] ~-~ \frac{\sqrt{2}}{1280} \eta_2(p^2)  
\right] (p^2)^2 \right] \NA ~+~ O(a) 
\end{eqnarray}
\begin{eqnarray}
\Pi^T_2(p^2) &=& \! \! \left[ -~ \frac{(p^2)^2}{160\epsilon} ~-~ 
\frac{C_A^2 \gamma^8 \sqrt{4C_A \gamma^4 - (p^2)^2}}{20 (p^2)^3} \tan^{-1} 
\left[ \frac{\sqrt{4 C_A \gamma^4 - (p^2)^2}}{p^2} \right] \right. \nonumber \\
&& \left. +~ \frac{\pi C_A^{5/2} \gamma^{10}}{20(p^2)^3} ~-~ 
\frac{C_A^2 \gamma^8}{20(p^2)^2} ~-~ \frac{C_A \gamma^4 
\sqrt{4C_A \gamma^4 - (p^2)^2}}{160 p^2} \tan^{-1} \left[ 
\frac{\sqrt{4 C_A \gamma^4 - (p^2)^2}}{p^2} \right] \right. \nonumber \\
&& \left. -~ \frac{p^2}{320} \sqrt{4C_A \gamma^4 - (p^2)^2} \tan^{-1} 
\left[ \frac{\sqrt{4 C_A \gamma^4 - (p^2)^2}}{p^2} \right] \right. \nonumber \\
&& \left. +~ \left[ -~ \frac{1}{160} ~+~ \frac{\sqrt{2}}{1920} \eta_2(p^2)  
\right] C_A \gamma^4 ~-~ \left[ \frac{\pi}{192} ~+~ \frac{\sqrt{2}}{960} 
\eta_1(p^2) \right] \sqrt{C_A} \gamma^2 p^2 \right. \nonumber \\
&& \left. ~~~+~ \left[ -~ \frac{9}{1600} ~+~ \frac{1}{320} \ln \left[ 
\frac{C_A \gamma^4}{\mu^2} \right] ~+~ \frac{\sqrt{2}}{1280} \eta_2(p^2)  
\right] (p^2)^2 \right] \NA ~+~ O(a) 
\end{eqnarray}
and 
\begin{eqnarray}
\Pi^T_3(p^2) &=& \! \! \left[ \frac{(p^2)^2}{120\epsilon} ~+~ 
\frac{2C_A^2 \gamma^8 \sqrt{4C_A \gamma^4 - (p^2)^2}}{5 (p^2)^3} \tan^{-1} 
\left[ \frac{\sqrt{4 C_A \gamma^4 - (p^2)^2}}{p^2} \right] \right. \nonumber \\
&& \left. -~ \frac{2\pi C_A^{5/2} \gamma^{10}}{5(p^2)^3} ~+~ 
\frac{2C_A^2 \gamma^8}{5(p^2)^2} ~+~ \frac{C_A \gamma^4 
\sqrt{4C_A \gamma^4 - (p^2)^2}}{120 p^2} \tan^{-1} \left[ 
\frac{\sqrt{4 C_A \gamma^4 - (p^2)^2}}{p^2} \right] \right. \nonumber \\
&& \left. +~ \frac{\pi C_A^{3/2}\gamma^6}{24 p^2} ~+~ \frac{p^2}{240} 
\sqrt{4C_A \gamma^4 - (p^2)^2} \tan^{-1} \left[ 
\frac{\sqrt{4 C_A \gamma^4 - (p^2)^2}}{p^2} \right] \right. \nonumber \\
&& \left. +~ \left[ -~ \frac{3}{40} ~+~ \frac{\sqrt{2}}{160} \eta_2(p^2) 
\right] C_A \gamma^4 ~-~ \frac{\sqrt{2}}{480} \sqrt{C_A} \gamma^2 p^2 
\eta_1(p^2) \right. \nonumber \\
&& \left. +~ \left[ \frac{17}{3600} ~-~ \frac{1}{240} \ln \left[ 
\frac{C_A \gamma^4}{\mu^2} \right] ~-~ \frac{\sqrt{2}}{960} \eta_2(p^2) \right]
(p^2)^2 \right] \NA ~+~ O(a) ~.
\end{eqnarray}
Again we have included the divergent parts of the correlation functions which
are absorbed by a contact renormalization. We note, though, that there is only
mixing for the first amplitude. Aside from several terms involving the factors 
$1/(p^2)^2$ and $1/(p^2)^3$ the actual functions appearing in each of the 
amplitudes are the same as for $\Pi_S(p^2)$. So, for example, the cut structure
is the same with a physical cut at $p^2$~$=$~$2 \sqrt{C_A} \gamma^2$. Repeating
the same moment calculation for the scalar operator of section $7$, we find 
that
\begin{equation}
\Pi^T_1(M^2) ~=~ ( \mbox{parton model} ) \left[ 1 ~-~ 
\frac{10C_A \gamma^4}{3M^4} ~+~ O \left( \frac{\gamma^8}{M^8} \right) \right]
\end{equation}
and there are no $O \left( \frac{C_A \gamma^4}{M^4} \right)$ corrections for
the other two scalar amplitudes. Further, if one considered the correlation 
function of the energy momentum tensor, as in \cite{82}, then the same power
correction as $\Pi^T_1(M^2)$ would emerge.


\begin{thebibliography}{99}
\bibitem{1} D.J. Gross \& F.J. Wilczek, Phys. Rev. Lett. {\bf 30} (1973), 1343.
\bibitem{2} H.D. Politzer, Phys. Rev. Lett. {\bf 30} (1973), 1346.
\bibitem{3} S. Perantonis \& C. Michael, Nucl. Phys. {\bf B347} (1990), 854.
\bibitem{4} C. Michael, Phys. Lett. {\bf B283} (1992), 103. 
\bibitem{5} G.S. Bali, K. Schilling \& A. Wachter, Phys. Rev. {\bf D56} (1997),
2566.
\bibitem{6} S. Mandelstam, Phys. Rept. {\bf 67} (1980), 109.
\bibitem{7} R. Anishetty, M. Baker, J.S. Ball, S.K. Kim \& F. Zachariasen,
Phys. Lett. {\bf B86} (1979), 52. 
\bibitem{8} S. Mandelstam, Phys. Rev. {\bf D20} (1979), 3223.
\bibitem{9} G.B. West, Phys. Lett. {\bf B115} (1982), 468.
\bibitem{10} V.N. Gribov, Nucl. Phys. {\bf B139} (1978), 1. 
\bibitem{11} D. Zwanziger, Nucl. Phys. {\bf B209} (1982), 336. 
\bibitem{12} D. Zwanziger, Nucl. Phys. {\bf B321} (1989), 591. 
\bibitem{13} D. Zwanziger, Nucl. Phys. {\bf B323} (1989), 513. 
\bibitem{14} G. Dell'Antonio \& D. Zwanziger, Nucl. Phys. {\bf B326}
(1989), 333. 
\bibitem{15} G. Dell'Antonio \& D. Zwanziger, Commun. Math. Phys. {\bf 138}
(1991), 291. 
\bibitem{16} D. Zwanziger, Nucl. Phys. {\bf B364} (1991), 127. 
\bibitem{17} D. Zwanziger, Nucl. Phys. {\bf B378} (1992), 525. 
\bibitem{18} D. Zwanziger, Nucl. Phys. {\bf B399} (1993), 477. 
\bibitem{19} D. Zwanziger, Nucl. Phys. {\bf B412} (1994), 657. 
\bibitem{20} D. Zwanziger, Phys. Rev. {\bf D65} (2002), 094039.
\bibitem{21} D. Zwanziger, Phys. Rev. {\bf D69} (2004), 016002.
\bibitem{22} A. Cucchieri \& T. Mendes, PoS LAT2007 (2007), 297.
\bibitem{23} I.L. Bogolubsky, E.M. Ilgenfritz, M. M\"{u}ller-Preussker \& A. 
Sternbeck, PoS LAT2007 (2007), 290.
\bibitem{24} A. Cucchieri \& T. Mendes, Phys. Rev. Lett. {\bf 100} (2008),
241601. 
\bibitem{25} A. Cucchieri \& T. Mendes, Phys. Rev. {\bf D 78} (2008), 094503. 
\bibitem{26} O. Oliveira \& P.J. Silva, Phys. Rev. {\bf D79} (2009), 031501.
\bibitem{27} Ph. Boucaud, J.P. Leroy, A.L. Yaounac, J. Micheli, O. P\`{e}ne \&
J. Rodr\'{\i}guez-Quintero, JHEP {\bf 06} (2008), 099.
\bibitem{28} A.C. Aguilar, D. Binosi \& J. Papavassiliou, Phys. Rev. {\bf D78}
(2008), 025010. 
\bibitem{29} C.S. Fischer, A. Maas \& J.M. Pawlowski, Annals Phys. {\bf 324}
(2009), 2408.
\bibitem{30} N. Maggiore \& M. Schaden, Phys. Rev. {\bf D50} (1994), 6616.
\bibitem{31} D. Dudal, R.F. Sobreiro, S.P. Sorella \& H. Verschelde, Phys. Rev.
{\bf D72} (2005), 014016.
\bibitem{32} J.A. Gracey, Phys. Lett. {\bf B632} (2006), 282.
\bibitem{33} J.A. Gracey, JHEP {\bf 05} (2006), 052.
\bibitem{34} T. Kugo \& I. Ojima, Prog. Theor. Phys. Suppl. {\bf 66} (1979), 1;
Prog. Theor. Phys. Suppl. {\bf 77} (1984), 1121.
\bibitem{35} T. Kugo, hep-th/9511033. 
\bibitem{36} L. Susskind, ``Coarse Grained Quantum Chromodynamics'' in `Les 
Houches 1976, Proceedings Weak and Electromagnetic Interactions at High 
Energy', (North-Holland Publishing Company, Amsterdam, 1977), 207. 
\bibitem{37} W. Fischler, Nucl. Phys. {\bf B129} (1977), 157.
\bibitem{38} T. Appelquist, M. Dine \& I.J. Muzinich, Phys. Rev. {\bf D17} 
(1978), 2074.
\bibitem{39} M. Peter, Phys. Rev. Lett. {\bf 78} (1997), 602.
\bibitem{40} M. Peter, Nucl. Phys. {\bf B501} (1997), 471.
\bibitem{41} Y. Schr\"{o}der, ``The Static Potential in QCD'' 
DESY-THESIS-1999-021.
\bibitem{42} Y. Schr\"{o}der, Phys. Lett. {\bf B447} (1999), 321.
\bibitem{43} B.A. Kniehl, A.A. Penin, Y. Schr\"{o}der, V.A. Smirnov \& M. 
Steinhauser, Phys. Lett. {\bf B607} (2005), 96.
\bibitem{44} A.V. Smirnov, V.A. Smirnov \& M. Steinhauser, Phys. Lett. 
{\bf B668} (2008), 293.
\bibitem{45} A.V. Smirnov, V.A. Smirnov \& M. Steinhauser, PoS RADCOR2007 
(2007), 024.
\bibitem{46} A.V. Smirnov, V.A. Smirnov \& M. Steinhauser, Nucl. Phys. Proc.
Suppl. {\bf 183} (2008), 308.
\bibitem{47} C. Anzai, Y. Kiyo \& Y. Sumino, arXiv:0911.4335 [hep-ph].
\bibitem{48} A.V. Smirnov, V.A. Smirnov \& M. Steinhauser, arXiv:0911.4742
[hep-ph].
\bibitem{49} Y.M. Makeenko, Surveys High Energ. Phys. {\bf 10} (1997), 1.
\bibitem{50} D. Zwanziger, arXiv:0904.2380 [hep-th].
\bibitem{51} J.C. Taylor, Nucl. Phys. {\bf B33} (1971), 436.
\bibitem{52} D. Dudal, S.P. Sorella, N. Vandersickel \& H. Verschelde, Phys.
Rev. {\bf D77} (2008), 071501. 
\bibitem{53} D. Dudal, J.A. Gracey, S.P. Sorella, N. Vandersickel \& H. 
Verschelde, Phys. Rev. {\bf D78} (2008), 065047. 
\bibitem{54} J.G.M. Gatheral, Phys. Lett. {\bf B133} (1983), 90.
\bibitem{55} J. Frenkel \& J.C. Taylor, Nucl. Phys. {\bf B246} (1984), 231. 
\bibitem{56} S. Titard \& F.J. Yndurain, Phys. Rev. {\bf D49} (1994), 6007.
\bibitem{57} J.A.M. Vermaseren, math-ph/0010025.
\bibitem{58} R.F. Sobreiro \& S.P. Sorella, JHEP {\bf 06} (2005), 054.
\bibitem{59} P. Nogueira, J. Comput. Phys. {\bf 105} (1993), 279. 
\bibitem{60} S.A. Larin \& J.A.M. Vermaseren, Phys. Lett. {\bf B303} (1993),
334.
\bibitem{61} A.D. Dolgov, A. Lepidi \& G. Piccinelli, JCAP {\bf 02} (2009), 
027.
\bibitem{62} G. Gabadadze \& R.A. Rosen, JCAP {\bf 02} (2009), 016.
\bibitem{63} P. Gaete \& E. Spallucci, Phys. Lett. {\bf B675} (2009), 145.
\bibitem{64} G. Gabadadze \& D. Pirtskhalava, JCAP {\bf 05} (2009), 017.
\bibitem{65} A.D. Dolgov, A. Lepidi \& G. Piccinelli, Phys. Rev. {\bf D80} 
(2009), 125009.
\bibitem{66} M. L\"{u}scher, Nucl. Phys. {\bf B180} (1981), 317.
\bibitem{67} J.M. Cornwall \& A. Soni, Phys. Lett. {\bf B120} (1983), 431.
\bibitem{68} W. Buchm\"{u}ller, G. Grunberg \& S.-H. H. Tye, Phys. Rev. Lett.
{\bf 45} (1980), 103;
Phys. Rev. Lett. {\bf 45} (1980), 587.
\bibitem{69} S.J. Brodsky, G.P. Lepage \& P.B. Mackenzie, Phys. Rev. {\bf D28}
(1983), 228. 
\bibitem{70} W. Celmaster \& R.J. Gonsalves, Phys. Rev. {\bf D20} (1979), 1420.
\bibitem{71} E. Braaten \& J.P. Leveille, Phys. Rev. {\bf D24} (1981), 1369.
\bibitem{72} P. Boucaud, G. Burgio, F. di Renzo, J.P. Leroy, J. Micheli, C.
Parrinello, O. P\`{e}ne, C. Pittori, J. Rodr\'{\i}guez-Quintero, C. Roiesnel \&
K. Sharkey, JHEP {\bf 04} (2000), 006.
\bibitem{73} P. Boucaud, F. de Soto, J.P. Leroy, A. Le Yaounac, J. Micheli, H.
Moutarde, O. P\`{e}ne \& J. Rodr\'{\i}guez-Quintero, Phys. Rev. {\bf D74} 
(2006), 034505.
\bibitem{74} P. Boucaud, J.P. Leroy, A. Le Yaounac, A.Y. Lokhov, J. Micheli, O.
P\`{e}ne, J. Rodr\'{\i}guez-Quintero \& C. Roiesnel, JHEP {\bf 01} (2006), 037.
\bibitem{75} J.A. Gracey, ``Recent results for Yang-Mills theory restricted
to the Gribov region'' in `Path Integrals - New Trends and Perspectives', 
(World Scientific, Singapore, 2008), 167.
\bibitem{76} F.A. Berends, M. B\"{o}hm, M. Buza \& R. Scharf, Z. Phys. 
{\bf C63} (1994), 227.
\bibitem{77} D. Dudal, S.P. Sorella, N. Vandersickel \& H. Verschelde, Phys.
Rev. {\bf D79} (2009), 121701. 
\bibitem{78} T. van Ritbergen, A.N. Schellekens \& J.A.M. Vermaseren, Int. J.
Mod. Phys. {\bf A14} (1999), 41.
\bibitem{79} A. Soni \& M.D. Tran, Phys. Lett. {\bf B109} (1982), 393. 
\bibitem{80} R. Akhoury \& V.I. Zakharov, Phys. Lett. {\bf B438} (1998), 165.
\bibitem{81} S. Narison \& V.I. Zakharov, Phys. Lett. {\bf B679} (2009), 355.
\bibitem{82} K.G. Chetyrkin, S. Narison \& V.I. Zakharov, Nucl. Phys. {\bf
B550} (1999), 353.
\bibitem{83} S. Narison, Nucl. Phys. {\bf B509} (1998), 312.
\bibitem{84} F.R. Graziani, Z. Phys. {\bf C33} (1987), 397.
\bibitem{85} V.P. Spiridonov \& K.G. Chetyrkin, Sov. J. Nucl. Phys. {\bf 47}
(1988), 522. 
\bibitem{86} J. Liu \& W. Wetzel, hep-ph/9611250.
\bibitem{87} H. Verschelde, Phys. Lett. {\bf B351} (1995), 242.
\bibitem{88} H. Verschelde, S. Schelstraete \& M. Vanderkelen, Z. Phys.
{\bf C76} (1997), 161.
\bibitem{89} H. Verschelde, K. Knecht, K. van Acoleyen \& M. Vanderkelen, Phys.
Lett. {\bf B516} (2001), 307.
\bibitem{90} T. van Ritbergen, J.A.M. Vermaseren \& S.A. Larin, Phys. Lett.
{\bf B400} (1997), 379.
\bibitem{91} A.J. Macfarlane, A. Sudbery \& P.H. Weisz, Commun. Math. Phys.
{\bf 11} (1968), 77.
\end{thebibliography}
\end{document}